\documentclass[11pt]{article}
\usepackage[margin=1in,letterpaper]{geometry}
\usepackage{amsmath,amssymb,amsthm,mathtools}
\usepackage{enumitem}
\usepackage{microtype}
\usepackage{needspace}
\usepackage{mdframed}
\usepackage{etoolbox}
\PassOptionsToPackage{table}{xcolor}
\usepackage{tikz}
\usepackage{hyperref}
\hypersetup{
 colorlinks=true,linkcolor=blue,citecolor=blue,urlcolor=blue,
 linktoc=all,bookmarksnumbered=true,
 pdftitle={Classical Verification of Quantum Computation Beyond the Polynomial Hierarchy},
 pdfauthor={Atul Mantri}
}
\numberwithin{equation}{section}

\newtheorem{theorem}{Theorem}[section]
\newtheorem{lemma}[theorem]{Lemma}
\newtheorem{claim}[theorem]{Claim}
\newtheorem{corollary}[theorem]{Corollary}
\theoremstyle{definition}
\newtheorem{definition}[theorem]{Definition}
\newenvironment{protocolenv}[1][]{\refstepcounter{theorem}\par\noindent\textbf{Protocol \thetheorem} (#1)\textbf{.}\par\nobreak\vspace{-0.5\baselineskip}\noindent\rule{\textwidth}{0.4pt}\par\nobreak\vspace{0.3\baselineskip}\noindent\ignorespaces}{\par}
\theoremstyle{definition}
\theoremstyle{remark}

\newcommand{\F}{\mathbb F_2}
\newcommand{\E}{\mathbb E}
\newcommand{\BQP}{\mathsf{BQP}}
\newcommand{\EQP}{\mathsf{EQP}}
\newcommand{\PH}{\mathsf{PH}}
\newcommand{\PSPACE}{\mathsf{PSPACE}}
\newcommand{\RE}{\mathsf{RE}}
\newcommand{\MA}{\mathsf{MA}}
\newcommand{\AM}{\mathsf{AM}}

\newcommand{\NP}{\mathsf{NP}}
\newcommand{\BPP}{\mathsf{BPP}}
\newcommand{\Pclass}{\mathsf{P}}
\newcommand{\AC}{\mathsf{AC}}

\newcommand{\SigO}[1]{\Sigma_{#1}^{\mathsf p,O}}

\newcommand{\PiO}[1]{\Pi_{#1}^{\mathsf p,O}}
\newcommand{\poly}{\operatorname{poly}}
\newcommand{\dist}{\operatorname{dist}}
\newcommand{\width}{\operatorname{width}}
\newcommand{\ket}[1]{\lvert #1\rangle}
\newcommand{\bra}[1]{\langle #1\rvert}
\newcommand{\norm}[1]{\lVert #1\rVert}
\newcommand{\cert}{\mathrm C}
\newcommand{\calT}{\mathcal T}
\newcommand{\calA}{\mathcal A}
\newcommand{\calL}{\mathcal L}
\newcommand{\calD}{\mathcal D}

\surroundwithmdframed[
 topline=true,bottomline=true,leftline=false,rightline=false,
 linewidth=0.6pt,linecolor=black,
 innerleftmargin=0pt,innerrightmargin=0pt,
 innertopmargin=4pt,innerbottommargin=4pt,
 skipabove=\baselineskip,skipbelow=\baselineskip,
 needspace=10\baselineskip
]{protocolenv}
\AtBeginEnvironment{protocolenv}{\interlinepenalty=10000\relax}

\title{Classical Verification of Quantum Computation Beyond the Polynomial Hierarchy}
\author{Atul Mantri\thanks{atulmantri@vt.edu}\\ Department of Computer Science, Virginia Tech, USA 24061}
\date{}

\begin{document}
\hypersetup{pageanchor=false}
\maketitle
\thispagestyle{empty}
\begin{abstract}
It is not known whether a classical polynomial-time verifier can verify arbitrary polynomial-time quantum computations with an efficient quantum honest prover without computational assumptions. We show that, relative to a classical oracle, such verification is possible for a language beyond the polynomial hierarchy that is decided exactly in quantum polynomial time. The protocol is public-coin, with perfect completeness and soundness error at most $1/4$. To our knowledge, no language with such a proof system was previously proved to lie outside $\AM$ relative to an oracle.

For every unbounded polynomial-time computable function, the oracle can be chosen so that the message count and the oracle query counts of the quantum algorithm, the verifier, and the honest prover are bounded by this function of the input length. The message count is optimal up to a constant factor, since for a suitable oracle every interactive proof system for the language uses at least a constant fraction of the messages of our protocol on infinitely many lengths, with a constant depending on its verifier.

The language is a variant of recursive Fourier sampling in which code tables encode secrets that address data tables. The tables are evaluated recursively, and the oracle stores only the leaves. The quantum algorithm recovers every secret exactly, and the verifier checks the prover's secrets recursively against random code entries. The recursion restricts the input in a correlated way, as the same secret fixes the code bits and chooses the free coordinate, so H{\aa}stad's switching lemma does not apply. We prove a switching theorem for these restrictions, which may be of independent interest. For instances of sufficient height, no circuit of constant depth and size $2^{\poly(n)}$ changes its output with probability above $2^{-\Omega(n)}$ when the answer of a random instance is flipped by changing a single leaf.
\end{abstract}

\clearpage
\pagenumbering{roman}
\hypersetup{pageanchor=true}
\pdfbookmark[1]{Contents}{contents}
{\small\tableofcontents}
\clearpage
\pagenumbering{arabic}
\section{Introduction}\label{sec:intro}

Interactive proofs allow a polynomial-time verifier to be convinced of statements that it cannot check on its own, provided that the prover is powerful enough. With a computationally unbounded prover, every language in $\PSPACE$ has an interactive proof, as Shamir~\cite{Sha92} showed building on the work of Lund, Fortnow, Karloff, and Nisan~\cite{LFKN92}. In the verification of quantum computation there is a further constraint. The honest prover is the quantum computer that performs the computation, so its strategy must itself run in quantum polynomial time. The classical verifier wishes to be convinced of the outcome of that computation, with soundness against computationally unbounded provers.
\begin{center}
\emph{Can a classical polynomial-time verifier be convinced of the outcome of a polynomial-time quantum computation by the quantum computer that runs it, without computational assumptions?}
\end{center}

The question was raised by Gottesman in 2004 (see Aharonov and Vazirani~\cite{AV13} and Mahadev~\cite{Mah18}), and positive answers are known in three modified settings. If the verifier can prepare or measure a few qubits at a time, every problem in $\BQP$ can be verified with soundness against computationally unbounded provers~\cite{ABEM17,Bro18,FK17,HM15}. If the verifier is classical but interacts with two entangled provers that cannot communicate, every problem in $\BQP$ can be verified~\cite{RUV13}, and, when the honest provers may be computationally unbounded, every language in $\RE$~\cite{JNVWY21}. Gheorghiu, Kapourniotis, and Kashefi~\cite{GKK19} survey protocols of these two kinds. If the verifier is classical and there is a single quantum prover, Mahadev's protocol~\cite{Mah18} verifies every problem in $\BQP$ under the assumption that learning with errors~\cite{Reg09} is hard for quantum algorithms. In the remaining setting, in which the verifier is classical, there is a single prover, and no assumption is made, no protocol for all of $\BQP$ with a polynomial-time quantum honest prover is known. Such a protocol cannot relativize. Bouland, Huang, Natarajan, Shalit, and Tal~\cite{BHNST26} constructed an oracle relative to which Forrelation has no interactive proof system at all, even with a computationally unbounded prover. The protocols known in this setting verify particular problems, which we review in Section~\ref{sec:related}. It is natural to ask which quantum problems have such proof systems, and in particular whether, relative to an oracle, such a problem can lie outside the polynomial hierarchy.

In this paper we show that it can. We construct a classical oracle and a language that is decided exactly by a polynomial-time quantum algorithm with access to the oracle, lies outside the polynomial hierarchy relative to it, and has a classical interactive proof system whose verifier and quantum honest prover run in polynomial time, with perfect completeness and soundness against computationally unbounded provers. In the notation of Section~\ref{sec:oracle-classes}, the language lies in $\EQP^O\setminus\PH^O$ (Theorem~\ref{thm:main}), where $\EQP^O$, the exact counterpart of $\BQP^O$, is the class of languages decided exactly by uniform families of polynomial-size quantum circuits with access to $O$, which accept every string in the language with probability one and every other string with probability zero. To our knowledge, no problem with a classical proof system and an efficient quantum honest prover was previously proved to lie outside $\AM$ relative to an oracle. Every problem known to have such a proof system lies in $\AM$ or is excluded only from $\MA$ relative to an oracle, which raises the question whether such classical verification implies membership in the polynomial hierarchy. Theorem~\ref{thm:main} shows that no relativizing argument can establish this. In the oracle model, efficient classical verification is compatible with quantum computation beyond every fixed level of the polynomial hierarchy. The same construction allows the number of messages, and the number of oracle queries made by the quantum algorithm, the verifier, and the honest prover, to grow as slowly as desired with the input length (Corollaries~\ref{cor:logstar} and~\ref{cor:querybudget}). For a suitable choice of the oracle, this number of messages is also necessary up to a constant factor (Theorem~\ref{thm:interaction}).

The language is built from a variant of recursive Fourier sampling~\cite{BV97}. An instance is a tree of tables whose leaves are stored in the oracle, of height at most logarithmic in $n$ for the input $1^n$, and an instance of positive height has $2^{\Theta(n)}$ bits in total. The value of a node is the entry of its data table at a secret address, the secret is encoded in its code table, and the entries of both tables are the values of the node's children. A quantum algorithm decodes the secrets exactly, and a classical verifier checks them with the help of the prover, so that quantum evaluation and classical verification both take time polynomial in $n$. The recursion acts on the input as a random restriction, but one in which the values of the fixed code bits and the positions of the free leaves are determined by the same secrets, so the standard switching lemma does not apply. Instead of counting the restrictions that keep a term alive, we order the terms of a DNF and bound, by a rank argument over the secrets, how often the first satisfied term contains the free leaves of many nodes, and each switching step applies the projections of several levels of the recursion at once. We prove a switching theorem for these restrictions and use it to show that, once the height exceeds a constant depending on the circuit depth and the size exponent, no constant-depth circuit of size $2^{\poly(n)}$ distinguishes instances with answers zero and one with more than $2^{-\Omega(n)}$ advantage. By the standard translation of alternating polynomial-length quantifiers into constant-depth circuits~\cite{FSS84,Yao85,Has86}, and a diagonalization over all machines, this gives the exclusion from $\PH^O$. Every instance stored in the oracle is valid, so the uniform quantum algorithm and proof system work at every input length.

\subsection{Our results}\label{sec:results}

Section~\ref{sec:prelim} defines the oracle classes and the proof systems. Throughout, an efficient honest prover is a uniform polynomial-time quantum algorithm with access to the same classical Boolean oracle as the verifier.

\begin{theorem}[Exact quantum separation with classical verification]\label{thm:main}
There are a classical Boolean oracle $O$ and a total unary language $\calL$ such that
\[
 \calL\in\EQP^O\qquad\text{and}\qquad\calL\notin\PH^O ,
\]
and such that $\calL$ has an interactive proof system relative to $O$ with a classical polynomial-time verifier, an efficient honest prover, perfect completeness, and soundness error at most $1/4$ against computationally unbounded provers. The proof system is public-coin and uses at most $\log n/4$ messages on input $1^n$.
\end{theorem}

Soundness holds against every prover, classical or quantum, that communicates classically with the verifier, and Theorem~\ref{thm:slow} gives the resource bounds in full. As in the formulation of recursive Fourier sampling verified by McKague~\cite{McK10}, the oracle stores only the leaves of the recursive construction and does not answer queries that test a proposed internal secret. If equality tests for all internal secrets were available, a certificate could list the secrets along the path from the root to the leaf that carries the root value, which would place the problem in $\NP^O\cap\mathsf{coNP}^O$. The oracle of Hallgren and Harrow~\cite[Definition~3 and footnote~3]{HH08} is stronger in this respect, since it returns the output bit of a node only on the correct secret and a failure symbol otherwise, and so tests proposed secrets.

\paragraph{Exact quantum computation beyond $\PH$.}
Exact secret recovery, by a Hadamard transform or by R\"otteler's hidden-shift algorithm~\cite{Rot10}, gives an exact quantum algorithm and perfect completeness. The Forrelation construction of Raz and Tal, which separates $\BQP$ from $\PH$ relative to an oracle, has a bounded-error one-query variant~\cite[Theorem~1.3 and Section~8.3]{RT19}. Here a single separating language combines exact quantum evaluation with classical verification with an efficient quantum honest prover and soundness against computationally unbounded provers.

\Needspace{8\baselineskip}
\begin{corollary}\label{cor:consequences}
Let $O$ and $\calL$ be as in Theorem~\ref{thm:main}.
\begin{enumerate}[label=(\roman*),nosep]
\item $\EQP^O\not\subseteq\PH^O$ and $\BQP^O\not\subseteq\PH^O$.
\item Relative to $O$, the language $\calL$ has no interactive proof system with a constant number of messages, even with a computationally unbounded honest prover.
\end{enumerate}
\end{corollary}

Part (i) follows from $\EQP^O\subseteq\BQP^O$. Green and Pruim~\cite{GP01} separated $\EQP$ from $\Pclass^{\NP}$ relative to an oracle, and part (i) excludes $\EQP^O$ from every level of $\PH^O$. Part (ii) follows from the relativizing transformations of Goldwasser and Sipser~\cite{GS86}, Babai~\cite{Bab85}, and Babai and Moran~\cite{BM88}, by which every language with an interactive proof system with a constant number of messages, even with an unbounded prover, is in $\AM^O\subseteq\PiO2$ (see~\cite{Gol08}). For every polynomial-time computable unbounded message bound, the oracle and the language can be chosen so that the language has a proof system within the bound with an efficient quantum honest prover (Corollary~\ref{cor:logstar}), while the argument for part (ii) excludes a constant number of messages for every language outside $\PH$ relative to its oracle. Relative to $O$, the language is decided with $2h_{\max}(n)\le\log n/4$ quantifier blocks (Section~\ref{sec:proofs-main}), where $h_{\max}(n)=\lfloor\log n/8\rfloor$ is the largest height of an instance for the input $1^n$ (Section~\ref{sec:problem}). The exclusion is from every fixed level of the hierarchy, and whether a matching lower bound on the number of alternations holds is open.

\paragraph{Slowly growing message and query bounds.}
The verifier checks all claims about nodes of the same height at once (Theorem~\ref{thm:batched}), so the number of messages is twice the height, while the lower bound needs only a fixed number of levels. This leaves the height free to grow as slowly as desired. A height function is a function $h$ from the positive integers to the nonnegative integers, computable from $1^n$ in polynomial time, that is unbounded and satisfies $h(n)\le h_{\max}(n)$ for every $n$. Unboundedness gives, for every constant $h_0$, infinitely many lengths $n$ with $h(n)\ge h_0$, which the oracle construction needs.

\begin{theorem}[Resource bounds by recursion height]\label{thm:slow}
For every height function $h$ there are a classical Boolean oracle $O_h$ and a total unary language $\calL_h$ with
\[
 \calL_h\in\EQP^{O_h}\qquad\text{and}\qquad\calL_h\notin\PH^{O_h} ,
\]
and with the following bounds on input $1^n$.
\begin{enumerate}[label=(\roman*),nosep]
\item A uniform quantum algorithm decides $\calL_h$ exactly with $3^{h(n)}$ queries and in time $3^{h(n)}\poly(n)$.
\item $\calL_h$ has a public-coin interactive proof system relative to $O_h$ with perfect completeness and soundness error at most $1/4$ against computationally unbounded provers, in which the verifier and the prover exchange $2h(n)$ messages, $h(n)$ from each party. The prover sends fewer than $n\cdot4^{h(n)}/3$ bits in total and the verifier fewer than $n\cdot4^{h(n)}$ random bits. The verifier makes at most $4^{h(n)}$ queries and runs in time $4^{h(n)}\poly(n)$.
\item The honest prover is a uniform quantum algorithm that makes $4^{h(n)}-3^{h(n)}$ queries and runs in time $4^{h(n)}\poly(n)$.
\end{enumerate}
\end{theorem}

Theorem~\ref{thm:main} is the case $h=h_{\max}$, for which $2h(n)\le\log n/4$. The oracle depends on the height function, and truncating a desired bound to the permitted height gives the following consequence.

\begin{corollary}[Arbitrary unbounded message bounds]\label{cor:logstar}
Let $f$ be an unbounded function from the positive integers to the nonnegative integers that is computable from $1^n$ in time polynomial in $n$, for example $f(n)=\log^*n$. There are a classical Boolean oracle $O_f$ and a total unary language $\calL_f\in\EQP^{O_f}\setminus\PH^{O_f}$ with a public-coin interactive proof system relative to $O_f$ with perfect completeness, soundness error at most $1/4$ against computationally unbounded provers, an efficient honest prover, and at most $f(n)$ messages on input $1^n$. The quantum algorithm, the verifier, and the honest prover make at most $2^{f(n)}$ queries each.
\end{corollary}

For $f(n)=\log^*n$ the protocol uses at most $\log^*n$ messages, and the three participants make at most $2^{\log^*n}$ queries each.

\begin{corollary}[Arbitrary unbounded query bounds]\label{cor:querybudget}
Let $Q$ be an unbounded function from the positive integers to the positive integers that is computable from $1^n$ in polynomial time. There are a classical Boolean oracle $O_Q$ and a total unary language $\calL_Q\in\EQP^{O_Q}\setminus\PH^{O_Q}$ with a public-coin interactive proof system relative to $O_Q$ having perfect completeness, soundness error at most $1/4$ against computationally unbounded provers, and an efficient honest prover, such that the exact quantum algorithm, verifier, and honest prover each make at most $Q(n)$ queries on input $1^n$. The proof system uses at most $\lfloor\log_2Q(n)\rfloor$ messages.
\end{corollary}

Corollary~\ref{cor:querybudget} follows from Corollary~\ref{cor:logstar} with $f(n)=\lfloor\log_2Q(n)\rfloor$ (Section~\ref{sec:proofs-main}). For $Q(n)=\max\{1,\log^*n\}$, all three participants make at most $\log^*n$ queries for $n\ge2$, and the protocol uses $O(\log\log^*n)$ messages. These are bounds on queries only, since writing the addresses and carrying out the computations still takes time polynomial in $n$.

Slowly growing interaction beyond $\PH$ itself is prior work. Aiello, Goldwasser, and H{\aa}stad~\cite[Theorem~8.3]{AGH90} proved that for every polynomial-time computable $f$ that is polynomially bounded and every integer-valued $g=o(f)$ there is an oracle $B$ with a unary language in $\AM^B[f]$, the class of languages with $f(n)$-message public-coin proof systems relative to $B$, but outside $\Sigma^{\mathsf p,B}_{g(n)}$. If $f(n)\to\infty$, choose $g=o(f)$ with $g(n)\to\infty$, such as $\lfloor\sqrt f\rfloor$. A language in $\Sigma^{\mathsf p,B}_k$ is accepted by a polynomial-time alternating machine with at most $k$ alternations, and $k\le g(n)$ for all sufficiently large $n$, while the finitely many shorter inputs can be hardwired. Their language is therefore outside $\PH^B$. The verifier checks a classical game transcript against the oracle, and no efficient honest prover is supplied. Our additional requirements are exact quantum evaluation and an efficient quantum honest prover for the same language. Our lower bound, which holds uniformly in the height, follows the same principle as their proof. A constant number of restrictions simplifies any fixed-depth circuit, while a sufficiently tall game or recursion keeps its root value undetermined.

For a suitable oracle, the protocol of Theorem~\ref{thm:slow} with $2h(n)$ messages is optimal up to a constant depending on the verifier, on infinitely many input lengths, even among proof systems with private coins and computationally unbounded honest provers. Theorem~\ref{thm:slow} remains valid for this oracle, with the following additional lower bound. Messages are counted as alternating communication slots, including slots containing the empty string, and the count on an input is the maximum over all provers and random strings (Section~\ref{sec:oracle-classes}).

\begin{theorem}[Matching lower bound on the number of messages]\label{thm:interaction}
For every height function $h$ there are a classical Boolean oracle $O'_h$ and a total unary language $\calL'_h$ with the properties of Theorem~\ref{thm:slow} such that the following holds. For every polynomial-time verifier $V$, with private coins allowed, such that $\calL'_h$ has an interactive proof system relative to $O'_h$ with verifier $V$, completeness at least $2/3$, and soundness error at most $1/3$ against computationally unbounded provers, there is a constant $\beta_V>0$ such that $V$ exchanges more than $\beta_Vh(n)$ messages on input $1^n$ for infinitely many $n$.
\end{theorem}

For $h=h_{\max}$ Theorem~\ref{thm:interaction} excludes every proof system that uses $o(\log n)$ messages at all sufficiently large lengths, while the protocol of Theorem~\ref{thm:slow} uses at most $\log n/4$. A public-coin protocol with few messages becomes, by sampling the verifier's challenges, a DNF of small width on the leaves, while each level of the recursion divides the width of any DNF by $n^{\Omega(1)}$, so a DNF that distinguishes the root values at height $h$ needs width $n^{\Omega(h)}$ and the number of messages is at least a constant fraction of $h$ (Section~\ref{sec:outline-messages} and Appendix~\ref{app:interaction}).

The exclusion from $\PH^O$ in every result above rests on a lower bound for constant-depth circuits, Theorem~\ref{thm:circuits} below, which we state after defining the distributions on instances.

\paragraph{The distributions and the coupling.}
An instance of size parameter $n$ (size $n$ for short) and height $h\le h_{\max}(n)$ is a tree with $2^{\Theta(n)}$ leaves when $h\ge1$ (Section~\ref{sec:problem}). At a node of dimension $m$, whose secret $s$ is a string in $\F^m$, a code table $L$ encodes $s$, a data table $R$, which is otherwise unconstrained, holds the node's value at entry $s$, and every entry of either table is the value of a smaller subtree. The subtrees whose values fill the two tables are the node's code children and data children, and the data child at entry $s$, which the secret selects, is the free child of the node. Only the leaves are input coordinates. Validity requires a codeword at every node, so the leaves determine all values uniquely.

For a prescribed root value $b$, the distribution $\mu_{n,h,b}$ is generated from the root down. At every node it chooses a uniform secret, prescribes its codeword to the code children, prescribes independent fair bits to the other data children, and prescribes the node's value to the free child. Each child then generates its own subtree independently in the same way. Using the same secrets and the same fair bits for both root values gives a coupled pair $(X_0,X_1)$ with $X_b\sim\mu_{n,h,b}$ that differs in exactly one leaf (Claim~\ref{claim:proj}).

\begin{theorem}[Coupling bound for constant-depth circuits]\label{thm:circuits}
Fix a code family of degree $d$ (Definition~\ref{def:code}), a depth $D$, and an exponent $A\ge1$. There are a constant $c>0$ and positive integers $h_0$ and $n_0$, depending only on the code family, $D$, and $A$, such that the following holds for every $n\ge n_0$ and every $h$ with $h_0\le h\le h_{\max}(n)$. Let $F$ be a circuit on the leaves of an instance of size $n$ and height $h$, with AND and OR gates of unbounded fan-in and input negations, of depth at most $D$ and size at most $2^{n^A}$. For the coupled pair $(X_0,X_1)$ of Section~\ref{sec:projections},
\begin{equation}\label{eq:coupling-stability}
 \Pr\bigl[F(X_0)\ne F(X_1)\bigr]\le2^{-cn}.
\end{equation}
In particular,
\[
 \bigl|\E_{\mu_{n,h,1}}F-\E_{\mu_{n,h,0}}F\bigr|\le2^{-cn}.
\]
\end{theorem}

The bound on the coupled pair controls the probability that the value of $F$ changes at all when the leaf changes, which is more than a bound on the difference of expectations. The proof uses only $h_0$ of the levels, so its constants do not depend on the full height $h$. Theorem~\ref{thm:slow} uses this uniformity in the height.

A polynomial-time oracle machine with $k$ alternating polynomial-length quantifiers becomes, on the instance of one size, a circuit of depth $k+2$ and size $2^{\poly(n)}$ (Section~\ref{sec:oracleproof}). If it computed the root value on every valid instance, its expectations under $\mu_{n,h,1}$ and $\mu_{n,h,0}$ would differ by one, so Theorem~\ref{thm:circuits} gives a valid instance on which it errs, and a diagonalization over all such machines constructs the oracle of Theorem~\ref{thm:main}.

Theorem~\ref{thm:circuits} is also a classical circuit lower bound for the recursive composition of the addressing function. Its switching argument uses the code family only through the degree bound of Definition~\ref{def:code} and does not require a quantum decoding algorithm. The tail bound for certificate complexity underlying the argument, Theorem~\ref{thm:certtail}, is more general, since it applies to independent block projections satisfying a uniform sensitivity bound, without reference to quantum computation or to a specific code.

\paragraph{Beyond the Hadamard code.}
Every code family of Definition~\ref{def:code} can be used, including nonlinear families whose codewords are the shifts of a bent function, among them a family of degree three (Appendix~\ref{app:codes}).

\paragraph{Noisy code tables.}
Errors in the code tables are also tolerated. Theorem~\ref{thm:robust} in Section~\ref{sec:noisyproof} gives, for instances in which every code table is within a fixed relative distance below $1/4$ of a codeword, a bounded-error quantum algorithm and a public-coin proof system with completeness at least $0.99$, soundness error at most $0.01$, and polynomially many messages, while the circuit lower bound is unchanged. The promise concerns the table of code-child values at every node and not arbitrary corruption of the leaves, and the lower bound holds already for the exact instances, which satisfy the promise.

\subsection{Related work}\label{sec:related}

Table~\ref{tab:cvqc} lists the problems known to have classical proof systems with an efficient quantum honest prover and soundness against computationally unbounded provers. Problems in $\NP\cap\mathsf{coNP}$ whose witnesses a polynomial-time quantum algorithm can compute, such as the decision versions of factoring~\cite{Sho97}, are verified by one message and lie in the polynomial hierarchy. Le Gall, Morimae, Nishimura, and Takeuchi~\cite[Theorem~1]{LGMNT18} gave a three-message protocol for computing the order of a solvable black-box group, a problem that Watrous~\cite{Wat01} solved in quantum polynomial time, and a two-message protocol when the prime factors of the order are given. Demarie, Ouyang, and Fitzsimons~\cite{DOF18} and Morimae, Takeuchi, and Nishimura~\cite{MTN18} gave one-message protocols for problems about quantum circuits with two basis changes, such as two layers of Hadamard gates, the second level of the Fourier hierarchy~\cite{Shi05}. McKague~\cite{McK10} gave a protocol with polynomially many messages for recursive Fourier sampling, the problem with which Bernstein and Vazirani~\cite{BV97} separated $\BQP$ from $\BPP$ relative to an oracle, and observed that it then lies outside $\MA$ relative to an oracle~\cite[Section~3]{McK10} (see also~\cite{Aar03}). Whether it lies outside $\PH$ relative to an oracle, in the formulation that McKague verifies, is open. McKague~\cite[Section~1]{McK10} records the conjecture that it does (see also~\cite{BHNST26}), and Remscrim~\cite[Section~5.4]{Rem16} identified the circuit lower bound that is missing.

\begin{table}[htbp]
\centering
\footnotesize
\renewcommand{\arraystretch}{1.25}
\begin{tabular}{@{}p{5.1cm}lp{3.5cm}p{3.7cm}@{}}
\hline
Problem & Reference & Messages & Excluded class, relative to an oracle \\
\hline
Factoring and other problems in $\NP\cap\mathsf{coNP}$ whose witnesses quantum algorithms compute & \cite{Sho97} & $1$ & none; these problems are in $\NP\cap\mathsf{coNP}$ \\
Order of a solvable black-box group & \cite{LGMNT18} & $3$, or $2$ given the prime factors & $\BPP$, in the black-box model~\cite[Section~1]{LGMNT18}; the problem is in $\AM$ \\
Problems on circuits with two basis changes & \cite{DOF18,MTN18} & $1$ & none; these problems are in $\MA$ \\
Recursive Fourier sampling & \cite{McK10} & polynomial & $\MA$~\cite{Aar03,McK10} \\
\rowcolor{gray!15} The variant of recursive Fourier sampling of this paper & Theorem~\ref{thm:main} & at most $\log n/4$; any unbounded $f(n)$ (Corollary~\ref{cor:logstar}) & $\PH$ \\
\hline
\end{tabular}
\caption{Classical verification without computational assumptions and with a single quantum prover. In every row the verifier is classical polynomial-time, the honest prover is a polynomial-time quantum algorithm, and soundness holds against every prover. Each row refers to the decision or promise version of the problem named in its first column, and the last column is the class from which that problem is proved to be excluded relative to an oracle. The shaded row is this paper.}
\label{tab:cvqc}
\end{table}

Two further results concern different models. Aaronson, Natarajan, Tal, and Vill\'anyi \cite{ANTV26} gave a relativizing classical multiprover proof system for every problem in $\BQP$, but its honest provers are not known to be efficient. Yamakawa and Zhandry~\cite{YZ22} gave a search problem, relative to a random oracle, whose solutions can be verified classically and can be found by quantum polynomial-time algorithms but not by classical probabilistic polynomial-time algorithms. Our result concerns a decision language verified by a single prover and excludes it from every level of $\PH^O$.

Oracle separations of quantum computation from the polynomial hierarchy come from a different line of work. Aaronson~\cite{Aar10} introduced Fourier Checking, now called Forrelation, for this purpose, and Raz and Tal~\cite{RT19} proved $\BQP\not\subseteq\PH$ relative to an oracle using a variant of it. Their separating language is decided with bounded error and is not known to have a classical proof system with an efficient quantum honest prover. Raz and Tal argue by Fourier analysis and a Gaussian approximation, and we argue by random projections and a switching argument, which is what allows a recursive construction with exactness, a proof system, and slowly growing message and query bounds. Bansal and Sinha~\cite{BS21} and Sherstov, Storozhenko, and Wu~\cite{SSW21} obtained optimal separations between quantum and randomized query complexity, using $k$-fold Forrelation and a variant of it. Recursive Fourier sampling has exact quantum evaluation~\cite{BV97}. Hallgren and Harrow~\cite{HH08} generalized its recursion to other quantum circuits, and R\"otteler~\cite[Theorems~6 and~8]{Rot10} to hidden shifts of bent functions, obtaining superpolynomial quantum-versus-classical query separations but no lower bounds against $\PH$. Green and Pruim~\cite{GP01} gave an oracle relative to which $\EQP\not\subseteq\Pclass^{\NP}$, and Watrous~\cite{Wat00} one relative to which $\BQP\not\subseteq\MA$. Girish and Servedio~\cite[Theorem~4]{GS26} proved an exact one-query separation from randomized queries for extremal Forrelation, whose Forrelation value is $\pm1$, and conjectured that it is hard for constant-depth circuits~\cite[Conjecture~9]{GS26}. Forrelation has also been used for oracle separations within the Fourier hierarchy~\cite{Shi05,Man26}.

Aiello, Goldwasser, and H{\aa}stad~\cite[Corollary~8.4]{AGH90} separated, relative to an oracle, public-coin proof systems with $f$ messages from those with $g=o(f)$ messages for every polynomial-time computable $f$ that is polynomially bounded, and we compare their result with ours after Corollary~\ref{cor:querybudget}.

\subsection{Techniques}\label{sec:techniques}

At each recursive step, the same secret determines both the fixed codeword and the data entry left free. Our main technical contribution is a switching theorem for random projections with this dependence. The addressing function at one node is due to Chakraborty, Chattopadhyay, Mande, and Paraashar~\cite[Definition~16]{CCMP20}. In recursive Fourier sampling the value of a node is the oracle bit at the recovered secret~\cite[Section~8.4]{BV97} or $g(s)$ for a fixed predicate $g$~\cite{McK10}. Here every data entry is the value of its own subtree, so the data table of a node is unconstrained by its code table, apart from the selected entry when the node's value is prescribed. The lower bound uses this freedom, which recursive Fourier sampling with a fixed predicate lacks, through an auxiliary experiment with independently chosen secrets and data tables, conditioned on the selected entries (Claim~\ref{claim:twostep}). Structured and correlated random projections appear already in the depth hierarchy theorem of H{\aa}stad, Rossman, Servedio, and Tan~\cite{HRST17}. The fixed codeword reveals the coordinate left free, so a single level can turn a CNF of small width into an AND of arbitrarily many free variables. Switching lemmas for independent restrictions~\cite{Ajt83,FSS84,Yao85,Has86} do not apply. The result of Meir and Wigderson~\cite[Proposition~1.7]{MW19} allows an arbitrarily correlated distribution of the fixed values, but the set of fixed positions is chosen independently of those values, whereas here the free position and the fixed code bits are determined by the same secret. We control this dependence through an algebraic bound on the first satisfied term of a DNF, whose index is the DNF index function of Lovett and Zhang~\cite[Definition~1.10]{LZ19}, and an argument from sensitivity to certificates, followed by a truncated decision-tree construction from certificates~\cite{BI87,BdW02,HH91,Tar89}, applied across several levels. The approximation is one-sided, with an error bound at each fixed assignment to the free variables, so the errors compose over a constant number of levels.

Section~\ref{sec:outline} outlines the proofs. Sections~\ref{sec:prelim} to~\ref{sec:mainproofs} contain the definitions, the recursive problem, the quantum algorithm and the proof system, the circuit lower bound, and the proofs of the main theorems, and Appendices~\ref{app:noisy} to~\ref{app:codes} treat noisy code tables, the message lower bound, and codes from bent functions.

\section{Proof outline}\label{sec:outline}

We describe the construction with the Hadamard code. The lower-bound estimates also apply to the codes of degree $d$ in Definition~\ref{def:code}.

\subsection{The problem and the distributions}\label{sec:outline-problem}

\begin{figure}[t]
\centering
\begin{tikzpicture}[x=1cm,y=1cm,font=\small]
 \node[draw,rounded corners,inner sep=4pt] (root) at (0,0) {node of height $j$: secret $s\in\F^m$, value $R_s$};
 \node[draw,rounded corners,inner sep=4pt] (code) at (-3.8,-1.7) {code table $L_x=s\cdot x$, $x\in\F^m$};
 \node[draw,rounded corners,inner sep=4pt] (pay) at (3.8,-1.7) {data table $R_y$, $y\in\F^m$};
 \draw[->] (root) -- (code);
 \draw[->] (root) -- (pay);
 \foreach \k in {0,1,2,3} {
   \draw (-5.6+\k*1.2,-2.5) -- (-6.0+\k*1.2,-3.5) -- (-5.2+\k*1.2,-3.5) -- cycle;
 }
 \node at (-3.8,-3.9) {$2^m$ subtrees of height $j-1$, one for each $x$};
 \foreach \k in {0,1,3} {
   \draw (2.0+\k*1.2,-2.5) -- (1.6+\k*1.2,-3.5) -- (2.4+\k*1.2,-3.5) -- cycle;
 }
 \draw[very thick] (4.4,-2.5) -- (4.0,-3.5) -- (4.8,-3.5) -- cycle;
 \node at (4.4,-3.75) {$y=s$};
 \node at (3.8,-4.2) {$2^m$ subtrees of height $j-1$, one for each $y$};
\end{tikzpicture}
\caption{A node of the recursion with the Hadamard code. The node hides a secret $s$. Its code table is the codeword of $s$, and its value is the entry of its data table at $s$ (the thick subtree). Every entry of both tables is the value of a smaller node, and only the leaves of the recursion are stored in the oracle.}
\label{fig:node}
\end{figure}
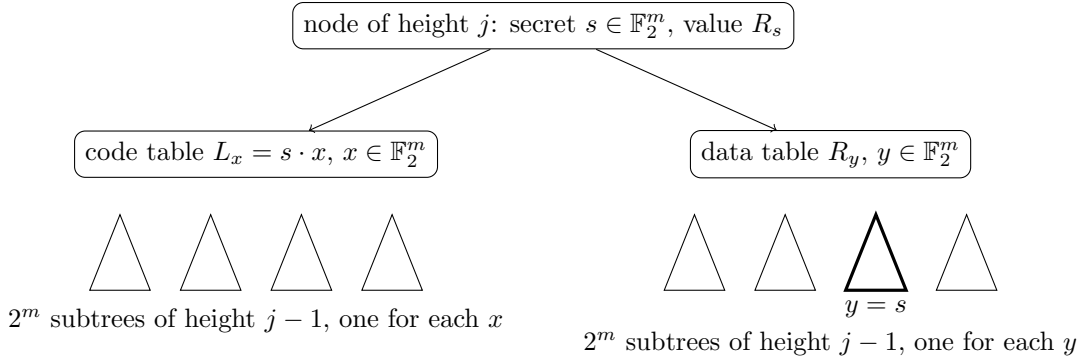

A node hides a secret $s\in\F^m$ (Figure~\ref{fig:node}). Its code table has entries $L_x=s\cdot x$, its data table $R$ is unconstrained, and its value is $R_s$, the entry of the data table at the secret encoded by the code table. Every entry of both tables is itself the value of a node of smaller height, down to the leaves. Only the leaves are stored in the oracle. The input is promised to be valid, meaning that every code table is a codeword. The parameter $n$ determines the dimensions, which decrease by a factor of sixteen per level towards the root, up to rounding to a multiple of eight, and the height satisfies $h\le h_{\max}(n)=\lfloor\log n/8\rfloor$.

At one node the random projection chooses a uniform secret, fixes its codeword and independent uniform values for the other data entries, and leaves $R_s$ free. The free entry copies the value prescribed to the node, and the random choices do not depend on that value. We compose the independent projections of the nodes at a fixed level and below it, which maps the vector $z\in\F^K$ of node values at that level to the leaves, and each coordinate is copied to exactly one free leaf along a path of free children. The composed projection $\rho$ is independent of the tree above that level and is defined on the whole cube $\F^K$, including vectors that are not the node values of a valid tree, and the coordinates of $z$ are its free variables. We use the same projections for both root values, so that $X_b=\rho(Z_b)$, where $Z_b$ is the vector of node values at that level under root value $b$, and $Z_0,Z_1$ differ in one coordinate while $X_0,X_1$ differ in one leaf (Section~\ref{sec:projections}).

\subsection{The quantum algorithm and the protocol}\label{sec:outline-upper}

We evaluate a node with three recursive calls. The first call, to the code children in superposition, prepares the phase state of the codeword, and a Hadamard transform recovers $\ket s$. The second call, to the data child $s$, adds its value to the output, and a third call to the code children, after the inverse transform, uncomputes $s$. Each call returns its work registers to their initial states, so every call is clean in the sense of Section~\ref{sec:exact} and can be made in superposition, and the recursion uses $3^h$ leaf queries (Theorem~\ref{thm:exact}).

To verify a claim that a node of positive height has a given value, the prover sends a claimed secret, and the verifier recursively checks three fresh random code entries against the codeword of the claimed secret and the data entry at the claimed secret against the claimed value (Protocol~\ref{prot:verify}). For a list of claims of one height, the prover sends all claimed secrets in one message before the verifier samples the challenges, which are independent for every entry, including repeated claims about the same node (Protocol~\ref{prot:batched}).

The soundness argument conditions on the history and on the prover's whole message of claimed secrets, after which the challenges are still fresh. A false claim with the right claimed secret has a false data claim among its children, and one with a wrong claimed secret has a number $Y$ of false code claims among its children with the binomial distribution with parameters $3$ and $1/2$, because distinct codewords differ in exactly half of the coordinates. Putting $Y=1$ in the first case, in both cases $\E[4^{-Y}]\le\max\{1/4,(5/8)^3\}=1/4$, so by induction on the height the verifier, continued from a list with $F$ false claims, accepts with probability at most $4^{-F}$, with no dependence on the height. Two code checks would not suffice, since $(5/8)^2>1/4$. The bound is conditional on the history, so it holds for every state of the prover's memory (Claim~\ref{claim:batched}). Batching by height gives a public-coin protocol with $2h$ messages and perfect completeness with the honest quantum prover (Theorem~\ref{thm:batched}).

\subsection{The constant-depth circuit lower bound}\label{sec:outline-lower}

We prove that a circuit of constant depth and size $2^{\poly(n)}$ changes its output on the coupled pair $(X_0,X_1)$ with probability at most $2^{-\Omega(n)}$, once the height is at least a constant $h_0$ depending on the depth and the size exponent. We follow H{\aa}stad's depth reduction~\cite{Has86}, which switches the bottom CNFs to DNFs, or DNFs to CNFs, under a projection, merges the new top gates with their parents, and repeats.

A rank bound for the first satisfied term of a DNF (Theorem~\ref{thm:winner}) gives a one-sided term deletion that preserves every rejection (Claim~\ref{claim:truncate}), deletion gives the sensitivity bound (Theorem~\ref{thm:flips}), sensitivity gives the tail bound for certificate complexity (Theorem~\ref{thm:certtail}), and the tail bound gives a switching theorem whose error is bounded at each fixed assignment (Theorem~\ref{thm:switch}). Pointwise switching, with the input at each stage generated by the fresh projections of the levels above those already applied, gives the depth reduction and the coupling bound (Theorem~\ref{thm:circuits}). The main step is a switching theorem for the correlated projections, which shows that for every constant $\alpha$, applying the projections of $\lfloor\alpha\rfloor+1$ levels switches every CNF or DNF of width at most $n^{\alpha}$, with an error at each fixed input small enough to sum over all bottom gates.

\paragraph{The obstruction at one level.}
Carmosino, Hoover, Impagliazzo, Kabanets, and Kolokolova prove a blockwise switching lemma~\cite[Definition~9 and Lemmas~10 and~15]{CHIKK21} for restrictions that leave one free variable per block and fix the other coordinates independently and uniformly, and it does not apply to our correlated distribution. The following example shows that one projection level can leave the certificate complexity of an accepting input as large as the number of blocks, because a CNF can contain a clause for every possible secret, and the projection keeps exactly the clause of the true secret. For a node with the Hadamard code let
\[
 \Gamma(L,R)=\bigwedge_{y\in\F^m}\Bigl(\,\bigvee_{i=1}^{m}[\,L_{e_i}\ne y_i\,]\ \vee\ R_y\Bigr),
\]
where $e_i$ are the unit vectors and $[P]$ is the literal asserting $P$. The clause with $y=s$ reduces to $R_s$ and every other clause is true, so $\Gamma$ has width $m+1$ and projects to the free variable with probability one, and $K$ copies, one for each of $K$ nodes, project to the AND of all $K$ free variables, whose accepting input needs a certificate of size $K$. Every switching step therefore exposes several levels, that is, applies the projections of several consecutive levels at once.

\paragraph{A rank bound independent of the number of terms.}
We order the terms of a width-$W$ DNF, and bound the probability that the first satisfied term, whose index is the DNF index function of Lovett and Zhang~\cite[Definition~1.10]{LZ19}, contains the free leaves of many nodes while testing few code entries of each of them. A node is light for a term if its free leaf occurs in the term and the term has at most $t$ literals on the node's code entries, its code literals. For one node, we fix a uniformly random data table first, including its entry at the secret, discard every term with a false data literal or a false literal on another node, and choose, for each remaining term with at most $t$ code literals that is first satisfied at some secret, one such secret. If the code literals of term $\lambda$ hold at the secret chosen for term $\xi$, then term $\lambda$ is satisfied there, as its other literals hold under the fixed tables, and so comes no earlier than term $\xi$ in the order. With the terms in order, the matrix $M$ whose entry $M_{\lambda\xi}$ indicates this is therefore lower triangular with ones on the diagonal, while each row evaluates a polynomial of degree at most $dt$ in the secret, so that, for three terms,
\[
 M=\begin{pmatrix}1&0&0\\ *&1&0\\ *&*&1\end{pmatrix},\qquad
 \text{number of rows of }M=\operatorname{rank}M\le\binom{m}{\le dt}.
\]
In words, at most $\binom{m}{\le dt}$ terms, the dimension of the space of multilinear polynomials of degree at most $dt$ in $m$ variables, are first satisfied at some secret at which the node is light, and each of them at no more than $W$ such secrets, as the free leaf is one of the at most $W$ data leaves in the term. Theorem~\ref{thm:winner} bounds the probability that at least $\ell$ of $K$ independent nodes of dimension $m$ are light for the first satisfied term, at any fixed assignment to the free variables, by
\[
 \binom K\ell\Bigl(\frac{2W}{\ell}\Bigr)^{\ell}\binom{m}{\le dt}^{\ell}2^{-m\ell},
\]
uniformly in the number of terms, where the factors correspond to the choice of the $\ell$ nodes, to the secret tuples at which a given term is first satisfied, with a factor two per node for the prescribed data entry at the secret, to the possible first satisfied terms, and to the probability of a given tuple of secrets. The dimensions decrease geometrically towards the root, which keeps the number of nodes at a level small compared with the secret space $\F^m$, and for $t$ a sufficiently small constant fraction of $m$ and polynomially bounded width the bound is exponentially small in $m$ per light node (Claim~\ref{claim:loss}).

\paragraph{Deletion and sensitivity.}
For an integer $J$, we delete every term for which more than $J$ nodes are light and project the rest, which gives a DNF on the whole cube $\F^K$ of width at most $\lfloor W/(t+1)\rfloor+J$. Every rejection is preserved, and acceptance at a fixed input is lost only when its first satisfied term is deleted (Claim~\ref{claim:truncate}).

Theorem~\ref{thm:flips} concerns $r$ independent blocks, each a node with its descendants over $j=\lfloor\alpha\rfloor+1$ levels and with one free leaf (Section~\ref{sec:projections}), and bounds the probability that a DNF of width at most $W\le n^{\alpha}$ accepts the input generated by the blocks and is sensitive there to the free leaf of every block, that is, rejects when any one root value is flipped under the same projections. Term deletion at each of the $j$ levels, with a threshold of about half of $r$ on the number of light nodes in a term, preserves every rejection, loses acceptance with probability exponentially small in $\nu r$ per level, and brings the width below $r$, so that no term contains all $r$ changed leaves, and the bound is exponentially small in $\nu r$, where $\nu$ is the dimension of the highest of the $j$ levels.

\paragraph{From sensitivity to certificates.}
Theorem~\ref{thm:certtail} turns the sensitivity bound into a tail bound for certificate complexity on the whole cube of assignments to the free variables. When the projection $H=\Psi\circ\rho$ of a width-$W$ CNF $\Psi$ through these $j$ levels accepts an assignment $z$ without a $1$-certificate of size at most $q$, there are disjoint sets of free variables of total size more than $q$ such that flipping any one set falsifies a clause of $\Psi$, and hence $H$, whereas flipping any proper subset leaves $H$ true. A coloring argument retains sets of total size at least $(q+1)/(2W+1)$ whose witnessing original clauses avoid the blocks of all other retained sets. For each family of sets fixed in advance, we condition on the projections outside its union and form, for each set, the CNF of all original clauses avoiding the other sets, with the outside leaves substituted. The event that $H$ accepts $z$ without a small certificate and yields this family implies the sensitivity event of Theorem~\ref{thm:flips} for the negations of these fixed CNFs. These events involve disjoint blocks, so their probabilities multiply. A union bound over the families yields the tail bound for certificate complexity, with no factor for the number of clauses (Section~\ref{sec:certtail}). A truncated version of the decision-tree construction from certificates~\cite{BI87,BdW02,HH91,Tar89}, which uses only certificates of size at most $q$ (Claim~\ref{claim:tree}), then gives a DNF of width polynomial in $W$ and $q$ that is at most $H$ everywhere and differs from $H$ at a fixed assignment with probability exponentially small in $\nu q/W$, and complementation gives the dual statement. This is the pointwise approximate switching of Theorem~\ref{thm:switch}.

\paragraph{Composition of pointwise errors.}
We put the circuit into alternating layers with bottom fan-in one, and at a stage with size at most $2^{n^a}$ and bottom fan-in at most $n^a$ we expose $\lfloor a\rfloor+1$ levels and choose the certificate bound $q$ polynomially large, so that a union bound over the gates controls the error of the stage. At a stage, three independent groups of projections determine the comparison. The levels below the stage fix the current formula, the levels of the stage give the replacement, and the levels above generate the input at which the replacement is compared with the projection of the current formula, so the errors at that input add over the gates and over the stages without conditioning on earlier successes (Claims~\ref{claim:stage} and~\ref{claim:telescope}). Switching and merging removes one layer, and the next stage has size at most $2^{n^{a'}}$ and bottom fan-in at most $n^{a'}$ for a larger constant $a'$, so a constant-depth circuit needs only a constant number $h_0$ of levels in total, independently of the remaining height, and the coupling bound is uniform for $h_0\le h\le h_{\max}(n)$. The last CNF or DNF has polynomial width, and the case $r=1$ of Theorem~\ref{thm:flips}, extended to taller blocks by Claim~\ref{claim:tall}, makes it take the same value on its two inputs except with probability $2^{-\Omega(n)}$ (Section~\ref{sec:circuits}).

\subsection{The oracle}\label{sec:outline-oracle}

We store the leaves of the instance with parameter $n$ at addresses beginning with $1^n0$, and we enumerate the polynomial-time oracle predicates with every fixed number $k$ of alternating polynomial-length quantifiers. At a fresh size $n$ with $h(n)\ge h_0$, we fix the other instances that the predicate can query to valid default trees. For every quantified tuple, the predicate is then an adaptive decision tree over the leaves of the instance of size $n$, and its accepting transcripts give an exact DNF of polynomial width and size $2^{\poly(n)}$. The quantifiers become AND and OR gates, giving depth $k+2$. Theorem~\ref{thm:circuits} supplies a valid instance on which this circuit errs. We then fix every oracle address that the predicate can query within its time bound, so that later stages preserve the error. The unary language contains $1^n$ exactly when the instance of size $n$ has root value one, and every stored instance is valid, so the quantum algorithm and the proof system apply at every length (Section~\ref{sec:oracleproof}). The construction for Theorem~\ref{thm:interaction} adds stages for verifiers whose number of messages is capped at a fraction of the height, which Section~\ref{sec:outline-messages} describes.

\subsection{The lower bound on the number of messages}\label{sec:outline-messages}

Messages are counted as in Section~\ref{sec:oracle-classes}. Sampling a complete tree of challenges of polynomial degree turns a public-coin protocol of a polynomial-time verifier with $r$ rounds, each a challenge followed by a reply, into a two-message protocol with $n^{O(r)}$ queries (Lemma~\ref{lem:collapse}). For a two-message verifier and a fixed random string, some reply is accepted exactly when a DNF on the leaves of width at most the number of queries accepts, so a correct protocol gives such a DNF whose expectations under the two root values differ by at least $1/3$ (Lemma~\ref{lem:am}). Repeated width reduction, by Theorem~\ref{thm:winner} with a single light node and Claim~\ref{claim:truncate} with deletion threshold zero at every level, so that every term with a light node is deleted, shows that a DNF that distinguishes the root values needs width $n^{\Omega(h)}$ (Theorem~\ref{thm:widthzero}), so the number of messages of a correct protocol with a polynomial-time verifier is at least a constant fraction of $h$, with a constant depending on its time bound (Theorem~\ref{thm:finite-messages}). Private coins are removed by the transformation of Goldwasser and Sipser~\cite{GS86}, and the oracle construction adds, for every verifier run under a polynomial clock, stages that store a valid instance on which its version cut off after a fraction of the height is not correct, so every proof system with that verifier exceeds the cap on infinitely many lengths (Theorem~\ref{thm:interaction}).

\subsection*{Acknowledgments and AI disclosure}
\addcontentsline{toc}{subsection}{Acknowledgments and AI disclosure}
We thank Matthew Coudron, Alexandru Gheorghiu, Dominik Hangleiter, and Ryan Sweke for extensive discussions of interactive proofs and oracle separations, and Umesh Vazirani for insightful examples involving the complement of Simon's problem. These conversations deepened our understanding of classical verification in the oracle model. The construction grew from changes to recursive Fourier sampling that we proposed. Discussions with OpenAI models (GPT-6.1 Sol and GPT-6 Astra) helped us understand classical lower-bound techniques and contributed substantially to developing the proof of Theorem~\ref{thm:circuits}, which underlies Theorem~\ref{thm:main}, including expanding sketches of several claims into detailed arguments. These tools also assisted with the matching-interaction argument in Appendix~\ref{app:interaction}. We subsequently revised and simplified the proofs to improve their clarity. Claude Opus assisted with drafting and revising the exposition. We take responsibility for the paper's content. This work is supported by our faculty startup grant from Virginia Tech.

\section{Preliminaries}\label{sec:prelim}

\paragraph{Notation.}
$\F$ is the field with two elements and $[m]=\{1,\dots,m\}$. Logarithms are to base two unless written $\ln$. The distance $\dist(u,v)$ between two Boolean tables indexed by the same set is their relative Hamming distance, the fraction of positions at which they differ. For integers $m\ge1$ and $t\ge0$,
\[
 \binom{m}{\le t}=\sum_{a=0}^{\min\{m,t\}}\binom ma
\]
is the number of subsets of $[m]$ of size at most $t$, and also the dimension of the space of multilinear polynomials over $\F$ in $m$ variables of degree at most $t$. For $0<\beta\le1/2$ we use the bound $\binom{m}{\le\lfloor\beta m\rfloor}\le2^{H_2(\beta)m}$, where $H_2(\beta)=-\beta\log\beta-(1-\beta)\log(1-\beta)$ is the binary entropy function, and $H_2(1/16)<0.34$.

For an integer $n\ge1$ let $\log^*n=\min\{k\ge0:n\le T_k\}$, where $T_0=1$ and $T_{k+1}=2^{T_k}$. It is computable in time polynomial in $n$ by repeated bit-length computation, since starting from $t=n$ and replacing $t$ by $\lceil\log t\rceil$, the bit length of $t-1$, while $t\ge2$ makes $\log^*n$ replacements.

For independent random variables $Y_1,\dots,Y_p$ with values in $[0,1]$ and $\delta>0$, Hoeffding's inequality~\cite{Hoe63} gives $\Pr[\sum_iY_i-\E\sum_iY_i\ge\delta p]\le\exp(-2\delta^2p)$, and the same bound for the lower tail.

\paragraph{Formulas and circuits.}
A literal is a variable or its negation. A term is a conjunction of literals and a clause is a disjunction of literals. The width of a term or a clause is its number of literals. A DNF is a disjunction of terms and a CNF is a conjunction of clauses. Its width is the maximum width of its terms or clauses, and its size is their number, which is unrestricted unless stated otherwise. An empty term is true and an empty clause is false. A DNF with no terms is false and a CNF with no clauses is true, and both have width zero.

We normalize a DNF by merging repeated literals in its terms and deleting every term that contains a variable together with its negation. We normalize a CNF by merging repeated literals in its clauses and deleting every clause that contains a variable together with its negation. After constants are substituted, we delete a term containing a false literal and delete true literals from the remaining terms, and dually we delete a clause containing a true literal and delete false literals from the remaining clauses.

A circuit consists of AND and OR gates of unbounded fan-in with negations at the inputs, and constants may also appear as inputs. Its depth is the number of gates on a longest path from an input to the output, and its size is the number of gates and wires. Counting only gates changes the exponent $A$ of Theorem~\ref{thm:circuits} by a constant factor (Claim~\ref{claim:normal}). Circuits of constant depth and polynomial size are $\AC^0$ circuits. The circuits of Theorem~\ref{thm:circuits} have constant depth and size $2^{\poly(n)}$ on $2^{\Theta(n)}$ inputs, which is quasi-polynomial in the number of inputs.

\paragraph{Decision trees and certificates.}
A decision tree over $K$ variables queries one variable at each internal node and outputs a bit at each leaf. Its depth is the maximum number of queries on a root-to-leaf path. For a Boolean function $H$ on $\F^K$ and an input $z$ with $H(z)=1$, a $1$-certificate of $H$ at $z$ is a set $S\subseteq[K]$ such that $H(z')=1$ for every $z'\in\F^K$ that agrees with $z$ on $S$. The certificate complexity $\cert_1(H,z)$ is the least size of a $1$-certificate of $H$ at $z$. The quantity $\cert_0(H,z)$ for inputs with $H(z)=0$ is defined dually. Certificates are always taken on the whole cube $\F^K$. For background on decision trees and certificates see Buhrman and de Wolf~\cite{BdW02}.

\paragraph{Projections.}
A projection $\rho$ from variables $z_1,\dots,z_K$ to variables $x_1,\dots,x_N$ substitutes a constant in $\F$ or a variable $z_i$ for every $x_j$. The $z_i$ are its free variables. An input coordinate $x_j$ mapped to one of them is free, and the other coordinates are fixed. An assignment $z\in\F^K$ therefore gives the assignment $\rho(z)\in\F^N$. For a Boolean function $F$ on $\F^N$, its projection is the function $F\circ\rho$ on $\F^K$. A projected DNF, after normalization, is a DNF of the same or smaller width, and dually for a CNF. A random projection is a random variable with values in the projections. Random projections of this kind, which map every coordinate to a constant or to a variable, appear in the depth hierarchy theorem of H{\aa}stad, Rossman, Servedio, and Tan~\cite{HRST17}.

\subsection{Oracle classes and interactive proof systems}\label{sec:oracle-classes}
A classical Boolean oracle $O$ is a function from binary strings to $\F$. A query maps $\ket a\ket b$ to $\ket a\ket{b+O(a)}$. A phase query maps $\ket a$ to $(-1)^{O(a)}\ket a$ and is implemented by one query with the answer register in the state $\ket-$. A uniform family of polynomial-size quantum circuits with oracle gates, over a fixed finite gate set containing the Hadamard, NOT, and Toffoli gates, decides a language $\calL$ exactly if it accepts every string in $\calL$ with probability one and every other string with probability zero. The class $\EQP^O$ consists of the languages decided exactly by such families~\cite{BV97}. The class $\BQP^O$ allows error probability at most $1/3$, and $\EQP^O\subseteq\BQP^O$. The gate set is closed under inverses and the oracle gate is its own inverse, so the inverse of a circuit is obtained by reversing the order of its gates and inverting each gate, and it has the same size.

For $k\ge0$, the class $\SigO k$ consists of languages of the form
\[
 \{\,w:\exists y_1\,\forall y_2\cdots Q_ky_k\,[V^O(w,y_1,\dots,y_k)=1]\,\},
\]
where $V$ is a deterministic polynomial-time oracle machine, $|y_i|=p_i(|w|)$ for polynomials $p_i$, and the $k$ quantifiers alternate, starting with $\exists$. For $k=0$ there are no quantifiers. The class $\PiO k$ is defined with the quantifiers starting with $\forall$, and $\PH^O$ is the union of all these classes~\cite{Sto76,Gol08}. The same classes are obtained from polynomial-time alternating oracle machines with $k$ alternating phases, starting existentially and universally, respectively~\cite{CKS81}.

An interactive proof system relative to $O$ consists of a randomized polynomial-time verifier with query access to $O$ that exchanges classical messages with a prover with query access to $O$ and then accepts or rejects~\cite{GMR89,Bab85}. Communication takes place in alternating message slots, with either party allowed to speak first. Every slot counts as one message, including a slot containing the empty string. The number of messages on an input is the maximum number of slots over all provers and random strings. An empty initial challenge or final reply inserted solely to express a protocol in verifier-first stage notation is not transmitted and is not counted as a slot.

For public-coin proof systems we use the standard fixed-schedule form~\cite{Bab85,BM88}, in which the verifier's messages are independent uniformly random strings and their lengths and the alternating schedule are determined by the input. There is no additional private randomness. The final decision is a deterministic function of the transcript and the oracle answers. Since the verifier's messages and schedule do not depend on oracle answers, its oracle queries and the computation of its decision may be deferred until the interaction ends. An early decision can be recorded while the remaining slots of the fixed schedule are completed, and these slots still count as messages.

The proof system has completeness $c$ and soundness error $s$ if some prover, the honest prover, makes the verifier accept every $w\in\calL$ with probability at least $c$, while no prover, classical or quantum, makes it accept any $w\notin\calL$ with probability more than $s$. The honest prover is efficient if it is a uniform polynomial-time quantum algorithm with query access to $O$. Unless stated otherwise, completeness is at least $2/3$ and soundness error at most $1/3$.

Conditional probabilities given a protocol history, that is, the transcript of the interaction so far, are taken only at histories of positive probability under the fixed oracle and prover strategy. Bounds on such conditional probabilities hold for every state of the prover's private memory that is consistent with the history.

\subsection{Codes with exact quantum decoders}\label{sec:codes}

The instances of Section~\ref{sec:problem} use a code to hide a secret. Besides being efficiently computable, the code must have two properties. A quantum algorithm recovers the secret exactly from the phase state of the codeword (Definition~\ref{def:code}), and every coordinate of the codeword is a polynomial of bounded degree in the secret. Exact decoding gives the quantum algorithm and the honest prover, and it implies the distance between codewords that the soundness proof uses. Bounded algebraic degree is used in the circuit lower bound.

\begin{definition}\label{def:code}
Fix an integer $d\ge1$. A code family of degree $d$ is a sequence $C=(C_m)$ of maps $C_m\colon\F^m\times\F^m\to\F$, one for every positive multiple $m$ of eight, with the following properties.
\begin{enumerate}[label=(\roman*),nosep]
\item The value $C_m(s,x)$ is computable in time polynomial in $m$, uniformly in $m$.
\item For every $x\in\F^m$, the coordinate $s\mapsto C_m(s,x)$ is a multilinear polynomial over $\F$ of degree at most $d$.
\item There is a uniform family of polynomial-size unitaries $U_m$, implemented exactly, together with their inverses, over the gate set of the definition of $\EQP^O$ above, such that
\[
 U_m\ket{C_s}=\ket s,\qquad\text{where}\qquad
 \ket{C_s}=2^{-m/2}\sum_{x\in\F^m}(-1)^{C_m(s,x)}\ket x .
\]
\end{enumerate}
We call $C_m(s,\cdot)$ the codeword of $s$ and $\ket{C_s}$ its phase state.
\end{definition}

Since $U_m$ maps the phase states of distinct secrets to orthogonal basis states, the phase states are orthogonal. The inner product of two phase states is $1-2\dist(C_m(s,\cdot),C_m(s',\cdot))$, so distinct codewords have distance exactly $1/2$, and in particular the code is injective.

The Hadamard code $C_m(s,x)=s\cdot x$ has degree one and the decoder $U_m=H^{\otimes m}$. Appendix~\ref{app:codes} gives nonlinear examples of degree three, whose codewords are the shifts of a bent function.

For $0\le\eta<1/4$, the addressing function of the code with noise rate $\eta$ is defined on pairs of tables $L,R\colon\F^m\to\F$ such that $L$ is within distance $\eta$ of the codeword of some secret $s$, and it takes the value
\[
 G^C_{m,\eta}(L,R)=R_s .
\]
In words, it returns the entry of $R$ at the secret $s$ determined by $L$. The secret is unique, because distinct codewords have distance $1/2>2\eta$. We write $G^C_m$ for the exact case $\eta=0$, in which $L$ is a codeword. For the Hadamard code and $\eta=0$ this is the addressing function of Chakraborty, Chattopadhyay, Mande, and Paraashar~\cite[Definition~16]{CCMP20}, in which a Hadamard codeword selects the address. Appendix~\ref{app:codes} expresses $(-1)^{R_s}$ in this case as a $3$-fold Forrelation in the sense of Aaronson and Ambainis~\cite{AA15}.

\section{The recursive problem and the distributions}\label{sec:problem}

This section defines the trees, fixes their dimensions, and introduces the random projections and the coupled pair of instances on which the lower bound rests.

\subsection{Trees and the promise}\label{sec:schedule}

Fix a code family $C$ of degree $d$ (Definition~\ref{def:code}). A tree of height zero is a single Boolean leaf, and its value is that bit. A tree of height $j\ge1$ consists of a node of some dimension $m$, a positive multiple of eight so that $C_m$ is defined, together with $2^m$ code children indexed by $x\in\F^m$ and $2^m$ data children indexed by $y\in\F^m$. Each child is a tree of height $j-1$, and the children are disjoint. Every tree of height zero is valid, and a tree of height $j\ge1$ is valid if every child is valid and the table of code-child values, $L_x$ for $x\in\F^m$, is the codeword of some secret $s$. We call $L$ the code table of the node and the table $R$ of data-child values, $R_y$ for $y\in\F^m$, its data table. The data table is unconstrained, and the value of the node is $R_s$, that is, the addressing function $G^C_m$ applied to the two tables of child values. Only the leaves are input coordinates. Since the code is injective, the leaves of a valid tree determine the values of all nodes and all secrets uniquely.

As a small illustration, we take a node of dimension $m=2$ and height one with the Hadamard code. If its secret is $s=(1,0)$, its code table is $L_{00}=L_{01}=0$ and $L_{10}=L_{11}=1$, its value is the data leaf $R_{10}$, and the other three data leaves are unconstrained. A node of height two can have four code children and four data children of this form. Its code and data tables then consist of the values of those children. The dimension-two example only illustrates the construction, and the family below uses multiples of eight.

For $n\ge1$ let
\begin{equation}\label{eq:schedule}
 h_{\max}(n)=\Bigl\lfloor\frac{\log n}{8}\Bigr\rfloor,\qquad
 m_i=8\Bigl\lfloor\frac{n}{8\cdot16^i}\Bigr\rfloor\quad(0\le i<h_{\max}(n)),
\end{equation}
so that $h_{\max}(n)=0$ for $n<256$. An instance of size parameter $n$ and height $h$, where $0\le h\le h_{\max}(n)$, is a valid tree of height $h$ in which nodes of height $j$ have dimension $m_{j-1}$. An instance of height zero is a single leaf. We continue to say ``size $n$'' for this parameter, although the table of leaf values itself has exponentially many bits when $h\ge1$.

Theorem~\ref{thm:main} uses the full height $h=h_{\max}(n)$, and Theorem~\ref{thm:slow} uses the heights $h(n)$ of a height function. Throughout Sections~\ref{sec:problem} to~\ref{sec:lower}, $h$ denotes the height under consideration, and bounds are uniform in $h\le h_{\max}(n)$ unless stated otherwise.

The dimensions are largest at the bottom and decrease towards the root. This keeps the length~\eqref{eq:address} of a leaf address linear in $n$, and it makes the logarithm of the number of nodes at each level small compared with the dimension of that level, as~\eqref{eq:blocks} shows. The factor sixteen is discussed after~\eqref{eq:blocks}, and the bound $h\le h_{\max}(n)$ keeps every dimension at least $8\sqrt n$.

The nodes just above the leaves have dimension $m_0\ge n-7$. For $h\ge1$ the root has dimension $m_{h-1}\ge8\lfloor2\sqrt n\rfloor\ge8\sqrt n$, since $16^{h-1}\le\sqrt n/16$. For every fixed $i$, $m_i=\Theta(n)$ with constants depending on $i$ and not on $h$. More precisely, $m_i\le n/16^i<m_i+8$, so for $0\le v<i<h_{\max}(n)$ and $0\le v<h'\le h_{\max}(n)$,
\begin{equation}\label{eq:geometric}
 m_i\le\frac{m_v+8}{16^{i-v}}
 \qquad\text{and}\qquad
 \sum_{i=v+1}^{h'-1}(m_i+1)\le\frac{m_v+8}{15}+h'-v-1 .
\end{equation}
A leaf is addressed by the path from the root. At a node of dimension $m_i$, one bit says whether to descend into a code child or a data child, and $m_i$ bits give the index of the child. Hence the address of a leaf has length
\begin{equation}\label{eq:address}
 a(n)=\sum_{i<h}(m_i+1)\le\frac{16n}{15}+h<2n,
\end{equation}
since $m_i\le n/16^i$ and $h\le\log n/8$, and the number of leaves is $2^{a(n)}$. The length $a(n)$ depends on $h$, which we suppress from the notation.

In the promise problem of size $n$ and height $h$, the input is a table of $2^{a(n)}$ bits indexed by the leaf addresses. The promise is that it is the leaf table of an instance of size $n$ and height $h$, and the answer is the value of the root.

\subsection{Random projections and the coupled pair}\label{sec:projections}

For a single node of dimension $m$, its random projection chooses a uniformly random secret $S\in\F^m$ and fixes every code entry to $L_x=C_m(S,x)$. It fixes every data entry $R_y$ with $y\ne S$ to an independent uniformly random bit, and it maps $R_S$ to a free variable $z$. The secret selects the data child $S$, which we call the free child of the node. The coordinate $R_S$, which carries the value of this child, is the free leaf of the node. It is the only free coordinate of the projection in the sense of Section~\ref{sec:prelim}, and Claim~\ref{claim:proj}(d) shows that a block of several levels, defined below, also has exactly one free leaf. The projection is sampled independently of $z$, and either value of $z$ gives a valid node of value $z$. The projection fixes everything about the node except the value of its free child, and the fixed code coordinates determine the secret and hence the position of the free child. This correlation between the fixed coordinates and the position of the free leaf sets these projections apart from random restrictions that fix the other coordinates independently (Section~\ref{sec:outline-lower}).

We number the levels of an instance from the bottom. The nodes of dimension $m_i$ form level $i$, for $0\le i<h$, and the leaves form level $-1$, so that a node at level $i$ has height $i+1$. The random projections at all nodes of the instance are independent, and $\rho_i$ applies those at the nodes of level $i$ simultaneously. Let $N_i$ be the number of nodes at level $i$. The projection $\rho_i$ maps the vector of node values at level $i$ to the vector of node values at level $i-1$. It fixes the children of every node at level $i$ as described and copies the value of the node into its free child. The composition
\begin{equation}\label{eq:canonical}
 X_b=\rho_0\bigl(\rho_1(\cdots\rho_{h-1}(b)\cdots)\bigr),\qquad b\in\{0,1\},
\end{equation}
is a random leaf table, and $\mu_{n,h,b}$ is its distribution. The table $X_b$ is the leaf table of an instance with root value $b$, since every node is valid by construction and the free child of every node carries the value of the node. Using the same projections $\rho_0,\dots,\rho_{h-1}$ for both values of $b$ gives the coupled pair $(X_0,X_1)$, whose marginals are $\mu_{n,h,0}$ and $\mu_{n,h,1}$.

For integers $u\ge0$ and $j\ge1$ with $u+j\le h$, a node at level $u+j-1$ together with its descendants at the levels $u+j-2,\dots,u-1$ is a block of height $j$ with dimensions $m_u,\dots,m_{u+j-1}$, and its leaves are its descendants at level $u-1$.

\begin{claim}\label{claim:proj}
The random projections have the following properties.
\begin{enumerate}[label=(\alph*),nosep]
\item The tables $X_0$ and $X_1$ of the coupled pair differ in exactly one leaf, the leaf reached from the root by descending into the free child at every level. All secrets and all other leaves coincide.
\item For $0\le v\le h$, let $Z_b=\rho_v\circ\cdots\circ\rho_{h-1}(b)$ be the vector of node values at level $v-1$ generated from the root value $b$, and let $\rho_{<v}=\rho_0\circ\cdots\circ\rho_{v-1}$, the identity for $v=0$. Then $X_b=\rho_{<v}(Z_b)$, and the projections $\rho_0,\dots,\rho_{v-1}$, and hence $\rho_{<v}$, are independent of $\rho_v,\dots,\rho_{h-1}$ and therefore of the pair $(Z_0,Z_1)$.
\item Each $\rho_i$ is defined on every $z\in\F^{N_i}$, whether or not $z$ is the vector of node values of a valid tree. Consequently, for $0\le v\le h$, $F\circ\rho_{<v}$ is a function on the whole cube $\F^{N_{v-1}}$ for every Boolean function $F$ on the leaves.
\item For a block of height $j$ with dimensions $m_u,\dots,m_{u+j-1}$, the random projections of its nodes compose to a projection with exactly one free leaf, and this projection is independent of the value prescribed to the root of the block. The two values of the root give a pair of inputs of the block that differ in that leaf. Fixing $z\in\F^{N_{u+j-1}}$ leaves independent blocks rooted at level $u+j-1$ with prescribed root values, one for each coordinate of $z$.
\end{enumerate}
\end{claim}

\begin{proof}
For (c), the projection of a node does not depend on the value prescribed to the node. The value is only copied into the free child, whose position is determined by the secret. Hence $\rho_i$ is defined on every $z\in\F^{N_i}$, whether or not $z$ is the vector of node values of a valid tree. It fixes the children of every node at level $i$ as described and copies the coordinate of $z$ at that node into its free child. The same holds for $\rho_{<v}$, which gives (c).

For (a), the same observation shows that changing the root value changes only the values copied along the path of free children from the root, hence only the leaf at the end of that path, which carries the root value.

For (b), the projections $\rho_0,\dots,\rho_{h-1}$ are chosen independently, so $\rho_{<v}$ is independent of $(\rho_v,\dots,\rho_{h-1})$. The pair $(Z_0,Z_1)$ is a function of the latter, since it is obtained by applying $\rho_{h-1},\dots,\rho_v$ to the two root values, and $X_b=\rho_{<v}(Z_b)$ by~\eqref{eq:canonical}.

For (d), the argument for (a) applied to a block shows that the projections of its nodes fix every leaf off the path of free children from its root to a constant. The composed projection therefore has exactly one free leaf, the end of that path, which carries the root value, and the two root values give inputs that differ exactly there. The blocks rooted at level $u+j-1$ are independent because the projections of distinct nodes are independent.
\end{proof}

The number of nodes at level $v$, for $0\le v\le h-1$, is
\begin{equation}\label{eq:blocks}
 N_v=2^{\sum_{i=v+1}^{h-1}(m_i+1)},\qquad
 \log N_v\le\frac{m_v}{15}+\frac{\log n}{8},
\end{equation}
because every node at a level $i>v$ has $2^{m_i+1}$ children, namely $2^{m_i}$ code children and $2^{m_i}$ data children. By~\eqref{eq:geometric} with $h'=h$ and $v\ge0$, the exponent is at most $(m_v+8)/15+h-v-1\le m_v/15+h-7/15$, and $h\le\log n/8$ gives the bound in~\eqref{eq:blocks}, which does not depend on $h$. We also write $N_{-1}=2^{a(n)}$ for the number of leaves.

The factor sixteen between consecutive dimensions is chosen for the lower bound. By~\eqref{eq:blocks}, a level of dimension $m_v$ has at most $2^{m_v/15+\log n/8}$ nodes, and the number of nodes of a level within a block is bounded in the same way by~\eqref{eq:geometric}, up to an additive $O(\log n)$ in the exponent. The union bound of Theorem~\ref{thm:winner}, as applied in Claim~\ref{claim:loss}, multiplies a bound of about $2^{-(1-H_2(1/16))m_v}$ for each node whose free leaf occurs in the first satisfied term together with few code literals of that node (a light block, Section~\ref{sec:rank}) by this number. The $1/16$ inside $H_2$ comes from the threshold on code literals in that claim, not from the ratio of the dimensions. The union bound of Theorem~\ref{thm:certtail}, applied in the proof of Theorem~\ref{thm:switch}, multiplies the bound $2^{-m_v/4}$ of Theorem~\ref{thm:flips} for each block by the square of this number. Both products are exponentially small in $m_v$.

\section{Exact quantum evaluation and classical verification}\label{sec:upper}

We first give an exact quantum algorithm, then the recursive verification protocol, and finally the form of the protocol batched by height, which Theorems~\ref{thm:main} and~\ref{thm:slow} use.

\subsection{The exact quantum algorithm}\label{sec:exact}

We fix a code family of Definition~\ref{def:code} and consider valid trees with the dimensions~\eqref{eq:schedule}. A quantum algorithm for a tree is clean if it adds the value of the tree to an output qubit and returns all of its work registers to their initial states, which is achieved by uncomputation~\cite{Ben73}. A clean algorithm can be called in superposition over several subtrees. The address of a child is the address of its parent node followed by a bit, $0$ for a code child and $1$ for a data child, and the index of the child. Thus $w0x$ is the code child $x$ and $w1y$ the data child $y$ of the node with address $w$. The algorithms below keep the address of the current node in a register that they never modify.

\begin{theorem}\label{thm:exact}
There is a uniform clean quantum algorithm that, on every valid tree of height $j$, computes the value of the tree exactly with $3^j$ queries to the leaves and in time $3^j\poly(n,j)$.
\end{theorem}

\begin{proof}
For $j=0$ one query adds the leaf to the output qubit. Let $\calA_{j-1}$ be the clean algorithm for height $j-1$, and consider a node of dimension $m$ at height $j$ with secret $s$. Prepare a uniform superposition over $x\in\F^m$ in a fresh register and a phase ancilla in the state $\ket-$. Call $\calA_{j-1}$ on the code child $x$ in superposition, with this ancilla as output qubit. Adding a bit $c$ to an ancilla in the state $\ket-$ multiplies the state by $(-1)^c$, so the call multiplies branch $x$ by $(-1)^{C_m(s,x)}$. Because $\calA_{j-1}$ is clean, on every branch its work registers return to their initial states and the ancilla returns to $\ket-$, so the branches differ only in this phase. The register storing $x$ therefore holds the phase state $\ket{C_s}$, which $U_m$ maps to $\ket s$ by Definition~\ref{def:code}(iii). If the work registers kept information about $x$, this register would be entangled with them and $U_m$ would not in general return $\ket s$.

Now call $\calA_{j-1}$ on the data child $s$ with the node's output qubit, adding $R_s$ to it. Apply $U_m^{-1}$ and call $\calA_{j-1}$ on the code children again with the phase ancilla. This call removes the code phases. Finally reverse the preparation of the uniform superposition and of the phase ancilla. Every work register returns to its initial state, and the node's address is unchanged, so the algorithm is clean.

There are three child calls per node, hence $3^j$ leaf queries. The recursion expands into fewer than $2\cdot3^j$ calls, with polynomial additional work per recursive call, which gives the time bound. At each height the same uniformly generated circuit works for every node, with its address supplied in the preserved register.
\end{proof}

For the Hadamard code with $m=2$ and the secret $s=(1,0)$ of Section~\ref{sec:schedule}, the phase query turns the uniform superposition into $\frac12(\ket{00}+\ket{01}-\ket{10}-\ket{11})=\ket-\ket+$, and $H\otimes H$ maps this state to $\ket{10}=\ket s$.

\subsection{Recursive verification}\label{sec:verify}

A claim $(w,b)$ asserts that the node with address $w$ has value $b$. The empty address names the root. Both parties know the size parameter, the height and dimensions of the instance, the fixed code family, and the claim. They query the same leaf table. The verifier's queries are classical, and the honest quantum prover may query in superposition. All messages are classical.

The current claim, the order of recursive calls, and the message lengths are determined by these public parameters and the transcript. No separate requests or node addresses are sent. The protocol follows the recursive secret-checking protocol of McKague~\cite{McK10}, with one additional recursive check for the data entry.

\begin{protocolenv}[Recursive verification]\label{prot:verify}
\textbf{Input.} A claim $(w,b)$ in a valid instance. Both parties have oracle access to the leaves as described above.

\begin{enumerate}[label=\arabic*.,leftmargin=*,itemsep=0.5em,topsep=0.5em]
\item \textbf{Leaf check.} If $w$ is a leaf, the verifier queries it and accepts if and only if the answer is $b$.
\item \textbf{Prover to verifier.} Otherwise, let $m$ be the dimension of $w$. The prover sends the $m$-bit string $s$ claimed to be its secret. The verifier retains $s$ for all subsequent checks.
\item \textbf{Code checks.} Three times in sequence, the verifier samples a fresh uniformly random $x\in\F^m$ and sends it to the prover. The parties run this protocol recursively on the claim that code child $w0x$ has value $C_m(s,x)$, a code claim.
\item \textbf{Data check.} The parties run this protocol recursively on the claim that data child $w1s$ has value $b$, the data claim.
\item \textbf{Decision.} The verifier accepts if and only if all four recursive checks accept. All four checks are executed in the prescribed order, including after a failed check.
\end{enumerate}
Messages must have the prescribed format and length, and otherwise the verifier rejects.
\end{protocolenv}

When consecutive challenges of the verifier are not separated by a message of the prover, as for the three code checks of a claim of height one, the verifier sends them in one message, so that the messages alternate as in Section~\ref{sec:oracle-classes}. Protocol~\ref{prot:verify} makes the four checks of a claim one after the other, so a claim about a node of height $j\ge1$ takes $\Theta(4^j)$ messages. Theorems~\ref{thm:main} and~\ref{thm:slow} use the form of the protocol batched by height, which the next subsection gives. We analyze soundness there, through one invariant for lists of claims (Claim~\ref{claim:batched}). Appendix~\ref{app:noisy} modifies Protocol~\ref{prot:verify} for noisy code tables and analyzes the modified protocol there.

\subsection{Verification batched by height}\label{sec:batched}
The checks of all claims about nodes of the same height can be made at once. The prover sends a claimed secret for every claim on a list in one message, and the verifier answers with all challenges in one message. The list of claims about the next height is determined by the transcript. Revealing the challenges of all claims at once gives the prover more information than it has in Protocol~\ref{prot:verify}, so the soundness analysis controls all false claims on a list jointly, through an invariant on the number of false claims (Claim~\ref{claim:batched}).

\begin{protocolenv}[Verification batched by height]\label{prot:batched}
\textbf{Input.} A claim $(w,b)$ about a node of height $j$ in a valid instance. Both parties have the same public parameters and oracle access to the leaves as in Protocol~\ref{prot:verify}.

\textbf{Initialization.} Form an ordered list containing just $(w,b)$. Throughout, repeated claims about the same node are kept as separate entries.

For the heights $j,j-1,\dots,1$ in turn, let $m$ be the common dimension of the nodes on the current list and perform the following.
\begin{enumerate}[label=\arabic*.,leftmargin=*,itemsep=0.5em,topsep=0.5em]
\item \textbf{Prover to verifier.} In one message, the prover sends an $m$-bit claimed secret $s$ for every entry $(w',b')$, in list order. The complete message is received before any challenges for this height are sampled.
\item \textbf{Verifier to prover.} For every entry, the verifier samples three fresh independent uniformly random strings $x_1,x_2,x_3\in\F^m$. It sends all triples in one message, in list order. Different entries receive independent triples even when they refer to the same node.
\item \textbf{Next list.} Both parties replace each entry $(w',b')$, with its secret $s$ and challenge triple, by the four claims
\[
 (w'0x_1,C_m(s,x_1)),\quad
 (w'0x_2,C_m(s,x_2)),\quad
 (w'0x_3,C_m(s,x_3)),\quad
 (w'1s,b'),
\]
in this order.
\end{enumerate}
\textbf{Final check.} When the list consists of leaves, the verifier queries every entry and accepts if and only if every claimed value matches the leaf value. At height zero this is the only check.

Messages must have the prescribed format and length, and otherwise the verifier rejects.
\end{protocolenv}

Both parties compute the list after every message from the transcript, so the verifier sends only random bits and the proof system is public-coin. The soundness analysis bounds the probability that the verifier accepts, when the protocol is continued from a list, in terms of the number of false claims on the list.

\begin{claim}[Joint soundness for lists of claims]\label{claim:batched}
Fix a valid tree and $j\ge0$. Consider any list of claims about nodes of height $j$ on which $F$ claims are false, any prover, classical or quantum, that exchanges classical messages with the verifier, and any history of the interaction before the list is processed. Then the conditional probability, given this history, that Protocol~\ref{prot:batched}, continued from this list, accepts is at most $4^{-F}$.
\end{claim}

\begin{proof}
We use induction on $j$. At height zero the verifier accepts exactly when no leaf claim is false, so the conditional probability is $0$ when $F\ge1$ and $1=4^{0}$ when $F=0$.

Let $j\ge1$. A malformed message is rejected, so we may assume that the prover's message gives a claimed secret for every entry on the list, and we condition on the history and on this entire message. The verifier samples the challenges afterwards, so they remain independent and uniform under this conditioning. This is where the order of the messages in Protocol~\ref{prot:batched} is used.

For each false claim $e=(w',b')$ on the list, we define a lower bound $Y_e$ on the number of false claims among its four child claims. Let $s$ be the claimed secret of this entry and $s^*$ the true secret of the node $w'$. If $s=s^*$, the data claim $(w'1s^*,b')$ is false, because the value of the node is $R_{s^*}\ne b'$, and we put $Y_e=1$. If $s\ne s^*$, we let $Y_e$ be the number of false claims among the three code claims. The codewords of $s$ and $s^*$ differ in exactly half of the coordinates (Section~\ref{sec:codes}), so each code claim is false with probability exactly $1/2$, and $Y_e$ has the binomial distribution with parameters $3$ and $1/2$. The variables $Y_e$ depend on disjoint sets of fresh challenges, so they are independent given the claimed secrets, even for claims about the same node.

Let $F'$ be the number of false claims on the new list. Child claims remain separate entries, so $F'$ is at least the sum of the $Y_e$. For every fixed outcome of the challenges, the induction hypothesis, applied to the new list and the history extended by the two messages of this height, bounds the conditional acceptance probability by $4^{-F'}$, and $4^{-F'}\le\prod_e4^{-Y_e}$, with the product over the false claims $e$ of the current list. Taking the expectation over the challenges, and using the independence of the $Y_e$, we obtain
\[
 \Pr[\text{accept}\mid\text{history, claimed secrets}]\le\prod_e\E\bigl[4^{-Y_e}\bigr].
\]
When $s=s^*$, the factor $\E[4^{-Y_e}]$ is $1/4$. When $s\ne s^*$, the three code claims are false independently with probability $1/2$ each, so the factor is $((1+1/4)/2)^3=(5/8)^3=125/512<1/4$. Hence the product is at most $4^{-F}$. Averaging over the claimed secrets gives the claim for every fixed history. The conditioning on the claimed secrets is legitimate for a quantum prover because its messages are classical. The induction hypothesis covers every continuation strategy and every state of the prover's private memory, so no independence between the prover's answers to different claims is assumed.
\end{proof}

We use three code checks because the inductive step requires $(5/8)^k\le1/4$, where $k$ is the number of code checks. This holds for $k=3$ and fails for $k=2$, since $(5/8)^2=25/64$.

\begin{theorem}\label{thm:batched}
Let the tree be an instance of size $n$. Let the honest prover be the uniform quantum algorithm that obtains every secret it sends by running the algorithm of Theorem~\ref{thm:exact} for the node up to its application of the decoder $U_m$ and measuring the register that holds $\ket s$. On a claim about a node of height $j$, Protocol~\ref{prot:batched} has the following properties.
\begin{enumerate}[label=(\roman*),nosep]
\item It accepts every true claim with probability one when the prover is honest.
\item For every false claim and every prover, classical or quantum, that exchanges classical messages with the verifier, it accepts with probability at most $1/4$.
\item It is public-coin, and the verifier and the prover exchange $2j$ messages, $j$ from each party. The prover sends fewer than $n\cdot4^j/3$ bits in total and the verifier fewer than $n\cdot4^j$ random bits. The verifier makes at most $4^j$ leaf queries and runs in time $4^j\poly(n,j)$.
\item The honest prover makes $4^j-3^j$ leaf queries and runs in time $4^j\poly(n,j)$.
\end{enumerate}
\end{theorem}

\begin{proof}
For (i), the secrets of the honest prover are correct, because the algorithm of Theorem~\ref{thm:exact} is exact. If a claim $(w',b')$ is true and $s$ is the true secret of $w'$, then every child claim is true, since the code child $x$ has value $C_m(s,x)$ and the data child $s$ has value $b'$. Hence every claim on every list is true, and the final queries accept. Part (ii) is Claim~\ref{claim:batched} with $F=1$, applied to the initial list and the empty history.

For (iii), the list at height $j-i$ has $4^i$ entries, each of dimension $m_{j-i-1}\le n$ when $i<j$. The prover sends one message at each positive height, of $4^im_{j-i-1}$ bits, so the total is at most $n\sum_{i<j}4^i=n(4^j-1)/3<n\cdot4^j/3$. Each verifier message has three times as many random bits as the prover's message preceding it. There is one message from each party at each of the heights $j,\dots,1$, and the protocol is public-coin by the remark after Protocol~\ref{prot:batched}. The verifier makes at most $4^j$ leaf queries. Computing $C_m(s,x)$ and handling addresses of length $O(n)$ gives its time bound.

For (iv), recovering a secret at height $j'$ uses one call of the algorithm of Theorem~\ref{thm:exact} for height $j'-1$ on the code children, followed by $U_m$ and a measurement, so it uses $3^{j'-1}$ leaf queries. There are $4^{j-j'}$ entries at height $j'$. The total is
\[
 \sum_{j'=1}^{j}4^{j-j'}3^{j'-1}=4^j-3^j.
\]
Each computation has only polynomial overhead per leaf query, so the total running time, including setup, is $4^j\poly(n,j)$.
\end{proof}

For an instance of size $n$ and height $h$, a claim about the root is verified with $2h\le\log n/4$ messages.

\section{The constant-depth circuit lower bound}\label{sec:lower}

This section proves Theorem~\ref{thm:circuits}. Section~\ref{sec:rank} bounds, by a rank argument, the probability that the first satisfied term of a DNF contains the free leaves of many blocks while testing few code bits of each of them. Section~\ref{sec:flips} turns this into a bound on the probability that a DNF accepts an input and is sensitive there to the free leaf of every block. Section~\ref{sec:certtail} derives a tail bound for certificate complexity, Section~\ref{sec:switch} the switching theorem, and Section~\ref{sec:circuits} the depth reduction and the coupling bound.

\subsection{A rank bound for the first satisfied term}\label{sec:rank}

Theorem~\ref{thm:winner} bounds the probability that the first satisfied term of a DNF contains the free leaves of many blocks and few code literals of each of them. Its proof counts secrets by a rank argument (Claim~\ref{claim:count}).

In Theorem~\ref{thm:winner} and Claims~\ref{claim:twostep} to~\ref{claim:truncate}, a block is a single node of dimension $m$ with its own random projection, hence a block of height one in the sense of Section~\ref{sec:projections}. We consider $K$ independent blocks, a DNF $\Phi$ on their leaves, and a fixed value $z\in\F^K$ of the free variables, the free values of the blocks, and write $\rho$ for their joint projection, so that $\rho(z)$ is the input with these prescribed values.

We normalize $\Phi$ (Section~\ref{sec:prelim}) and give its terms a fixed order, indexing them by distinct integers. At an accepting input we consider the first satisfied term. Its index is the value of the DNF index function of Lovett and Zhang~\cite[Definition~1.10]{LZ19}, and we study this term under the random projections, whose code tables determine the positions of the free leaves. For a block of height one, a code literal is a literal on one of its code children and a data literal is a literal on one of its data children. Fix an integer $t\ge0$. We say that a block is \emph{light} for a term if the term contains the block's free leaf and at most $t$ code literals in that block. The secret determines which leaf is free, so the set of light blocks depends on $\rho$.

For a single term $T$ of width at most $W$, the event that $T$ contains the free leaf of a given block has probability at most $W2^{-m}$. This holds because the free leaf is the data leaf at the uniformly random secret and $T$ contains at most $W$ data leaves of the block. The number of terms of $\Phi$ is not bounded, so a union bound over all terms is not useful. Theorem~\ref{thm:winner} considers only the first satisfied term. In the auxiliary experiment of Claim~\ref{claim:twostep} below, we fix the data tables of $\ell$ chosen blocks, including the entries at the secrets, and the projections and free values of all other blocks. In this experiment the free leaf of a chosen block is still its data leaf at the secret, but the bit there is fixed by the table, so changing the secret changes which leaf is free and leaves every entry of the table unchanged. The rank argument in the proof of Claim~\ref{claim:count} shows that at most $\binom{m}{\le dt}^{\ell}$ terms are the first satisfied term at some secret tuple at which all $\ell$ chosen blocks are light for it, independently of the number of terms of $\Phi$.

\begin{theorem}[Light blocks in the first satisfied term]\label{thm:winner}
Let $\Phi$ be a DNF of width at most $W$ on $K$ independent blocks of dimension $m$ with a code family of degree $d$, let $t\ge0$ be an integer, and let $z\in\F^K$. For every integer $1\le\ell\le W$,
\begin{multline}\label{eq:winner-tail}
 \Pr_\rho\bigl[\Phi(\rho(z))=1\ \text{and at least $\ell$ blocks are light for the first satisfied term}\bigr]\\
 \le\binom K\ell\Bigl(\frac{2W}{\ell}\Bigr)^{\ell}\binom{m}{\le dt}^{\ell}2^{-m\ell}.
\end{multline}
If $\ell>K$ the event is empty.
\end{theorem}

The proof has two parts. Claim~\ref{claim:twostep} replaces the projections of $\ell$ chosen blocks by an experiment in which their data tables are uniformly random, at the cost of a factor $2^{\ell}$. Claim~\ref{claim:count} bounds, for fixed data tables, the number of secret tuples at which the $\ell$ chosen blocks are light for the first satisfied term.

\Needspace{14\baselineskip}
\begin{claim}\label{claim:twostep}
Fix a set $I$ of $\ell$ blocks, and fix the projections and the free values of all other blocks. Consider the experiment in which every block of $I$ receives a uniformly random secret and a uniformly random data table, all independent, and its code table is the codeword of its secret. In this experiment the data entry at the secret is sampled uniformly, without being constrained to equal the prescribed free value.

Conditioned on the event that in every block of $I$ the data entry at the secret equals the value prescribed by $z$, the experiment has the law of the random projections of the blocks of $I$ with free values $z$. The conditioning event has probability exactly $2^{-\ell}$. Hence, for every event $E$ that is determined by the secrets of the blocks of $I$ and by the resulting input,
\[
 \Pr_{\text{projections}}[E]\le2^{\ell}\Pr_{\text{experiment}}[E].
\]
\end{claim}

\begin{proof}
Given the secrets, the data entries at the secrets are independent uniform bits, so the conditioning event has probability $2^{-\ell}$. Conditioning leaves the secrets uniform and all other data entries independent and uniform, which gives the projection law. The inequality follows from conditional probability. It compares probabilities after averaging over the data tables and does not assert the inequality for each fixed table.
\end{proof}

\begin{claim}[Counting secret tuples by rank]\label{claim:count}
In the experiment of Claim~\ref{claim:twostep}, fix the data tables $f=(f_i)_{i\in I}$ of the blocks of $I$, where $f_i\colon\F^m\to\F$. Let $P(f)$ be the set of secret tuples $\tau=(\tau_i)_{i\in I}$ for which $\Phi$ accepts the resulting input and every block of $I$ is light for the first satisfied term. Then
\[
 |P(f)|\le\underbrace{\Bigl(\frac W\ell\Bigr)^{\ell}}_{\text{tuples with a given first satisfied term}}\;\underbrace{\binom{m}{\le dt}^{\ell}}_{\text{possible first satisfied terms}} .
\]
\end{claim}

\begin{proof}
We partition $P(f)$ according to the index $\lambda$ of the first satisfied term $T_\lambda$, and bound the size of each class and then the number of classes.

Let $\tau$ be in the class of $\lambda$ and let $i\in I$. The free leaf of block $i$ is its data leaf with index $\tau_i$, which carries the bit $f_i(\tau_i)$. Since block $i$ is light for $T_\lambda$, the term $T_\lambda$ contains this leaf, so $\tau_i$ is one of the indices of data leaves of block $i$ that occur in $T_\lambda$. If $T_\lambda$ contains $k_i$ data literals of block $i$, the class of $\lambda$ therefore has at most $\prod_{i\in I}k_i$ elements. Since $\sum_{i\in I}k_i\le W$, the inequality between the arithmetic and the geometric means gives
\[
 \prod_{i\in I}k_i\le\Bigl(\frac{W}{\ell}\Bigr)^{\ell}.
\]

To count the classes, we choose a representative $\tau_\lambda$ in every nonempty class and write every code literal of a term in a block of $I$ as a triple $(i,x,\beta)$, which asserts that the code entry $L_x$ of block $i$ equals the bit $\beta$. For an index $\lambda$ let
\[
 g_\lambda(\tau)=\prod_{(i,x,\beta)}\bigl(1+C_m(\tau_i,x)+\beta\bigr),
\]
where the product is over the code literals of $T_\lambda$ in the blocks of $I$. Each factor is the indicator, over $\F$, that its literal holds. Thus $g_\lambda(\tau)=1$ exactly when all these code literals hold. With the data tables and the leaves of the blocks outside $I$ fixed, only the code tables in $I$ vary with $\tau$. We form the matrix $M_{\lambda\xi}=g_\lambda(\tau_\xi)$ on the nonempty classes. Its diagonal entries are one, since $T_\lambda$ is satisfied at $\tau_\lambda$.

If $M_{\lambda\xi}=1$, the code literals of $T_\lambda$ in $I$ hold at $\tau_\xi$. Its data literals hold at its representative $\tau_\lambda$ and keep the same values at $\tau_\xi$ because the tables $f$ are fixed. Its literals on blocks outside $I$ also keep their values, because those leaves are fixed. Thus $T_\lambda$ is satisfied at $\tau_\xi$, where $T_\xi$ is the first satisfied term, so $\xi\le\lambda$. With classes ordered by index, $M$ is lower triangular with diagonal one and has linearly independent rows over $\F$.

Each row of $M$ evaluates its function $g_\lambda$ at the representatives. A linear relation among these functions would give the same relation among the rows. Their number is therefore at most the dimension of a function space containing them.

Every factor has degree at most $d$ in its secret, by Definition~\ref{def:code}(ii), and every block in $I$ has at most $t$ code literals in $T_\lambda$. The product has degree at most $dt$ in each secret $\tau_i$. Reducing powers by $\tau_{i,a}^2=\tau_{i,a}$ gives the multilinear polynomial that agrees with it on Boolean inputs, which preserves its values there and does not increase the degree in any $\tau_i$. These multilinear polynomials lie in the space of multilinear polynomials of degree at most $dt$ in each $\tau_i$, which is spanned by the products of one monomial of degree at most $dt$ in each $\tau_i$ and has dimension $\binom{m}{\le dt}^{\ell}$. This bounds the number of classes.
\end{proof}

\begin{proof}[Proof of Theorem~\ref{thm:winner}]
If at least $\ell$ blocks are light for the first satisfied term, then there is a set $I$ of $\ell$ blocks that are all light for the first satisfied term. Fix $I$, and fix the projections and the free values of the other blocks. By Claim~\ref{claim:twostep}, the probability under the projections that every block of $I$ is light for the first satisfied term is at most $2^{\ell}$ times its probability in the experiment. In the experiment the secret tuple is uniform on $(\F^m)^I$ and independent of the data tables $f$, so the latter probability equals
\[
 \E_f\Bigl[\frac{|P(f)|}{2^{m\ell}}\Bigr]\le\Bigl(\frac W\ell\Bigr)^{\ell}\binom{m}{\le dt}^{\ell}2^{-m\ell},
\]
where the inequality is Claim~\ref{claim:count}. The bound holds for every value of the projections of the blocks outside $I$, hence after averaging over them, and summing over the $\binom K\ell$ choices of $I$ gives~\eqref{eq:winner-tail}. No conditioning on the identity of the first satisfied term or on acceptance was used.
\end{proof}

\subsection{Term deletion and sensitivity to the free leaves}\label{sec:flips}

For an integer $J\ge0$, we delete every term for which more than $J$ blocks are light and project the remaining terms. Deleting whole terms preserves every rejection on the entire cube of free values, and it can lose acceptance at a fixed input only by deleting the first satisfied term there. Claim~\ref{claim:truncate} bounds the width of the result, and Theorem~\ref{thm:winner} with $\ell=J+1$ bounds the probability of this loss when $J+1\le W$. When $J+1>W$, no term is deleted, since a term for which more than $J$ blocks are light would contain the free leaves of more than $J$ distinct blocks and so would have width more than $W$.

\begin{claim}[Term deletion and width reduction]\label{claim:truncate}
In the setting of Theorem~\ref{thm:winner}, fix an integer $J\ge0$. Let $\Phi^-_\rho$ be the DNF on $z$ obtained from $\Phi$ by deleting every term for which more than $J$ blocks are light and projecting the remaining terms by $\rho$. Then, for every outcome of $\rho$ and simultaneously for all $z\in\F^K$,
\[
 \Phi^-_\rho(z)\le\Phi(\rho(z))
 \qquad\text{and}\qquad
 \width(\Phi^-_\rho)\le\Bigl\lfloor\frac{W}{t+1}\Bigr\rfloor+J .
\]
The DNF $\Phi^-_\rho$ is determined by $\Phi$ and $\rho$ and does not depend on $z$. If $\Phi(\rho(z))=1$ and $\Phi^-_\rho(z)=0$, then more than $J$ blocks are light for the first satisfied term of $\Phi$ at $\rho(z)$.
\end{claim}

\begin{proof}
Deleting whole terms can only turn acceptance into rejection, and projection preserves each retained term's value. For the width, consider a retained term whose projection is satisfiable. Its remaining literals are on free variables. Distinct literals belong to distinct blocks, because each block has only one free leaf and the term is normalized. At most $J$ of these blocks are light. Every other such block contributes at least $t+1$ original code literals, on disjoint coordinates across blocks, so there are at most $\lfloor W/(t+1)\rfloor$ of them.

The deletion rule uses the secrets and original term supports, not $z$. If the first satisfied term at $\rho(z)$ is retained, its projected term is still satisfied at $z$, proving the final assertion.
\end{proof}

Now let the blocks be trees of several levels. Consider $r$ independent blocks with prescribed root values, generated by their random projections, and let $X$ be the resulting leaf table. For $1\le i\le r$, let $X^{(i)}$ be the leaf table in which block $i$ alone is generated from the opposite root value, with the same projections. By Claim~\ref{claim:proj}(d), $X$ and $X^{(i)}$ differ in exactly one leaf, the free leaf of block $i$. A block of height zero is a single coordinate carrying its prescribed value. For a DNF $\Phi$ on the leaves of the blocks, we consider the event
\begin{equation}\label{eq:sensitive}
 \Phi(X)=1\qquad\text{and}\qquad\Phi(X^{(i)})=0\quad\text{for all }1\le i\le r ,
\end{equation}
that $\Phi$ accepts $X$ and is sensitive at $X$ to the free leaf of every block, in the usual sense of the sensitivity of a Boolean function to a coordinate~\cite{CDR86,Nis91}. If the width of $\Phi$ is less than $r$, this event is empty, since a term satisfied at $X$ must contain, for every $i$, the leaf at which $X^{(i)}$ differs from $X$, and these $r$ leaves are distinct. Section~\ref{sec:certtail} applies Theorem~\ref{thm:flips} below to negated CNFs. When a projected CNF accepts an input without a small $1$-certificate, flipping all coordinates of any one of certain disjoint sets falsifies it, while flipping all but one coordinate of that set does not (Claims~\ref{claim:minimal} and~\ref{claim:family}).

\begin{theorem}[Sensitivity to every block]\label{thm:flips}
Fix a code family of degree $d$, a level $u\ge0$, and an exponent $\alpha\ge0$, all independent of $n$, and put $j=\lfloor\alpha\rfloor+1$ and $\nu=m_{u+j-1}$. For all sufficiently large $n$, every DNF of width at most $W$ on the leaves of $r$ independent blocks of height $j$ with dimensions $m_u,\dots,m_{u+j-1}$, where $1\le r\le W\le n^{\alpha}$, accepts $X$ and is sensitive at $X$ to the free leaf of every block with probability at most $2^{-\nu r/4}$, uniformly over the prescribed root values of the blocks.
\end{theorem}

The dimensions $m_u,\dots,m_{u+j-1}$ are defined when $u+j\le h_{\max}(n)$, and the height of an instance that contains the blocks plays no role. All thresholds on $n$ depend only on $u$, $\alpha$, and $d$.

The proof exposes the levels of the blocks from the bottom up and applies Theorem~\ref{thm:winner} and Claim~\ref{claim:truncate} at every level. Throughout the proof, $r$ counts the original blocks of height $j$. At step $p<j$, let $K_p$ be the total number of nodes at level $u+p$ in these $r$ trees. The two results are applied to these $K_p$ nodes, each regarded as a block of height one and dimension $m_{u+p}$.

The thresholds are
\[
 J=\lfloor r/2\rfloor,\qquad \ell=J+1,\qquad t_i=\lfloor m_i/(16d)\rfloor\quad(u\le i<u+j),
\]
where $t_i$ is the value of $t$ in the definition of light blocks at the level of dimension $m_i$. The threshold $J$ is the maximum number of light blocks allowed in a retained term, and $\ell=J+1$ is the first number that triggers deletion. Choosing $J=\lfloor r/2\rfloor$ leaves the final width below $r$ (Claim~\ref{claim:width}). A deleted first satisfied term has at least $\ell$ light blocks, and $\ell>r/2$, so Claim~\ref{claim:loss} bounds the probability of losing acceptance at each level by $2^{-(7/25)\nu r}$.

Write $\Phi_0=\Phi$. For $0\le p<j$, let $\Phi_{p+1}$ be the DNF $(\Phi_p)^-_{\rho_{u+p}}$ of Claim~\ref{claim:truncate}, applied to the $K_p$ blocks of height one with $t=t_{u+p}$ and deletion threshold $J$. Here $\rho_{u+p}$ denotes the independent projections at that level in the forest of the $r$ original trees.

For every $0\le p\le j$, the domain of $\Phi_p$ consists of the node values at level $u+p-1$. In particular, $\Phi_j$ is a DNF on the $r$ root values. The terms of each $\Phi_p$ inherit the order and the indices of the terms of $\Phi$. A literal of $\Phi_p$ on the free leaf of a block at level $u+p$ survives the projection as a literal on the value of that block's root node, and at the next step this node is itself a code child or a data child of its parent and its literal is classified accordingly.

To track the widths, we define the width bounds $W_0=W$ and
\[
 W_{p+1}=\Bigl\lfloor\frac{W_p}{t_{u+p}+1}\Bigr\rfloor+J\qquad(0\le p<j),
\]
so that $\width(\Phi_p)\le W_p$ for all $p\le j$ by Claim~\ref{claim:truncate}.

\begin{claim}\label{claim:width}
For all sufficiently large $n$, uniformly in $1\le r\le W\le n^{\alpha}$, we have $W_p\le W$ for all $p\le j$ and $W_j<r$. For $r=1$, $W_j=0$.
\end{claim}

\begin{proof}
For fixed $u$, $j$, and $d$, each $t_{u+p}+1$ is $\Theta(n)$, so their product over $p<j$ is $\Theta(n^j)$, and each is at least four for large $n$. Unwinding the recurrence,
\[
 W_j\le\frac{W}{\prod_{p<j}(t_{u+p}+1)}+J\Bigl(1+\frac14+\frac1{16}+\cdots\Bigr)
 \le O(n^{\alpha-j})+\frac43J\le O(n^{\alpha-j})+\frac23r ,
\]
and $O(n^{\alpha-j})<r/3$ for large $n$ since $j>\alpha$ and $r\ge1$. Hence $W_j<r$. For $r=1$ we have $J=0$, so the integer $W_j$ is at most $O(n^{\alpha-j})<1$. Finally $W_{p+1}\le W_p/4+r/2\le W_p/4+W/2\le W$ by induction, since $r\le W$.
\end{proof}

\begin{claim}\label{claim:loss}
Fix $p<j$, condition on the projections of the levels $u,\dots,u+p-1$, and fix the node values $z$ at level $u+p$. For all sufficiently large $n$, uniformly in $1\le r\le W\le n^{\alpha}$, in the exposed projections, and in $z$,
\begin{multline*}
 \Pr\bigl[\Phi_p(\rho_{u+p}(z))=1\ \text{and at least $\ell$ blocks are light for the first satisfied term of $\Phi_p$ at $\rho_{u+p}(z)$}\bigr]\\
 \le2^{-(7/25)\nu r}.
\end{multline*}
\end{claim}

\begin{proof}
Write $m=m_{u+p}$ and $t=t_{u+p}$. Conditional on the exposed projections, $\Phi_p$ is a fixed DNF of width at most $W_p$ on the leaves of the $K_p$ nodes at level $u+p$ of the original $r$ trees. These nodes are independent blocks of height one and dimension $m$ with free values $z$, and they are the blocks counted as light in this claim. Each of the $r$ trees has $2^{\sum_{a=u+p+1}^{u+j-1}(m_a+1)}$ nodes at level $u+p$, so by~\eqref{eq:geometric} and $\log r\le\alpha\log n$,
\[
 \log K_p\le\alpha\log n+\frac{m+8}{15}+j-1=\frac{m}{15}+O_{u,\alpha}(\log n).
\]
If $W_p<\ell$ the event is empty, since a term for which at least $\ell$ blocks are light contains the free leaves of $\ell$ distinct blocks and so has at least $\ell$ literals. Otherwise Theorem~\ref{thm:winner} with $K=K_p$, width $W_p$, threshold $t$, and $\ell=J+1$ bounds the probability by
\[
 \binom{K_p}{\ell}\Bigl(\frac{2W_p}{\ell}\Bigr)^{\ell}\binom{m}{\le dt}^{\ell}2^{-m\ell}
 \le\Bigl(K_p\cdot2n^{\alpha}\cdot2^{H_2(1/16)m}\cdot2^{-m}\Bigr)^{\ell},
\]
using $\binom{K_p}{\ell}\le K_p^{\ell}$, $W_p\le W\le n^{\alpha}$, and $dt\le m/16$ with the entropy bound of Section~\ref{sec:prelim}. The logarithm of the base is at most
\begin{align*}
 &\frac{m}{15}+O_{u,\alpha}(\log n)+1+\alpha\log n+H_2\Bigl(\frac1{16}\Bigr)m-m\\
 &\qquad\le-\Bigl(1-0.34-\frac1{15}\Bigr)m+O_{u,\alpha}(\log n)\\
 &\qquad\le-\frac{14}{25}m
\end{align*}
for large $n$, since $1-0.34-1/15>0.593>0.56=14/25$ and $m=\Theta(n)$. In words, the exponent $m$ in the probability $2^{-m}$ of a given value of a uniform secret is reduced by $H_2(1/16)m$ for the number of possible first satisfied terms, by $m/15+O(\log n)$ for the number of blocks, and by $O(\log n)$ for the width, and more than $(14/25)m$ remains for each light block. Hence the probability is at most $2^{-(14/25)m\ell}\le2^{-(7/25)\nu r}$, because $m\ge\nu$ and $\ell\ge r/2$.
\end{proof}

\begin{proof}[Proof of Theorem~\ref{thm:flips}]
For $0\le p\le j$ let $Y_p$ and $Y_p^{(i)}$ be the vectors of node values at level $u+p-1$ of $X$ and of the $X^{(i)}$. Since $X$ and the $X^{(i)}$ are generated by the same projections,
\[
 Y_0=X,\qquad Y_0^{(i)}=X^{(i)},\qquad Y_p=\rho_{u+p}(Y_{p+1}),\qquad Y_p^{(i)}=\rho_{u+p}(Y_{p+1}^{(i)})\quad(p<j).
\]
The vectors $Y_p$ and $Y_p^{(i)}$ differ in exactly one coordinate, which lies in block $i$. For $p<j$ this is Claim~\ref{claim:proj}(d) applied to the part of block $i$ above level $u+p-1$, and for $p=j$ the vectors consist of the root values of the blocks. Hence the $r$ coordinates at which the $Y_p^{(i)}$ differ from $Y_p$ are distinct. Let $E_p$ be the event
\[
 \Phi_p(Y_p)=1\qquad\text{and}\qquad\Phi_p(Y_p^{(i)})=0\quad\text{for all }1\le i\le r .
\]
Then $E_0$ is the event~\eqref{eq:sensitive} of the theorem. The event $E_j$ is empty, since $\width(\Phi_j)\le W_j<r$ by Claim~\ref{claim:width}, while a term of $\Phi_j$ satisfied at $Y_j$ would contain the $r$ distinct coordinates at which the $Y_j^{(i)}$ differ from $Y_j$.

Fix $p<j$ and condition on the projections of the levels $u,\dots,u+p-1$. Then $\Phi_p$ is fixed and $\rho_{u+p}$ is fresh. The vectors $Y_{p+1}$ and $Y_{p+1}^{(i)}$ are generated by the prescribed root values and the projections at levels above $u+p$. Thus they are independent of the projections of the levels $u,\dots,u+p$ taken together, by the argument of Claim~\ref{claim:proj}(b).

Since $\Phi_{p+1}\le\Phi_p\circ\rho_{u+p}$ by Claim~\ref{claim:truncate}, we have $\Phi_{p+1}(Y_{p+1}^{(i)})\le\Phi_p(Y_p^{(i)})$, so every rejection persists.

The acceptance $\Phi_p(Y_p)=1$ persists to $\Phi_{p+1}(Y_{p+1})=1$ unless more than $J$ blocks are light for the first satisfied term of $\Phi_p$ at $Y_p=\rho_{u+p}(Y_{p+1})$, by the last assertion of Claim~\ref{claim:truncate}. Let $\widetilde E_p$ be the event of Claim~\ref{claim:loss} with $z=Y_{p+1}$. Since $\ell=J+1$, the only possible loss of acceptance is included in this event, and hence
\[
 E_p\subseteq E_{p+1}\cup\widetilde E_p .
\]

The independence above allows us to apply Claim~\ref{claim:loss} for every fixed value of $Y_{p+1}$ and of the exposed projections. It bounds the conditional probability of $\widetilde E_p$ by $2^{-(7/25)\nu r}$, so the same bound holds after averaging over them. Iterating the inclusion from $p=0$ to $p=j-1$ and using that $E_j$ is empty gives $E_0\subseteq\bigcup_{p<j}\widetilde E_p$, and it follows that
\[
 \Pr[E_0]\le\sum_{p<j}\Pr[\widetilde E_p]\le j\,2^{-(7/25)\nu r}\le2^{-\nu r/4}
\]
for all sufficiently large $n$, since $\nu=\Theta(n)$ and $j$ is a constant. All thresholds on $n$ depend only on $u$, $\alpha$, and $d$.
\end{proof}

\begin{claim}\label{claim:tall}
In the setting of Theorem~\ref{thm:flips}, the same bound holds for $r$ independent blocks of any height at least $j$ whose bottom level has dimension $m_u$.
\end{claim}

\begin{proof}
If the height is $j$ there is nothing to prove. Otherwise we expose the projections of all levels of the blocks above level $u+j-1$. In each block, the exposed projections determine the secrets above that level, hence the path of free children from the root down to that level, and the subtree of height $j$ at the end of this path. We also expose the projections of the nodes of all other subtrees of height $j$ of the blocks. The value prescribed to the subtree on the path is the root value of the block, since values are copied along the path of free children. The table $X^{(i)}$ uses the same projections and differs from $X$ only in this value for block $i$. The projections above level $u+j-1$ determine which $r$ subtrees lie on these paths, and they are independent of the projections inside the subtrees of height $j$, which are independent across subtrees (Section~\ref{sec:projections}).

Conditional on the exposed projections, the $r$ remaining subtrees are therefore independent blocks of height $j$ with dimensions $m_u,\dots,m_{u+j-1}$ and prescribed root values. The DNF with the exposed leaves substituted is a fixed DNF of width at most $W$ on their leaves, and the event~\eqref{eq:sensitive} for the original DNF is the same event for this DNF on the remaining subtrees. Theorem~\ref{thm:flips} applies.
\end{proof}

\subsection{A tail bound for certificate complexity}\label{sec:certtail}

This subsection and the next turn the sensitivity bound of Theorem~\ref{thm:flips} into the switching theorem that Section~\ref{sec:circuits} applies at every stage of the depth reduction. The hypotheses of Theorem~\ref{thm:certtail} do not involve the code family or the recursion, since the theorem applies to any projection of blocks in the sense of the next paragraph that satisfies the uniform sensitivity bound in its statement. The recursive construction supplies that bound through Theorem~\ref{thm:flips}. Theorem~\ref{thm:certtail} shows that, under a projection for which the events of Theorem~\ref{thm:flips} are unlikely, for each fixed input $z$ it is unlikely that the projected CNF accepts $z$ and has no small $1$-certificate at $z$. Claim~\ref{claim:tree} converts small certificates into a shallow decision tree, and Theorem~\ref{thm:switch} combines the two with Theorem~\ref{thm:flips}.

In this subsection and the next, a projection $\rho$ of $K$ blocks is a product of $K$ independent random projections, one per block. Each fixes every leaf except one free leaf carrying the block's free variable, and is sampled independently of all free values. Setting that variable to zero or one gives two inputs generated by the same projection and differing only in the free leaf.

Sensitivity to a block is defined as in Section~\ref{sec:flips}, with these free values in place of prescribed root values. The input $X$ has the given free values, and $X^{(i)}$ flips the value of block $i$ alone. Claim~\ref{claim:proj}(d) supplies projections of this kind, with each block's root value as its free value.

\Needspace{12\baselineskip}
\begin{theorem}[Tail bound for certificate complexity]\label{thm:certtail}
Let $\rho$ be a projection of $K$ blocks as above, and suppose that for every integer $1\le r\le W$ and every DNF of width at most $W$ on the leaves of $r$ of the blocks, the DNF accepts $X$ and is sensitive at $X$ to every one of the $r$ blocks with probability at most $2^{-\gamma r}$, uniformly over the free values. Then for every CNF $\Psi$ of width at most $W$ on the leaves of the blocks, every $z\in\F^K$, and every integer $q\ge0$, the projected CNF $H_\rho=\Psi\circ\rho$ satisfies
\begin{equation}\label{eq:cert-tail}
 \Pr_\rho\bigl[H_\rho(z)=1\ \text{and}\ \cert_1(H_\rho,z)>q\bigr]
 \le\sum_{\Lambda=\lceil(q+1)/(2W+1)\rceil}^{K}2^{-(\gamma-2\log K)\Lambda}.
\end{equation}
An empty sum is zero.
\end{theorem}

The proof finds, for every projection in the event of~\eqref{eq:cert-tail}, a family of disjoint sets of coordinates. Flipping all coordinates of one set falsifies $H_\rho$, and flipping any one of them back restores it, as in the event of Theorem~\ref{thm:flips}. The family, and the clauses $c_g$ chosen for its sets in Claim~\ref{claim:minimal}, depend on the projection, so the hypothesis cannot be applied after conditioning on their selection. Instead, Claim~\ref{claim:coloring} extracts a subfamily whose clauses $c_g$ avoid the blocks of the other sets, and Claim~\ref{claim:family} bounds an event fixed in advance for each family of sets, defined from all clauses of $\Psi$ that avoid the other sets. A union bound over the families completes the proof.

Fix a projection $\rho$ with $H_\rho(z)=1$. We normalize $H_\rho$ as in Section~\ref{sec:prelim} and call its surviving clauses the remaining clauses. Every remaining clause $c$ is true at $z$. Let
\[
 S(c)=\{\,i\in[K]:\ \text{$c$ contains a literal on $z_i$ that is true at $z$}\,\}
\]
be the set of coordinates on which $c$ is satisfied at $z$, a nonempty subset of $[K]$. Each $S(c)$ has at most $W$ elements, since every leaf of a block is a constant or the free leaf of that block, so the projected clause has at most $W$ literals.

We also write $S(c)$ for an original clause $c$ of $\Psi$ whose projection is a remaining clause, meaning the set $S(c')$ of its projection $c'$. Such an original clause has no literal fixed to true by $\rho$.

\begin{claim}\label{claim:transversal}
A set $A\subseteq[K]$ is a $1$-certificate of $H_\rho$ at $z$ if and only if it meets $S(c)$ for every remaining clause $c$. Hence $\cert_1(H_\rho,z)$ is the least size of a set that meets every $S(c)$.
\end{claim}

\begin{proof}
If $A$ meets every $S(c)$, fixing the coordinates of $A$ to their values at $z$ fixes a true literal in every clause, so every completion satisfies $H_\rho$. If $A$ misses $S(c)$ for some remaining clause $c$, then every literal of $c$ on a coordinate of $A$ is false at $z$. Keeping the values of $A$, we set the variables of $c$ outside $A$ to the values that make their literals false and complete the assignment arbitrarily. Then $c$ is false, so $H_\rho=0$, and $A$ is not a certificate.
\end{proof}

\begin{claim}\label{claim:minimal}
Let $U_1,\dots,U_M$ be a maximal family of pairwise disjoint sets that are inclusion-minimal among the sets $S(c)$, and for each $g$ choose an original clause $c_g$ of $\Psi$ with $S(c_g)=U_g$, the witnessing clause of $U_g$. If $\cert_1(H_\rho,z)>q$, then $\sum_g|U_g|\ge q+1$.

For each $g$, flipping all coordinates of $U_g$ in $z$ falsifies $c_g$, while flipping any proper subset of $U_g$ leaves every remaining clause true.
\end{claim}

\begin{proof}
Maximality makes the union a certificate, and inclusion-minimality gives the second assertion. Every $S(c)$ contains an inclusion-minimal one, and by maximality the union of the family meets every minimal one. Hence the union meets every $S(c)$, so it is a $1$-certificate by Claim~\ref{claim:transversal} and has at least $q+1$ elements.

Flipping all of $U_g$ makes every literal of the projection of $c_g$ false, since its true literals are exactly those on $U_g$. The original clause $c_g$ is therefore false as well, since all its fixed literals are false. Suppose that flipping a set $P\subsetneq U_g$ falsified a remaining clause $c$. Then every true literal of $c$ lies on $P$ and no false literal of $c$ lies on $P$, so $S(c)\subseteq P\subsetneq U_g$, which contradicts the minimality of $U_g$.
\end{proof}

To obtain a product bound, we arrange that the events of different sets involve disjoint blocks, which requires that the witnessing clause $c_g$ of each $U_g$ contain no literal on a leaf of a block of another set, including leaves that $\rho$ fixes to constants. For an original clause $c$ of $\Psi$, let $B(c)\subseteq[K]$ be the set of indices $i$ such that $c$ contains a literal on a leaf of block $i$, the block support of $c$. It is determined by $\Psi$ alone and includes the blocks that $c$ meets only at leaves fixed to constants by $\rho$, whereas the projection of $c$ has literals only on the free variables of the blocks whose free leaf occurs in $c$. Hence $B(c)$ may be larger than the support of the projected clause, and $S(c)\subseteq B(c)$. The requirement is that $B(c_g)\cap U_{g'}=\emptyset$ for all distinct $g,g'$, and the sets $S(c)$ do not show whether it holds. For example, let $a_i$ be the free leaf of block $i$ for $i=1,\dots,4$, let $a_3'$ be another leaf of block $3$, and let $\rho$ map $a_i$ to $z_i$ and $a_3'$ to $0$. Then
\[
 (a_1\vee a_2\vee a_3')\wedge(a_3\vee a_4)\qquad\text{projects to}\qquad(z_1\vee z_2)\wedge(z_3\vee z_4),
\]
and at $z=(1,1,1,1)$ the two sets $S(c)$, namely $\{1,2\}$ and $\{3,4\}$, are disjoint, while the first original clause has block support $\{1,2,3\}$, which meets the second set. The value of $a_3'$ is determined by the projection of block $3$, which is random in the bound for the event of the set $\{3,4\}$ in the proof of Claim~\ref{claim:family}. A CNF for $\{1,2\}$ that contains the first clause would therefore depend on the projection of block $3$, and the two events would not be independent. The next claim therefore works with the block supports of the original clauses and extracts a large subfamily in which $B(c_g)\cap U_{g'}=\emptyset$ for all distinct $g,g'$.

\Needspace{7\baselineskip}
\begin{claim}\label{claim:coloring}
If $\cert_1(H_\rho,z)>q$, there is a subfamily of the $U_g$ of total size $\Lambda\ge(q+1)/(2W+1)$ such that $B(c_g)\cap U_{g'}=\emptyset$ for all distinct $g,g'$ in the subfamily.
\end{claim}

\begin{proof}
We put a directed edge from $g$ to $g'\ne g$ if $B(c_g)$ meets $U_{g'}$. Since $|B(c_g)|\le W$, as $c_g$ has at most $W$ literals, and the sets $U_{g'}$ are disjoint, every vertex has outdegree at most $W$. Hence every induced subgraph of the underlying undirected graph on a vertex set $V'$ has at most $W|V'|$ edges and average degree at most $2W$. The graph is therefore $2W$-degenerate, and it has a proper coloring with $2W+1$ colors, obtained by repeatedly removing a vertex of degree at most $2W$ and then coloring the vertices greedily in the reverse order of removal.

We weight $g$ by $|U_g|$ and take a heaviest color class. Its total weight is at least $\sum_g|U_g|/(2W+1)\ge(q+1)/(2W+1)$ by Claim~\ref{claim:minimal}, and no edge joins two of its members.
\end{proof}

We keep $\Psi$ and $z$ fixed and now let $\rho$ vary. The subfamily just constructed depends on $\rho$, as do its witnessing clauses $c_g$. To apply the hypothesis of Theorem~\ref{thm:certtail}, we instead fix a family of sets in advance and define an event that does not refer to the construction.

Fix a family $\mathcal U$ of disjoint nonempty subsets of $[K]$, each of size at most $W$, whose union has size $\Lambda$. For each $U\in\mathcal U$, we take all clauses $c$ of the original CNF $\Psi$ with $B(c)\cap U'=\emptyset$ for every $U'\in\mathcal U$ other than $U$. This condition is determined by $\Psi$ and $\mathcal U$ and does not depend on the projection. In particular, a clause with a literal on a leaf of a block of another set is excluded even when $\rho$ fixes that leaf to a constant. We substitute the values in $\rho(z)$ of the leaves outside the union of the family and call the resulting CNF $H_U$. Leaves of blocks in $U$ are not substituted, and by the condition on the clauses no other unsubstituted leaf occurs, so $H_U$ is a CNF of width at most $W$ on the leaves of the blocks in $U$, not on the free variables.

Its negation $\neg H_U$ is a DNF of width at most $W$ on those blocks. Including all original clauses that satisfy the condition makes $H_U$ fixed once the projections of the blocks outside the union of $\mathcal U$ are fixed, before the projections in $U$ are sampled.

Let $E_{\mathcal U}$ be the event that for every $U\in\mathcal U$, the DNF $\neg H_U$ accepts the input whose free values on $U$ are opposite to those of $z$ and is sensitive there to every block of $U$.

\begin{claim}[Families of sets fixed in advance]\label{claim:family}
For every fixed family $\mathcal U$ as above, $\Pr_\rho[E_{\mathcal U}]\le2^{-\gamma\Lambda}$.

Every $\rho$ with $H_\rho(z)=1$ for which Claims~\ref{claim:minimal} and~\ref{claim:coloring} can produce the subfamily $\mathcal U$ lies in $E_{\mathcal U}$.
\end{claim}

\begin{proof}
We first bound the event for a fixed family $\mathcal U$ and condition on the projections of all blocks outside its union. Each $H_U$ is now a fixed CNF of width at most $W$ on the leaves of the blocks in $U$, determined only by $\Psi$, the fixed family, and the projections of the blocks outside the union. Its negation is a width-$W$ DNF on $|U|\le W$ blocks. A clause of $H_U$ that comes from an original clause $c$ with $B(c)\cap U=\emptyset$ is now constant. If it is false it makes the event impossible, and if it is true it has no effect.

The events for different $U$ now depend on disjoint blocks with independent projections. The hypothesis of Theorem~\ref{thm:certtail}, uniform over the free values, bounds each event by $2^{-\gamma|U|}$. Their conditional probabilities multiply. The resulting bound does not depend on the projections outside the union, so averaging gives
\[
 \Pr[E_{\mathcal U}]\le\prod_{U\in\mathcal U}2^{-\gamma|U|}
 =2^{-\gamma\Lambda}.
\]

We next check that every projection producing $\mathcal U$ belongs to this event. Suppose $H_\rho(z)=1$ and the construction yields $\mathcal U$. Fix $U\in\mathcal U$ and its witnessing clause $c_g$ with $S(c_g)=U$. By Claim~\ref{claim:coloring}, $B(c_g)$ is disjoint from every other set of $\mathcal U$, so $c_g$ is one of the clauses used to form $H_U$. At the input specified in $E_{\mathcal U}$, all free values in $U$ are opposite to $z$. This clause is then false by Claim~\ref{claim:minimal}, so $\neg H_U$ accepts this input.

We now restore one coordinate $i\in U$ to its value at $z$ and extend the free values to the full assignment that equals $z$ outside $U$, which does not affect $H_U$. This assignment differs from $z$ exactly on the proper subset $U\setminus\{i\}$, and for $|U|=1$ the input is $\rho(z)$. Every remaining clause is true by Claim~\ref{claim:minimal}, and each clause deleted during normalization is identically true under $\rho$. Hence every original clause of $\Psi$ is true, so $H_U$ is true and $\neg H_U$ rejects the input with this coordinate restored. This holds for each $U\in\mathcal U$, proving $\rho\in E_{\mathcal U}$.

The probability bound is for a fixed family, and no probability is conditioned on acceptance or on the family being produced.
\end{proof}

\Needspace{8\baselineskip}
\begin{proof}[Proof of Theorem~\ref{thm:certtail}]
Let $\rho$ be in the event of the theorem. By Claims~\ref{claim:minimal} and~\ref{claim:coloring}, the construction yields a family $\mathcal U$ of disjoint nonempty sets of size at most $W$ with union of size $\Lambda$, where $(q+1)/(2W+1)\le\Lambda\le K$. By Claim~\ref{claim:family}, this $\rho$ lies in $E_{\mathcal U}$, an event of probability at most $2^{-\gamma\Lambda}$. The number of such families with union of size $\Lambda$ is at most $K^{\Lambda}\Lambda^{\Lambda}\le K^{2\Lambda}$, since the union can be chosen as a sequence of $\Lambda$ coordinates and then partitioned. A union bound over all families gives~\eqref{eq:cert-tail}. The union bound ranges over families of sets of blocks only, and no choice of witnessing clauses is enumerated, which is why no factor for the number of clauses appears.
\end{proof}

\subsection{Switching under correlated projections}\label{sec:switch}

Claim~\ref{claim:tree} converts small certificates into a decision tree. The standard simulation bounds the decision-tree depth of a function by the product of its $0$- and $1$-certificate complexities~\cite{BI87,BdW02,HH91,Tar89}, and we truncate it so that it uses only certificates of size at most $q$.

\begin{claim}\label{claim:tree}
For every CNF $H$ of width at most $W$ on $\F^K$ and every integer $q\ge1$ there is a decision tree $\calT$ of depth at most $Wq$ such that $\calT(z)\le H(z)$ for every $z$, and $\calT(z)=1$ whenever $H(z)=1$ and $\cert_1(H,z)\le q$. The tree is determined by $H$ and $q$.
\end{claim}

\begin{proof}
We fix in advance an order of all partial assignments to at most $q$ of the $K$ variables, so that the tree depends only on $H$ and $q$. The tree proceeds in rounds. In each round, the residual CNF is $H$ with the answers to the queries made so far substituted and normalized (Section~\ref{sec:prelim}), a CNF on the variables not yet queried. If the residual function is constant, the tree returns its value. Otherwise the tree takes the first partial assignment in the order that is on at most $q$ of the variables not yet queried and forces the residual CNF to one, and rejects if there is none. If there is one, the tree queries its coordinates, accepts if the answers agree with it, and otherwise continues with the next round.

Every residual clause contains a literal that the chosen partial assignment makes true, for otherwise the clause could be falsified by an assignment extending it. After the coordinates of the partial assignment are queried, such a clause either becomes true and is deleted, or that literal receives the opposite value and is deleted from the clause. Hence the maximum width of the residual clauses decreases in every round, there are at most $W$ rounds of at most $q$ queries each, and the depth is at most $Wq$.

The tree accepts only when the answers agree with a partial assignment that forces the residual CNF to one or when the residual function is constantly one, so $\calT\le H$. Let $H(z)=1$ and let $Q$ be a $1$-certificate of $H$ at $z$ of size at most $q$. Along the path of $z$, the values of $z$ on the part of $Q$ not yet queried form a partial assignment, on at most $q$ of the variables not yet queried, that forces the residual CNF to one in every round. Indeed, every assignment to the variables not yet queried that extends it agrees with $z$ on $Q$, since the queried variables carry the values of $z$, so it satisfies $H$. Hence the tree never rejects $z$ for lack of a partial assignment and never returns the value of a residual function that is constantly zero, and it accepts $z$.
\end{proof}

The accepting paths of $\calT$ give a DNF with at most $2^{Wq}$ terms of width at most $Wq$ with the same two properties. Applied to the negation of a DNF of width at most $W$, the claim gives a CNF with at most $2^{Wq}$ clauses of width at most $Wq$. This CNF is at least the DNF everywhere, and it agrees with the DNF at every input that the DNF rejects with a $0$-certificate of size at most $q$. The construction need not be efficient, since the lower bound is non-uniform.

\begin{theorem}[Switching with pointwise error]\label{thm:switch}
Fix a level $u\ge0$ and an exponent $\alpha\ge0$, and put $j=\lfloor\alpha\rfloor+1$ and $\nu=m_{u+j-1}$. Consider an instance of size $n$ and height $h$ with $u+j\le h\le h_{\max}(n)$. Let $\rho=\rho_u\circ\cdots\circ\rho_{u+j-1}$ be the projection of the levels $u,\dots,u+j-1$, which maps the node values at level $u+j-1$ to the node values at level $u-1$.

For all sufficiently large $n$ the following holds. For every $W$ with $1\le W\le n^{\alpha}$, every CNF $\Psi$ of width at most $W$ on the node values at level $u-1$, and every integer $q\ge1$, there is a random DNF $\widetilde\Psi_\rho$ on the node values at level $u+j-1$. It has at most $2^{Wq}$ terms of width at most $Wq$ and is determined by $\Psi$, $q$, and $\rho$. For every $z$ it satisfies $\widetilde\Psi_\rho(z)\le\Psi(\rho(z))$.

For every fixed $z$, the probability of disagreement satisfies
\begin{equation}\label{eq:switch-error}
 \Pr_\rho\bigl[\widetilde\Psi_\rho(z)\ne\Psi(\rho(z))\bigr]\le2\cdot2^{-\nu(q+1)/(30W)} .
\end{equation}
The dual statement switches a DNF of width at most $W$ to a CNF that is at least the projected DNF everywhere, with the same bounds. The threshold on $n$ depends only on $u$, $\alpha$, and $d$, and not on $h$.
\end{theorem}

The guarantee holds for each fixed $z$ separately, with high probability over $\rho$. No single projection is claimed to simplify $\Psi$ on all $z$ at once, and no union bound over all $z$ is taken in the next section.

\begin{proof}
The node values at level $u+j-1$ have $K=N_{u+j-1}$ coordinates, each the root of a block of height $j$ with dimensions $m_u,\dots,m_{u+j-1}$. By Claim~\ref{claim:proj}(d), $\rho$ is a projection of $K$ blocks in the sense of this section, and Theorem~\ref{thm:flips} gives the hypothesis of Theorem~\ref{thm:certtail} with $\gamma=\nu/4$. By~\eqref{eq:blocks},
\[
 2\log K\le\frac{2\nu}{15}+\frac{\log n}{4}\le\frac{3\nu}{20},
 \qquad\text{so}\qquad
 \gamma-2\log K\ge\frac{\nu}{4}-\frac{3\nu}{20}=\frac{\nu}{10}
\]
for large $n$, uniformly in $h$, since $\nu=m_{u+j-1}=\Theta(n)$ with constants depending only on $u$ and $j$. The sum in~\eqref{eq:cert-tail} is then a geometric series with ratio at most $2^{-\nu/10}\le1/2$, and its first index is at least $(q+1)/(3W)$ since $2W+1\le3W$. Hence
\[
 \Pr_\rho\bigl[\Psi(\rho(z))=1\ \text{and}\ \cert_1(\Psi\circ\rho,z)>q\bigr]\le2\cdot2^{-\nu(q+1)/(30W)} .
\]
Let $\widetilde\Psi_\rho$ be the DNF of the accepting paths of the tree of Claim~\ref{claim:tree} for $\Psi\circ\rho$ and $q$. It is at most $\Psi\circ\rho$ everywhere, and it differs from $\Psi\circ\rho$ at $z$ only in the event just bounded. The tree depends only on $\Psi$, $q$, and $\rho$. The dual statement follows by applying the argument to the negation of the DNF.
\end{proof}

\subsection{Depth reduction and the coupling bound}\label{sec:circuits}

Fix a code family of degree $d$, a depth $D$, and an exponent $A\ge1$, and let $F$ be a circuit as in Theorem~\ref{thm:circuits} over the leaves of an instance of size $n$ and height $h\le h_{\max}(n)$. We bound the probability that $F$ has different values on the pair $(X_0,X_1)$ generated by the same projections. The constant $h_0$ of the theorem is determined after Claim~\ref{claim:stage}, and $h\ge h_0$ is assumed from there on.

The argument follows H{\aa}stad's depth reduction~\cite{Has86}. The bottom two layers of the layered formula of Claim~\ref{claim:normal} are CNFs or DNFs of small width. The projection of the next few levels switches each of them to a DNF or a CNF of small width, with an error at every fixed input far below the reciprocal of the number of gates, and the new top gates merge with their parents, which lowers the depth by one. In symbols, a bottom $\mathrm{OR}\circ\mathrm{AND}\circ\mathrm{OR}$ becomes $\mathrm{OR}\circ\mathrm{OR}\circ\mathrm{AND}$ by approximate switching and then $\mathrm{OR}\circ\mathrm{AND}$ by exact merging. After constantly many stages the formula has depth two and polynomial width. Theorem~\ref{thm:flips} with $r=1$, extended by Claim~\ref{claim:tall}, then shows that this formula distinguishes its two inputs, which differ in one coordinate, only with probability $2^{-\Omega(n)}$.

\begin{claim}\label{claim:normal}
There are constants $D'$ and $a_0\ge1$, depending only on $D$ and $A$, such that for large $n$ the circuit $F$ computes the same function as a layered formula of depth $D'$ with alternating layers of AND and OR gates, of size at most $2^{n^{a_0}}$, whose bottom two layers form CNFs or DNFs of width at most one.
\end{claim}

\begin{proof}
We push the negations to the inputs and unroll the circuit, a directed acyclic graph of depth at most $D$, into a formula, which raises the size to at most a fixed power depending on $D$. If the size convention counts only gates, we first include the $2^{a(n)}\le2^{2n}$ input coordinates, whose number is bounded uniformly in $h$ by~\eqref{eq:address}. After repeated wires are removed, the number of wires is at most quadratic in the number of gates and inputs, still $2^{\poly(n)}$.

We merge adjacent gates of the same type and insert unary identity gates, so that the formula is layered with alternating layers and all inputs at the bottom. Finally we insert on every bottom input, a literal or a constant, a unary gate of the type opposite to that of its parent. An AND of literals and constants of arbitrary fan-in becomes an AND of unary ORs, a CNF of width at most one, and an OR of literals and constants becomes a DNF of width at most one. These operations increase the depth by at most a constant factor and the size by at most a fixed power. A constant formula, and a formula of depth at most one, is padded to depth two by unary gates of alternating types. In the same way, a formula whose depth is less than $D'$ is padded at its output by unary gates of alternating types to depth exactly $D'$.
\end{proof}

We now use the normal form of Claim~\ref{claim:normal} for $F$. We reduce its depth to two in $D'-2$ stages, indexed by $i=0,\dots,D'-3$. At stage $i$, the current formula $F_i$ is layered with alternating layers, is on the node values at level $u_i-1$, and has depth $D'-i$. We use deterministic bounds $S_i\le2^{n^{a_i}}$ on its size and $W_i\le n^{a_i}$ on its bottom fan-in, the width of its bottom CNFs or DNFs, uniform over the projections exposed so far, for a constant $a_i\ge1$. Initially $u_0=0$, $F_0=F$, $S_0=2^{n^{a_0}}$, and $W_0=1$, by Claim~\ref{claim:normal}.

Only the projections below level $u_i$ have been exposed. They determine $F_i$, and the projections at level $u_i$ and above remain independent of them by Claim~\ref{claim:proj}(b). To apply the switching theorem at the current width, we expose $j_i$ more levels and use the certificate-size bound $q_i$, where
\[
 q_i=\bigl\lceil n^{2a_i+2}\bigr\rceil,\qquad j_i=\lfloor a_i\rfloor+1.
\]
The value of $j_i$ is the number of levels required by Theorem~\ref{thm:switch}. The next formula is on the node values at level $u_{i+1}-1$, where $u_{i+1}=u_i+j_i$. We put $\nu_i=m_{u_{i+1}-1}$, which is $\Theta_i(n)$. The bound $q_i$ is chosen so that the error of one replacement, summed over the at most $S_i$ replacements of a stage, remains exponentially small (Claim~\ref{claim:stage}). The projection of these levels maps the node values at level $u_{i+1}-1$ to those at level $u_i-1$:
\[
 \sigma_i=\rho_{u_i}\circ\cdots\circ\rho_{u_{i+1}-1}.
\]

\begin{claim}\label{claim:stage}
Let $F_i$ have depth at least three and let $u_{i+1}\le h$. Condition on the projections below level $u_i$, so that $F_i$ is fixed.

There is a layered formula $F_{i+1}$ on the node values at level $u_{i+1}-1$, determined by $F_i$ and $\sigma_i$, of depth $D'-i-1$. Its size is at most $S_i2^{2W_iq_i+4}$ and its bottom fan-in is at most $W_iq_i\le2n^{3a_i+2}$.

For every fixed assignment $z$ to the node values at level $u_{i+1}-1$,
\[
 \Pr_{\sigma_i}\bigl[F_i(\sigma_i(z))\ne F_{i+1}(z)\bigr]\le\theta_i,
\]
where
\[
 \theta_i=2S_i\,2^{-\nu_i(q_i+1)/(30W_i)}
 \le2^{-\Omega_i(n^{a_i+3})}.
\]
In particular, the bounds $S_{i+1}=S_i2^{2W_iq_i+4}$ and $W_{i+1}=W_iq_i$ for the next stage satisfy $S_{i+1}\le2^{n^{a_{i+1}}}$ and $W_{i+1}\le n^{a_{i+1}}$ for $a_{i+1}=3a_i+4$ and large $n$.
\end{claim}

\begin{proof}
We apply Theorem~\ref{thm:switch} with $u=u_i$, exponent $a_i$, and $q_i$ to every bottom two-layer subformula of $F_i$, using the same projection $\sigma_i$, and switch each CNF to a DNF, or each DNF to a CNF, according to the type of the bottom layer. There are at most $S_i$ such subformulas, each of width at most $W_i\le n^{a_i}$.

Fix $z$. If every replacement agrees there with the projection of its original subformula, the whole formula agrees there. A union bound therefore bounds the disagreement probability by $\theta_i$, without requiring independence between replacement errors. Since $\nu_i=\Theta_i(n)$, $q_i\ge n^{2a_i+2}$, and $W_i\le n^{a_i}$,
\[
 \log\theta_i\le1+n^{a_i}-\frac{\nu_i\,n^{2a_i+2}}{30\,n^{a_i}}=-\Omega_i(n^{a_i+3}).
\]
The top gate of each replacement has the type of the parent of the subformula it replaces, so merging the two reduces the depth by one. We retain unary gates and both layers of each replacement, including a path of unary gates for a constant replacement, so that $F_{i+1}$ remains layered.

Write $T_i=W_iq_i$. Each replacement has at most $2^{T_i}$ terms or clauses of width at most $T_i$. Including its top gate, it has at most $1+(T_i+2)2^{T_i}\le2^{2T_i+3}$ gates and wires, and a constant represented by a path of unary gates also satisfies this bound. The part of $F_i$ above the replaced subformulas has at most $S_i$ gates and wires. Hence the new size is at most $S_i2^{2T_i+3}+S_i\le S_i2^{2T_i+4}$, and its bottom fan-in is at most
\[
 T_i\le n^{a_i}(n^{2a_i+2}+1)\le2n^{3a_i+2}.
\]
Finally, $\log S_{i+1}\le n^{a_i}+4n^{3a_i+2}+4\le n^{3a_i+4}$ and $2n^{3a_i+2}\le n^{3a_i+4}$ for large $n$, which proves the next size and fan-in bounds.
\end{proof}

Since $D'$ is a constant, the exponents $a_i$ are constants, and the total number $u_*=u_{D'-2}$ of levels exposed by the $D'-2$ stages is a constant depending on $D$ and $A$. Let $F_*=F_{D'-2}$ be the final formula, of depth two and width at most $W_*=W_{D'-2}\le n^{a_*}$, where $a_*=a_{D'-2}$ is the exponent of the final formula, and put $j_*=\lfloor a_*\rfloor+1$ and
\[
 h_0=u_*+j_* ,
\]
the number of levels that the stages and the final estimate use. We assume $h_0\le h\le h_{\max}(n)$. Then every stage has enough remaining levels, since $u_{i+1}\le u_*<h_0$ for $i<D'-2$, and $h_0$ depends only on $D$ and $A$.

\Needspace{9\baselineskip}
\begin{claim}\label{claim:telescope}
For each root value $b$, let $Z^{(i)}_b$ be the vector of node values at level $u_i-1$ of the instance with root value $b$, so that $Z^{(0)}_b=X_b$ and $Z^{(i)}_b=\sigma_i(Z^{(i+1)}_b)$. Write $Z^*_b=Z^{(D'-2)}_b$ for the vector at level $u_*-1$. Then
\[
 \Pr\bigl[F(X_b)\ne F_*(Z^*_b)\bigr]\le\sum_i\theta_i .
\]
\end{claim}

\begin{proof}
At stage $i$, three disjoint sets of projection choices determine the objects in the comparison. The levels below $u_i$ determine $F_i$. The levels $u_i,\dots,u_{i+1}-1$ determine $\sigma_i$. The levels above $u_{i+1}-1$ generate $Z^{(i+1)}_b$. These three sets of choices are independent by Claim~\ref{claim:proj}(b).

We condition on the projections below level $u_i$ and on $Z^{(i+1)}_b$. The formula is now fixed, and the input at which it is evaluated is fixed independently of the still fresh $\sigma_i$, so Claim~\ref{claim:stage} applies. Averaging over the projections below level $u_i$ and $Z^{(i+1)}_b$ gives
\[
 \Pr\bigl[F_i(\sigma_i(Z^{(i+1)}_b))\ne
 F_{i+1}(Z^{(i+1)}_b)\bigr]\le\theta_i.
\]
Since $Z^{(i)}_b=\sigma_i(Z^{(i+1)}_b)$, this is the probability of disagreement between the evaluations $F_i(Z^{(i)}_b)$ and $F_{i+1}(Z^{(i+1)}_b)$ of two consecutive stages. If $F(X_b)=F_0(Z^{(0)}_b)$ and $F_*(Z^*_b)=F_{D'-2}(Z^{(D'-2)}_b)$ differ, then some two consecutive stages differ, and a union bound over the $D'-2$ stages proves the claim. We have not conditioned on earlier stages succeeding, so no independence between their errors is needed. The direction of the approximation may alternate between stages.
\end{proof}

\begin{claim}\label{claim:final}
There is a constant $c_*>0$, depending on $D$ and $A$, such that $\Pr[F_*(Z^*_0)\ne F_*(Z^*_1)]\le2^{1-c_*n}$ for large $n$.
\end{claim}

\begin{proof}
We condition on the exposed projections, so that $F_*$ is a fixed DNF or CNF of width at most $W_*$ on the node values at level $u_*-1$. The remaining levels $u_*,\dots,h-1$ form one block of height $h-u_*\ge j_*$ with bottom dimension $m_{u_*}$. Using the same projections of these levels with root values zero and one generates the pair $(Z^*_0,Z^*_1)$.

Suppose first that $F_*$ is a DNF, and put $\nu_*=m_{u_*+j_*-1}=\Theta(n)$. Theorem~\ref{thm:flips} with $r=1$, $u=u_*$, and exponent $a_*$, extended by Claim~\ref{claim:tall} if $h-u_*>j_*$, gives
\[
 \Pr[F_*(Z^*_1)=1,\ F_*(Z^*_0)=0]\le2^{-\nu_*/4}.
\]
The same bound holds with the two root values exchanged. If $F_*$ is a CNF, the same bounds apply to its negation.

Summing the two orientations bounds disagreement by $2^{1-\nu_*/4}$. Since $\nu_*=m_{h_0-1}\ge n/16^{h_0}$ for large $n$, this is at most $2^{1-c_*n}$ with $c_*=16^{-h_0}/4$, as claimed.
\end{proof}

\begin{proof}[Proof of Theorem~\ref{thm:circuits}]
Let $h_0$ be as above, and let $n_0$ be the largest of the thresholds on $n$ in Claim~\ref{claim:normal}, in Claim~\ref{claim:stage} for each of the $D'-2$ stages, and in Claim~\ref{claim:final}. These thresholds depend only on $D$, $A$, and $d$, because the thresholds of Theorems~\ref{thm:flips} and~\ref{thm:switch} depend only on $u$, the exponent, and $d$. Fix $n\ge n_0$ and a height $h$ with $h_0\le h\le h_{\max}(n)$. By Claim~\ref{claim:normal} we may assume that $F$ has the normal form, and Claims~\ref{claim:stage} to~\ref{claim:final} apply. For the pair generated by the same projections,
\begin{align*}
 \Pr[F(X_0)\ne F(X_1)]
 &\le\Pr[F(X_0)\ne F_*(Z^*_0)]+\Pr[F_*(Z^*_0)\ne F_*(Z^*_1)]+\Pr[F_*(Z^*_1)\ne F(X_1)]\\
 &\le2\sum_i\theta_i+2^{1-c_*n},
\end{align*}
by Claims~\ref{claim:telescope} and~\ref{claim:final}, where the first and third terms are the errors of the depth reduction under the two root values and the middle term is the final estimate. The right-hand side is at most $2^{-cn}$ for a constant $c>0$ depending on $D$, $A$, and the code family, after increasing $n_0$ if necessary. This proves the coupling bound~\eqref{eq:coupling-stability}. Since the marginals of this pair are $\mu_{n,h,0}$ and $\mu_{n,h,1}$,
\[
 \bigl|\E_{\mu_{n,h,1}}F-\E_{\mu_{n,h,0}}F\bigr|
 \le\E\bigl[|F(X_1)-F(X_0)|\bigr]
 =\Pr[F(X_0)\ne F(X_1)],
\]
which gives the distinguishing bound. Neither $c$, $h_0$, nor $n_0$ depends on $h$.
\end{proof}

\section{Proofs of the main theorems}\label{sec:mainproofs}

In this section we build the oracle by diagonalizing against all polynomial-time oracle machines with a constant number of alternating quantifiers, following the method of Furst, Saxe, and Sipser~\cite{FSS84}, Yao~\cite{Yao85}, H{\aa}stad~\cite{Has86}, and, for random oracles, Cai~\cite{Cai89}, and applying Theorem~\ref{thm:circuits} at every stage. We then prove the other results of Section~\ref{sec:results}, using Theorems~\ref{thm:exact} and~\ref{thm:batched} and Appendix~\ref{app:noisy}.

\subsection{Constructing the oracle}\label{sec:oracleproof}

We fix a code family of Definition~\ref{def:code}. The Hadamard code suffices for Theorems~\ref{thm:main} and~\ref{thm:slow}. We also fix a height function $h$ as defined in Section~\ref{sec:results}, that is, an unbounded function with $h(n)$ computable from $1^n$ in polynomial time and $h(n)\le h_{\max}(n)$ for every $n\ge1$. The function $h_{\max}$ is itself a height function and gives Theorem~\ref{thm:main}.

At size $n$ we store one instance of height $h(n)$. Its leaves occupy the addresses $1^n0p$, where $p$ ranges over the root-to-leaf paths of the tree, which are strings of the length $a(n)$ in~\eqref{eq:address} for this height. All addresses of an instance have the same length $n+1+a(n)$, greater than $n$ and at most $3n$. The first zero determines $n$, so instances of different sizes are disjoint. An instance of height zero, which is the case at every $n<256$, is one leaf at $1^n0$. Every other string is malformed and receives answer zero. There are no addresses for internal secrets or node values.

For every size $n$, every height $j\le h_{\max}(n)$, and every root value $b$ there is a valid default tree $\calD^b_j$, defined by induction on $j$, with the dimensions~\eqref{eq:schedule} of size $n$. The notation suppresses the size. At height zero it is the single leaf $b$. At an internal node of dimension $m$ we choose the secret zero, we give the code child $x$ a default tree of value $C_m(0,x)$, we give the data child indexed by zero a default tree of value $b$, and we give all other data children default trees of value zero. The code table is then the codeword of the secret zero and the data child at this secret has value $b$, so the tree is valid with value $b$. For the Hadamard code, $\calD^0_j$ is the all-zero leaf table.

We fix a deterministic polynomial-time oracle machine $V$ with $k$ alternating quantifiers over strings of lengths $p_1(n),\dots,p_k(n)$. On input $1^n$ it accepts if and only if
\[
 Q_1y_1\in\{0,1\}^{p_1(n)}\ \cdots\ Q_ky_k\in\{0,1\}^{p_k(n)}\ \bigl[V^O(1^n,y_1,\dots,y_k)=1\bigr].
\]
We call a tuple $(y_1,\dots,y_k)$ of strings of these lengths a complete tuple. Let $P(n)\ge4n$ be a polynomial bound on the running time of $V$ on input $1^n$ and on the lengths of its queries, over all complete tuples and all sequences of oracle answers, and on the lengths $p_1(n),\dots,p_k(n)$.

\begin{claim}\label{claim:transcript}
Fix $n$, a complete tuple $(y_1,\dots,y_k)$, and the oracle answers on all strings other than the leaves of the instance of size $n$. Then there is a DNF of width at most $P(n)$ with at most $2^{P(n)}$ terms over the leaves of the instance that agrees with $V^O(1^n,y_1,\dots,y_k)$ on every leaf table, valid or not.
\end{claim}

\begin{proof}
With the other answers fixed, the computation of $V$ on $(1^n,y_1,\dots,y_k)$ is an adaptive decision tree over the leaves of depth at most $P(n)$. We take the disjunction, over the accepting root-to-leaf paths of this tree, of the conjunction of the recorded answers along the path. This DNF has width at most $P(n)$ and at most $2^{P(n)}$ terms. A table that satisfies the conjunction of a path reproduces that path, by induction on the queries, since the recorded answers determine the next query address. Hence the DNF is one exactly on the accepting tables.
\end{proof}

This DNF, and hence the circuit of Claim~\ref{claim:expand} below, is an ordinary Boolean function on the full Boolean cube of leaf assignments. Theorem~\ref{thm:circuits} applies to such circuits, and its proof evaluates them on projected leaf tables that need not be valid (Claim~\ref{claim:proj}(c)). The diagonalization below tests the circuit on valid tables only.

\begin{claim}\label{claim:expand}
Fix the oracle answers on all strings other than the leaves of the instance of size $n$. Then the acceptance of $V$ on $1^n$ is computed on these leaves by a circuit of depth at most $k+2$ and size at most
\[
 O\bigl((k+P(n)+1)\,2^{p_1(n)+\cdots+p_k(n)+P(n)}\bigr)=2^{\poly(n)} ,
\]
with a fixed polynomial in the exponent determined by $V$.
\end{claim}

\begin{proof}
We replace each existential quantifier by an OR gate and each universal quantifier by an AND gate over the $2^{p_i(n)}$ values of $y_i$, and each leaf of the quantifier tree, which corresponds to one complete tuple, by the DNF of Claim~\ref{claim:transcript} for that tuple. The $k$ quantifiers give $k$ layers of gates and the DNF gives two more, which is the depth bound. The size bound counts the $2^{p_1(n)+\cdots+p_k(n)}$ complete tuples, each with a DNF of at most $2^{P(n)}$ terms of width at most $P(n)$, together with the quantifier gates and their wires.
\end{proof}

We construct $O_h$ in stages, one for each machine. The stage of a machine $V$ chooses a fresh size $n$ and assigns all unassigned strings of length at most $P(n)$ so that $V$ errs on $1^n$. We enumerate all machines $V$, all finite quantifier counts $k$, both starting quantifiers, and all polynomial time bounds and length bounds. Each enumerated machine is run under a clock given by its enumerated polynomial time bound and rejects when the clock runs out, as in Appendix~\ref{app:interaction}. This enumeration covers every language in $\PH^{O_h}$, however the remaining values of $O_h$ are chosen. We maintain a length bound $B$ such that the assigned strings are exactly those of length at most $B$, all assignments are permanent, and every assigned instance is valid. Initially the empty string is assigned zero and $B=0$.

At the stage of $V$, we choose an exponent $A\ge1$ such that the size bound in Claim~\ref{claim:expand} is at most $2^{n^A}$ for all large $n$. Let $h_0,n_0$ be the constants of Theorem~\ref{thm:circuits} for the code family, depth $k+2$, and exponent $A$. We choose $n$ with $h(n)\ge h_0$ that exceeds $B$ and $n_0$ and is large enough that the size bound is at most $2^{n^A}$. Such an $n$ exists because $h(n)\ge h_0$ for infinitely many $n$, since otherwise $h$ would be bounded. We call the instance of size $n$ the target.

We assign to every other unassigned instance whose addresses have length at most $P(n)$ its default tree of root value zero, with the height and dimensions of its own size, and we assign zero to the unassigned malformed strings of length at most $P(n)$. With these defaults, every instance that $V$ can read other than the target is valid, as the quantum algorithm and the protocol require at every size, and the acceptance of $V$ on $1^n$ depends only on the target leaves. The target is entirely unassigned, since its addresses have length at least $n+1>B$, and all of it lies within the new length bound, since its address length is at most $3n\le P(n)$. No instance is partly assigned, since all of its leaves have the same address length.

The machine queries only strings of length at most $P(n)$, so the remaining longer strings do not affect it. By Claim~\ref{claim:expand}, its acceptance on $1^n$ is now computed by a circuit $F$ on the target leaves of depth at most $k+2$ and size at most $2^{n^A}$. Since $n\ge n_0$ and $h_0\le h(n)\le h_{\max}(n)$, Theorem~\ref{thm:circuits} applies to $F$. If $F$ computed the root value on every valid target table, its expectations under $\mu_{n,h(n),1}$ and $\mu_{n,h(n),0}$ would differ by one, contrary to that theorem. We choose a valid target table on which $F$ errs and assign the entire instance. We assign zero to every remaining unassigned string of length at most $P(n)$, and we replace $B$ by $P(n)$.

Any instance that remains unassigned after all stages receives its default tree of root value zero, and any remaining string receives zero. We define $\calL_h$ to contain $1^n$, for $n\ge1$, exactly when the instance of size $n$ has root value one, and to contain no other strings. This defines $O_h$ and $\calL_h$. For $h=h_{\max}$ we write $O$ and $\calL$. The choice of a sufficiently large size at each stage is part of the oracle construction. The quantum algorithm, the verifier, and the honest prover do not compute these sizes or the constants $h_0,n_0$. They use only the input, the fixed height function, and oracle access.

\begin{claim}\label{claim:freeze}
Every machine $V$ of the enumeration errs on the string $1^n$ of its stage, with respect to the final oracle $O_h$.
\end{claim}

\begin{proof}
At the stage of $V$, every string that $V$ can query on input $1^n$, on any branch of the quantified computation and for any sequence of oracle answers, has length at most $P(n)$ and is assigned at the end of the stage. Later stages assign only strings of length greater than $P(n)$. Hence the acceptance of $V$ on $1^n$ is the same for the final oracle as for the oracle at the end of the stage, where $V$ errs.
\end{proof}

\subsection{The separation and the resource bounds}\label{sec:proofs-main}

\begin{proof}[Proof of Theorems~\ref{thm:main} and~\ref{thm:slow}]
Let $O_h$ and $\calL_h$ be as constructed for a height function $h$. By Claim~\ref{claim:freeze}, every machine of the enumeration errs on some input, so $\calL_h\notin\PH^{O_h}$. Every stored instance is valid, including those at sizes never chosen at a stage.

The quantum algorithm rejects the empty string and every nonunary input. On input $1^n$ it computes $h(n)$ and the dimensions~\eqref{eq:schedule} in polynomial time, evaluates the instance of size $n$ with the algorithm of Theorem~\ref{thm:exact} for height $h(n)$, and accepts exactly when the root value is one. For $h(n)=0$ it makes one query to the leaf $1^n0$. The algorithm is exact, makes $3^{h(n)}$ queries, and runs in time $3^{h(n)}\poly(n)$, which is polynomial because $3^{h(n)}\le3^{\log n/8}\le n^{1/4}$. Hence $\calL_h\in\EQP^{O_h}$, with the bounds of part (i) of Theorem~\ref{thm:slow}.

The verifier likewise rejects every input other than $1^n$ with $n\ge1$, and on input $1^n$ it runs Protocol~\ref{prot:batched} on the claim that the root of the instance of size $n$ has value one. Theorem~\ref{thm:batched} with $j=h(n)$ gives perfect completeness with its uniform quantum honest prover, soundness error at most $1/4$ against every prover, classical or quantum, a public-coin proof system with $2h(n)$ messages, $h(n)$ from each party, and the bit and query bounds of parts (ii) and (iii) of Theorem~\ref{thm:slow}. The verifier and the honest prover run in time $4^{h(n)}\poly(n)$, which is polynomial since $4^{h(n)}\le n^{1/4}$. This proves Theorem~\ref{thm:slow}. Taking $h=h_{\max}$ proves Theorem~\ref{thm:main}, including its message bound $2h_{\max}(n)\le\log n/4$.
\end{proof}

\begin{proof}[Proof of Corollary~\ref{cor:logstar}]
We put $h(n)=\lfloor\min\{f(n),2h_{\max}(n)\}/2\rfloor$. This function is computable from $1^n$ in polynomial time and is at most $h_{\max}(n)$. It is also unbounded. For every $H$, the unbounded function $f$ satisfies $f(n)\ge2H$ for infinitely many $n$, and $h_{\max}(n)\ge H$ for all but finitely many $n$, so $h(n)\ge H$ for some $n$. Hence $h$ is a height function. Theorem~\ref{thm:slow} with this $h$ gives the oracle $O_f=O_h$ and the language $\calL_f=\calL_h$. The proof system has $2h(n)\le f(n)$ messages, and the query counts $3^{h(n)}$, $4^{h(n)}$, and $4^{h(n)}-3^{h(n)}$ of the quantum algorithm, the verifier, and the honest prover are at most $4^{f(n)/2}=2^{f(n)}$. The function $\log^*n$ is computable in polynomial time (Section~\ref{sec:prelim}) and is unbounded.
\end{proof}

\begin{proof}[Proof of Corollary~\ref{cor:querybudget}]
We put $f(n)=\lfloor\log_2 Q(n)\rfloor$. This is a nonnegative integer, computable in polynomial time as one less than the bit length of the positive integer $Q(n)$, and $f$ is unbounded because $Q$ is. We apply Corollary~\ref{cor:logstar} with this $f$. Each of the three query counts is at most $2^{f(n)}\le Q(n)$, and the number of messages is at most $f(n)$. In particular, if $Q(n)=1$ then $f(n)=0$ and the height function $h(n)=\lfloor\min\{f(n),2h_{\max}(n)\}/2\rfloor$ of Corollary~\ref{cor:logstar} is zero, so the quantum algorithm and the verifier each make one query, the honest prover makes none, and no messages are exchanged.
\end{proof}

Theorems~\ref{thm:main} and~\ref{thm:slow} exclude every fixed level of the polynomial hierarchy. The number $h_0$ of levels used by the lower-bound proof depends on that level and on the polynomial time bound. There is also a direct procedure with at most $2h(n)$ quantifier blocks that decides $\calL_h$ relative to $O_h$. It guesses the secret $s$ of a node existentially, chooses universally either a code child $x$, claimed to have value $C_m(s,x)$, or the data child $s$, claimed to have the value of the node, and recurses on that claim, with polynomial work on every path. A leaf is checked by one query. A true claim with the true secret makes every child claim true. A false claim fails on some branch for every guessed secret, because with the true secret the data child has the true value of the node, and with a wrong secret some code child disagrees with the codeword of the wrong secret. Induction on the height completes the argument. The universal choice ranges over all code positions, rather than only the three positions sampled by the verifier. This procedure uses at most $\log n/4$ quantifier blocks for Theorem~\ref{thm:main} and at most $f(n)$ for Corollary~\ref{cor:logstar}. A lower bound of $\Omega(h(n))$ on the number of alternations needed for $\calL_h$ relative to $O_h$ is not proved.

\subsection{Excluding constant-message interactive proofs}

\begin{proof}[Proof of Corollary~\ref{cor:consequences}]
Part (i) follows from Theorem~\ref{thm:main} and the containment $\EQP^O\subseteq\BQP^O$. For part (ii), we suppose that $\calL$ has an interactive proof system relative to $O$ with a constant number of messages, completeness at least $2/3$, and soundness error at most $1/3$, and we derive a contradiction. Since the verifier and all messages are classical, an optimal prover can be taken classical and deterministic. The system is therefore an interactive proof system in the classical sense, and the transformation of Goldwasser and Sipser~\cite{GS86} gives a public-coin proof system with a constant number of messages. The results of Babai~\cite{Bab85} and of Babai and Moran~\cite{BM88} reduce the number of messages to two, which gives an Arthur--Merlin proof system. The standard containment $\AM^O\subseteq\PiO2$ then applies~\cite{BM88,Gol08}. All these steps relativize, and none of them requires an efficient honest prover. Hence $\calL\in\AM^O\subseteq\PiO2\subseteq\PH^O$, contradicting Theorem~\ref{thm:main}.
\end{proof}

\subsection{Extension to noisy code tables}\label{sec:noisyproof}

Theorems~\ref{thm:main} and~\ref{thm:slow} use exact code tables, in which the code table of every node is a codeword. In this subsection every code table may be within a fixed relative distance $\eta<1/4$ of a codeword. The circuit lower bound uses the code family only through its degree bound and the random projections. Since the exact instances satisfy the weaker promise, the lower bound carries over unchanged. The quantum algorithm and the protocol also use the decoder and the distance between codewords, and Appendix~\ref{app:noisy} adapts both to noisy code tables. The oracle construction uses only exact valid trees, namely the default trees and the target tables of Section~\ref{sec:oracleproof}, and is unchanged.

For $0\le\eta<1/4$, we define validity with noise rate $\eta$ from the leaves up. Every child must be valid with noise rate $\eta$, and at an internal node the table of code-child values must be within distance $\eta$ of a codeword. The secret is unique, since distinct codewords have distance $1/2>2\eta$, and the node value is the value of the data child at that secret. This condition concerns tables of child values at every node. It does not describe arbitrary corruption of terminal leaves. Every exact valid tree is valid with noise rate $\eta$, with the same value.

The following theorem extends the evaluation and verification results to such instances, with bounded error in place of exactness and perfect completeness.

\begin{theorem}[Evaluation and verification with noisy code tables]\label{thm:robust}
Fix a code family of Definition~\ref{def:code}, a noise rate $0\le\eta<1/4$, and a height function $h$. Instances of size $n$ and height $h(n)$ in which every code table is within relative distance $\eta$ of a codeword have a uniform polynomial-time bounded-error quantum algorithm and a public-coin interactive proof system with completeness at least $0.99$, soundness error at most $0.01$ against computationally unbounded provers, an efficient honest prover, and polynomially many messages. No family of constant-depth circuits of size $2^{\poly(n)}$ computes the root value on every such instance.
\end{theorem}

\begin{proof}[Proof of Theorem~\ref{thm:robust}]
The quantum algorithm is the algorithm of Theorem~\ref{thm:noisyquantum} in Appendix~\ref{app:noisy}, and the proof system is that of Theorem~\ref{thm:robustip}, run on the claim that the root has value one. Let $C_\eta$ and $t_\eta$ be the constants of these theorems. For $h=h(n)$, the factors $C_\eta^h$, $(2t_\eta)^h$, and $(2t_\eta C_\eta)^h$ in the time bounds of the algorithm, the verifier, and the honest prover are polynomial in $n$, since $\eta$ is fixed and $h(n)\le\log n/8$. The communication, and hence the number of messages, is bounded likewise.

For the lower bound, every circuit that is correct on all instances of size $n$ and height $h(n)$ that are valid with noise rate $\eta$ is correct on all exact valid instances of this size and height. A family of constant-depth circuits of size $2^{\poly(n)}$ has some depth $D$, and its size is at most $2^{n^A}$ for some $A\ge1$ and all large $n$. Let $h_0,n_0$ be the constants of Theorem~\ref{thm:circuits} for the code family, $D$, and $A$. For every $n\ge n_0$ with $h(n)\ge h_0$ at which the size bound holds, the circuit of the family is not correct on all exact valid instances of size $n$ and height $h(n)$, since otherwise its expectations under $\mu_{n,h(n),1}$ and $\mu_{n,h(n),0}$ would differ by one, contrary to that theorem. There are arbitrarily large such $n$ because $h$ is unbounded. The lower bound thus uses only the exact valid instances. It does not assert a correlation estimate for a distribution of corrupted code tables.

Finally, let $O$ be any oracle that stores, at every size $n$ and in the format of Section~\ref{sec:oracleproof}, an instance of height $h(n)$ that is valid with noise rate $\eta$. The quantum algorithm and the protocol apply at every size, so the language of root values of $O$ is in $\BQP^O$ and has the stated proof system. In particular, the oracle $O_h$ constructed in Section~\ref{sec:oracleproof} for this height function and this code family stores exact valid instances and has these properties.
\end{proof}

\paragraph{Open problems.}
Two questions remain open. First, in the formulation of recursive Fourier sampling verified by McKague~\cite{McK10}, the value of a node is $g(s)$ for a fixed polynomial-time predicate $g$. Our lower bound uses the fact that the data table is unconstrained by the code table apart from the data entry at the secret. This freedom lets Claim~\ref{claim:count} fix the data tables in the auxiliary experiment of Claim~\ref{claim:twostep}, in which the secrets are independent of the data tables, and count secrets. Recursive Fourier sampling with a fixed predicate lacks this freedom, and whether it is outside $\PH$ relative to some oracle is open. Second, the language $\calL_h$ is decided with $2h(n)$ quantifier blocks (Section~\ref{sec:proofs-main}), while Theorem~\ref{thm:slow} excludes every fixed number of alternations. Whether there are an oracle $O$ and a problem in $\BQP^O$ that has an interactive proof system with an efficient quantum prover and information-theoretic soundness but is not decided with a logarithmic number of alternations relative to $O$ is open.

\appendix
\section{Quantum evaluation and verification with noisy code tables}\label{app:noisy}

In this appendix we prove the evaluation and verification guarantees of Theorem~\ref{thm:robust}. We fix a code family of Definition~\ref{def:code} and a noise rate $0\le\eta<1/4$, and we put $\kappa=1/4-\eta>0$. Constants may depend on $\eta$ and on the code family. A tree that is valid with noise rate $\eta$ is valid with every larger noise rate below $1/4$, so the algorithms may use any fixed rational $\eta'$ with $\eta\le\eta'<1/4$ in place of $\eta$ in all parameter choices below. Throughout, trees have the dimensions~\eqref{eq:schedule} of an instance of size $n$ and are valid with noise rate $\eta$ in the sense of Section~\ref{sec:noisyproof}. Claim~\ref{lem:decoder} recovers the secret of one node from its noisy code table, Claim~\ref{lem:clean} turns a bounded-error algorithm into a unitary close to a clean oracle, Theorem~\ref{thm:noisyquantum} assembles the recursion, and Theorem~\ref{thm:robustip} gives the protocol.

For a node of dimension $m$ with code table $L$, the true secret is the unique $s^*$ whose codeword is within distance $\eta$ of $L$. The agreement of a secret $s$ is the fraction of positions $x\in\F^m$ with $L_x=C_m(s,x)$, that is, the relative agreement of its codeword with $L$. The true secret has agreement at least $1-\eta=3/4+\kappa$. Every other secret has agreement at most $1/2+\eta=3/4-\kappa$, since its codeword agrees with the true codeword in exactly half of the positions (Section~\ref{sec:codes}) and $L$ differs from the true codeword in at most an $\eta$ fraction of all positions. This gap is used in the validation step of Claim~\ref{lem:decoder} and in the soundness argument of Theorem~\ref{thm:robustip}.

\subsection{Decoding and evaluation with noisy code tables}

\begin{claim}\label{lem:decoder}
For every fixed $0<\zeta<1$ there is a constant $q_{\eta,\zeta}$ such that the secret of a node whose code table is within distance $\eta$ of a codeword can be recovered with error probability at most $\zeta$ by a quantum algorithm that makes $q_{\eta,\zeta}$ queries to the code table and runs in polynomial time. Consequently the value of such a node is computed with error at most $\zeta$ by a constant number of queries to its tables.
\end{claim}

\begin{proof}
Let $s^*$ be the true secret and $L$ the code table. One query to $L$ prepares its phase state $\ket L=2^{-m/2}\sum_x(-1)^{L_x}\ket x$. Since $U_m^{-1}$ maps $\ket{s^*}$ to the phase state $\ket{C_{s^*}}$ of the codeword of $s^*$ (Definition~\ref{def:code}(iii)), the amplitude of $U_m\ket L$ at $\ket{s^*}$ is the inner product of $\ket{C_{s^*}}$ and $\ket L$. This inner product is
\[
 1-2\dist\bigl(L,C_m(s^*,\cdot)\bigr)\ge1-2\eta .
\]
Measuring $U_m\ket L$ therefore returns $s^*$ with probability at least $(1-2\eta)^2>1/4$, which may be far from one. To reduce the error, the algorithm produces a constant-size list of candidates that contains $s^*$ with probability at least $1-\zeta/2$ and validates each candidate by its agreement. We put
\[
 M=\bigl\lceil4\ln(2/\zeta)\bigr\rceil,\qquad
 p=\Bigl\lceil\frac{\ln(2M/\zeta)}{2\kappa^2}\Bigr\rceil .
\]
Here $M$ is the length of the candidate list and $p$ is the number of positions read to validate a candidate. The algorithm repeats the measurement $M$ times independently, which gives a list of $M$ candidate secrets. The true secret is absent from the list with probability at most $(3/4)^M\le e^{-M/4}\le\zeta/2$.

To validate a candidate $s$, the algorithm reads $L$ at $p$ uniformly random positions, chosen independently of the list and of the positions used for the other candidates, and accepts $s$ if $L_x=C_m(s,x)$ at a fraction at least $3/4$ of these positions. It returns the first accepted candidate, and an arbitrary secret if no candidate is accepted. A validation decision is wrong if it rejects the true secret or accepts another secret. Since the true secret has agreement at least $3/4+\kappa$ and every other secret at most $3/4-\kappa$, Hoeffding's inequality~\cite{Hoe63} bounds the probability of each wrong decision by $\exp(-2p\kappa^2)\le\zeta/(2M)$. A union bound over the $M$ decisions bounds the probability of any wrong decision by $\zeta/2$. If the true secret is on the list and every validation decision is correct, the output is the true secret. Hence the output is wrong with probability at most $\zeta$.

The algorithm makes one query to $L$ for each of the $M$ measurements and $p$ queries for each validation, in total at most $M(1+p)$ queries. This number is a constant, which we take as $q_{\eta,\zeta}$. One further query to the data table at the recovered secret gives the value of the node. All classical processing takes polynomial time. Implementing the algorithm unitarily by deferred measurement~\cite[Section~4.4]{NC00} gives a circuit that retains its work registers.
\end{proof}

\begin{claim}\label{lem:clean}
Fix a height. An address is valid if it is the address of a node of this height in a tree that is valid with noise rate $\eta$. Suppose that a uniform unitary algorithm uses the address register only as a control and that, on every valid address with its work registers in the zero state, a designated output bit of the algorithm equals the value of the node at that address with probability at least $9/10$. Then for every fixed $\delta>0$ there is a uniformly generated unitary circuit $W$ with
\[
 \norm{(W-\mathcal E)\Pi_0}\le\delta ,
\]
where $\mathcal E$ is the clean oracle that adds the value of the node at the address to an output qubit on which the given algorithm does not act, and $\Pi_0$ is the orthogonal projector onto the subspace of valid addresses with all work registers in the zero state. The norm is the operator norm, so the bound states that $\norm{(W-\mathcal E)\psi}\le\delta\norm\psi$ for every $\psi$ in the range of $\Pi_0$. The time, the number of queries, and the space of $W$ exceed those of the given algorithm by at most a constant factor depending only on $\delta$. The bound holds on superpositions of valid addresses with an arbitrary joint state of the output qubit and of any reference system, and $W$ uses the address register only as a control.
\end{claim}

\begin{proof}
Let $A$ be the unitary that runs an odd number $\ell$ of copies of the algorithm coherently, each on fresh work registers, and computes the majority of their output bits reversibly into a fresh qubit, the majority qubit. At a fixed valid address with zero work registers, the output bits of the copies are independent and each is wrong with probability at most $1/10$. The majority is wrong only if at least $\ell/2$ of them are wrong, which has probability at most $e^{-2\ell(2/5)^2}$ by Hoeffding's inequality.

Let $W=A^{-1}\,\mathrm{CNOT}\,A$, where $\mathrm{CNOT}$ is the controlled-NOT gate with the majority qubit as control and the output qubit as target. The inverse $A^{-1}$ reverses the circuit of $A$ and inverts each gate, so $W$ uses the address register only as a control. Both $A$ and $\mathcal E$ use the address register only as a control. Apart from this control, $A$ acts on the work registers and the majority qubit, and $\mathcal E$ acts only on the output qubit, on which $A$ does not act. Hence $A$ commutes with $\mathcal E$, and $W-\mathcal E=A^{-1}(\mathrm{CNOT}-\mathcal E)A$. We fix a valid address $a$ and a state $\psi$ with this address, zero work registers, and an arbitrary joint state of the output qubit and of a reference system, and we write $A\psi=\psi_{\mathrm{good}}+\psi_{\mathrm{bad}}$ according to whether the majority qubit holds the value of the node at $a$. On $\psi_{\mathrm{good}}$ the operators $\mathrm{CNOT}$ and $\mathcal E$ add the same bit to the output qubit, so $(\mathrm{CNOT}-\mathcal E)\psi_{\mathrm{good}}=0$. The norm of $\psi_{\mathrm{bad}}$ is $\norm\psi$ times the square root of the probability that the majority is wrong, hence at most $e^{-\ell(2/5)^2}\norm\psi$, and the difference of two unitaries has norm at most two. Hence
\[
 \norm{(W-\mathcal E)\psi}\le2e^{-\ell(2/5)^2}\norm\psi ,
\]
and a fixed odd $\ell$, depending only on $\delta$, makes the factor $2e^{-\ell(2/5)^2}$ at most $\delta$. Different addresses occupy orthogonal subspaces that are invariant under both $W$ and $\mathcal E$, so the same bound holds on superpositions of valid addresses, which is the statement. By the gate-set convention of Section~\ref{sec:oracle-classes}, the circuit of $A^{-1}$ has the same size as that of $A$, hence the same time, number of queries, and space. Since $W$ runs $2\ell$ copies of the given algorithm or of its inverse together with the majority computation and one gate, its time, number of queries, and space exceed those of the given algorithm by a factor depending only on $\ell$, hence only on $\delta$.
\end{proof}

\begin{theorem}\label{thm:noisyquantum}
There is a uniform quantum algorithm that, on every tree of height $j$ that is valid with noise rate $\eta$, computes its value with error at most $1/10$, with at most $C_\eta^j$ leaf queries and in time $C_\eta^j\poly(n,j)$, for a constant $C_\eta$. For every fixed $\delta>0$ and $\zeta>0$, a unitary within $\delta$ of the clean oracle for the values of such trees in the sense of Claim~\ref{lem:clean}, and a decoder of their secrets with error at most $\zeta$, have resource bounds of the same form.
\end{theorem}

\begin{proof}
We construct algorithms $B_j$ with error at most $1/10$, one per height, each keeping the address of its node in a register that it uses only as a control. At height zero, $B_0$ is a leaf query. At height $j\ge1$, we take the one-node algorithm of Claim~\ref{lem:decoder} with error at most $1/100$. It makes a fixed number $q=q_{\eta,1/100}+1$ of table queries, namely $q_{\eta,1/100}$ queries to the code table and one query to the data table. Each table entry is the value of a child, which is a node of height $j-1$, so a table query is a query to the values of nodes of height $j-1$. We apply Claim~\ref{lem:clean} to $B_{j-1}$ with $\delta=1/(100q)$ to obtain a unitary $W$ within $\delta$ of the clean oracle $\mathcal E$ for these values, and we let $B_j$ be the one-node algorithm with every table query replaced by a call of $W$. Each call of $W$ uses its own fresh work registers, which are in the zero state before the call. Since the one-node algorithm and $W$ use the address of the node only as a control, so does $B_j$.

We compare $B_j$ with the one-node algorithm with exact table queries by a hybrid argument. For $0\le i\le q$, let $H_i$ be the algorithm that answers the first $i$ table queries by $W$ and the remaining ones by $\mathcal E$, so that $H_0$ is the one-node algorithm and $H_q=B_j$. The algorithms $H_{i-1}$ and $H_i$ differ only in the $i$th query, and their states before this query coincide. Before this query, the work registers of the call are fresh and therefore in the zero state. The address register of the call holds a superposition of addresses of children of the node, and every child of a valid node is valid. All other registers, including the work registers of the earlier calls of $W$, form the reference system. By Claim~\ref{lem:clean} the states after this query differ in norm by at most $\delta$, and the later operations are the same unitaries in both algorithms, so the final states of $H_{i-1}$ and $H_i$ differ in norm by at most $\delta$. Hence the final states of $B_j$ and of the one-node algorithm differ in norm by at most $q\delta=1/100$, and every output probability changes by at most $2q\delta=1/50$. The error of $B_j$ is therefore at most $1/100+1/50<1/10$. The argument uses only the bound $1/10$ for $B_{j-1}$, so by induction on $j$ every $B_j$ has error at most $1/10$.

The number $\ell$ of copies in Claim~\ref{lem:clean} is a constant, since $q$ and $\delta$ depend only on $\eta$. For a constant $K_\eta$ and a fixed polynomial $\poly(n,j)$, the numbers of queries $Q_j$, the times $T_j$, and the retained work space $S_j$ satisfy
\[
 Q_j\le K_\eta Q_{j-1},\qquad
 T_j\le K_\eta\bigl(T_{j-1}+\poly(n,j)\bigr),\qquad
 S_j\le K_\eta S_{j-1}+\poly(n,j),
\]
because each of the $q$ calls of $W$ runs $\ell$ copies of $B_{j-1}$, each on its own work space, and then their inverses. These recurrences give the stated bounds after enlarging $C_\eta$.

All subtrees of one height use the same circuit, which is called in superposition by appending to the address of the node the bit $0$ or $1$ and the index of the child, as in Section~\ref{sec:upper}. The description of the circuit is generated recursively within the time bound.

A unitary within $\delta$ of the clean oracle is obtained from $B_j$ by Claim~\ref{lem:clean}. To decode a secret with error $\zeta$, we run the algorithm of Claim~\ref{lem:decoder} with error $\zeta/2$, which makes $q'=q_{\eta,\zeta/2}$ queries, and we replace each query by a call of the unitary of Claim~\ref{lem:clean} for the children with $\delta=\zeta/(4q')$, again on fresh work registers. By the hybrid argument above, this replacement adds at most $2q'\delta=\zeta/2$ to the error. All these quantities are constants, so the resource bounds have the stated form.
\end{proof}

\subsection{Verification with noisy code tables}

\begin{theorem}\label{thm:robustip}
For every fixed $0\le\eta<1/4$ there is a classical public-coin interactive proof system for claims about nodes of trees that are valid with noise rate $\eta$, with completeness at least $0.99$, soundness error at most $0.01$, and a uniform quantum honest prover. There is a constant $t_\eta$ with the following bounds for a claim about a node of height $j$. The verifier makes at most $(2t_\eta)^j$ leaf queries and runs in time $(2t_\eta)^j\poly(n,j)$, and the communication is bounded likewise. The honest prover runs in time $(2t_\eta C_\eta)^j\poly(n,j)$, after enlarging $C_\eta$ if necessary.
\end{theorem}

\begin{proof}
We maintain a fixed error bound at every height by keeping the success probabilities of honest trials and of trials on false claims on opposite sides of $3/4$. We put
\[
 \epsilon=\min\{1/100,\kappa/8\},
\]
and let $t=t_\eta$ be a fixed positive multiple of four, so that $3t/4$ is an integer, large enough that
\[
 \exp\bigl(-2t(5\kappa/8)^2\bigr)\le\epsilon .
\]
We modify Protocol~\ref{prot:verify} as follows, with the same claims, public parameters, and oracle access as in Section~\ref{sec:verify}. A leaf claim is checked by one query. For a claim $(w,b)$ about an internal node of dimension $m$, the parties run $t$ trials in sequence. In each trial the prover first sends a secret $s\in\F^m$. The verifier then sends a fresh uniformly random $x\in\F^m$, and the parties run the protocol recursively on the claim that code child $w0x$ has value $C_m(s,x)$ and then on the claim that data child $w1s$ has value $b$. The trial succeeds if both recursive checks accept, and the verifier accepts the claim $(w,b)$ if at least $3t/4$ trials succeed.

A malformed secret message makes its trial fail, and the verifier uses the secret zero in the recursive calls of that trial. Both recursive checks are run even after a failure, so the number and the order of the recursive calls do not depend on the prover's messages.

Both parties derive the child claims from the transcript, and every verifier message consists only of fresh random bits. The protocol is therefore public-coin.

Both directions use the following tail bound, which requires no independence between the trials. Let $Y_1,\dots,Y_t$ be Bernoulli variables, and let the past of trial $i$ be the transcript before that trial, which determines $Y_1,\dots,Y_{i-1}$. Suppose that $\E[Y_i\mid\text{past}]\le\gamma$ for every $i$. Then for every $\lambda>0$ the conditional exponential moment satisfies
\[
 \E\bigl[e^{\lambda Y_i}\mid\text{past}\bigr]\le1-\gamma+\gamma e^{\lambda} ,
\]
and iterating this inequality over the trials gives $\E[e^{\lambda\sum_iY_i}]\le(1-\gamma+\gamma e^{\lambda})^t$, which is the exponential moment of a sum of $t$ independent Bernoulli variables with mean $\gamma$. Markov's inequality and Hoeffding's lemma $1-\gamma+\gamma e^{\lambda}\le e^{\lambda\gamma+\lambda^2/8}$, with $\lambda=4\theta$, then give $\Pr[\sum_iY_i\ge t(\gamma+\theta)]\le e^{-2t\theta^2}$ for every $\theta>0$. The lower tail bound, for conditional means at least $\gamma$, follows by applying the same argument to $1-Y_i$.

We prove by induction on the height, uniformly over all valid subtrees and all previous histories, that a false claim is accepted with probability at most $\epsilon$ against every prover, and that a true claim is accepted with probability at least $1-\epsilon$ when the prover is the honest prover described below. At leaves the verifier reads the value, so a false leaf claim is rejected and a true leaf claim is accepted with probability one.

For soundness, we consider a false claim $(w,b)$ about a node of height $j\ge1$ with true secret $s^*$, and we fix the history before one of its trials. A malformed secret message makes the trial fail, so we may assume that the prover sends a secret $s\in\F^m$. If $s=s^*$, the claim that data child $w1s^*$ has value $b$ is false, since the value of the node is the value of this data child and differs from $b$. By induction this claim is accepted with probability at most $\epsilon$, and the trial succeeds with at most this probability. If $s\ne s^*$, the agreement of $s$ is at most $1/2+\eta$, so the claim that code child $w0x$ has value $C_m(s,x)$ is true with probability at most $1/2+\eta$ over the choice of $x$. When this claim is false, it is accepted with probability at most $\epsilon$ by induction, conditional on everything before. Hence the trial succeeds with probability at most
\[
 \frac12+\eta+\epsilon\le\frac34-\frac{7\kappa}8 ,
\]
because the trial succeeds only if the code check accepts, whatever the data check does. Averaging over the prover's choice of $s$ preserves the bound. The outcomes of the $t$ trials are therefore Bernoulli variables whose conditional success probabilities, given the past, are at most $3/4-7\kappa/8$. By the tail bound above with $\theta=7\kappa/8$ and the choice of $t$, at least $3t/4$ of them succeed with probability at most $\exp(-2t(7\kappa/8)^2)\le\exp(-2t(5\kappa/8)^2)\le\epsilon$. The argument conditions only on the transcript, so it applies to quantum provers that send classical messages.

For completeness, the honest prover computes the secret of the node in every trial with the decoder of Theorem~\ref{thm:noisyquantum} with error at most $\epsilon$, on fresh work registers for every run, and sends the result. It follows this policy at every invocation, including invocations on false child claims, which arise at corrupted code positions and after decoding errors. We consider a true claim $(w,b)$ and one of its trials, with the history fixed. Since the decoder runs on fresh work registers, the recovered secret is wrong with probability at most $\epsilon$, conditional on the history. If it is correct, the data claim is true, and the code claim is false only if $x$ is one of the at most $\eta$ fraction of positions at which the code table differs from the codeword. When both child claims are true, each is rejected with probability at most $\epsilon$ by induction, conditional on the history. Hence an honest trial succeeds with probability at least
\[
 1-\eta-3\epsilon\ge\frac34+\frac{5\kappa}8 ,
\]
where $3\epsilon$ is the sum of the bounds for one decoding error and two recursive rejections. The lower tail bound above with $\theta=5\kappa/8$ then shows that fewer than $3t/4$ trials succeed with probability at most $\exp(-2t(5\kappa/8)^2)\le\epsilon$ by the choice of $t$, so the claim is rejected with probability at most $\epsilon$.

Each invocation at an internal node makes $2t$ recursive invocations, so a claim about a node of height $j$ causes at most $(2t)^j$ leaf queries and at most $2(2t)^j$ invocations in total. Each invocation takes the verifier time $\poly(n,j)$ and exchanges at most $2tn$ bits, which gives the bounds for the verifier and the communication. The honest prover runs the decoder of Theorem~\ref{thm:noisyquantum} $t$ times per invocation, each run in time at most $C_\eta^j\poly(n,j)$, and multiplying by the number of invocations gives its bound after enlarging $C_\eta$. Since $\epsilon\le1/100$, the completeness and the soundness error stated in the theorem follow.
\end{proof}

\section{A matching lower bound on the number of messages}\label{app:interaction}

This appendix proves Theorem~\ref{thm:interaction}. Throughout, instances have size $n$ and height $h\le h_{\max}(n)$ with the dimensions~\eqref{eq:schedule}, levels are numbered from the bottom as in Section~\ref{sec:projections}, $\rho_i$ is the projection at level $i$, $N_i$ is the number of nodes at level $i$, and $(X_0,X_1)$ is the coupled pair of leaf tables with root values $0$ and $1$. The code family has degree $d$. We put
\[
 t_i=\Bigl\lfloor\frac{m_i}{16d}\Bigr\rfloor\quad(0\le i<h),\qquad
 P(n,h)=\prod_{i<h}(t_i+1),
\]
with $P(n,0)=1$. The value $t_i$ is the threshold in the definition of light blocks at level $i$, as in the proof of Theorem~\ref{thm:flips}.

The argument has five steps. Theorem~\ref{thm:widthzero} shows that a DNF of width less than $P(n,h)$ cannot distinguish the two root values, by applying Theorem~\ref{thm:winner} with $\ell=1$ and Claim~\ref{claim:truncate} with $J=0$ at every level. Unlike Theorems~\ref{thm:flips} and~\ref{thm:switch}, which are stated for a fixed level and applied at a constant number of bottom levels, this theorem is uniform over all levels up to the full height $h\le h_{\max}(n)$. Lemma~\ref{lem:am} turns this into a query lower bound for two-message public-coin protocols. Lemma~\ref{lem:collapse} reduces a public-coin protocol with $r$ rounds to a two-message one at the cost of a factor $\Delta^r$ in the number of queries, where $\Delta$ is polynomial in $r$ and in the total length of the prover's replies. For a polynomial-time verifier this factor is $n^{O(r)}$. Theorem~\ref{thm:finite-messages} combines the three. Finally, we prove Theorem~\ref{thm:interaction} by adding stages for verifiers to the oracle construction of Section~\ref{sec:oracleproof}.

\subsection{DNFs of small width cannot distinguish the root values}

For $0\le i<h\le h_{\max}(n)$ we have $16^{i+1}\le16^{h}\le\sqrt n$, so $n/16^i\ge16\sqrt n$ and
\begin{equation}\label{eq:mi-bounds}
 m_i=8\Bigl\lfloor\frac{n}{8\cdot16^i}\Bigr\rfloor\ge\frac{n}{16^i}-8\ge\max\Bigl\{\frac{n}{2\cdot16^i},\,8\sqrt n\Bigr\}.
\end{equation}
Hence $t_i+1>m_i/(16d)\ge n/(32d\cdot16^i)$, and since $2(h-1)\le2h_{\max}(n)\le\log n/4$,
\begin{equation}\label{eq:P-lower}
 \log P(n,h)\ge\sum_{i<h}\bigl(\log n-4i-\log(32d)\bigr)\ge\frac34\,h\log n-h\log(32d).
\end{equation}
In the other direction $P(n,h)\le(n+1)^h$, since $t_i+1\le m_i+1\le n+1$.

\begin{theorem}[Width reduction to a constant]\label{thm:widthzero}
There is a threshold on $n$, depending only on $d$, beyond which the following holds for every $h\le h_{\max}(n)$. Every DNF $\Phi$ on the leaves of an instance of size $n$ and height $h$ with width less than $P(n,h)$ satisfies
\[
 \bigl|\E\,\Phi(X_1)-\E\,\Phi(X_0)\bigr|\le2^{-2\sqrt n}.
\]
\end{theorem}

\begin{proof}
We normalize $\Phi$ and write $\Phi_0=\Phi$ and $W_0=\width(\Phi)$. For $0\le i<h$ let $\Phi_{i+1}=(\Phi_i)^-_{\rho_i}$ be the DNF of Claim~\ref{claim:truncate} applied to the $N_i$ nodes at level $i$, each a block of height one and dimension $m_i$, with $t=t_i$ and $J=0$. Thus $\Phi_{i+1}$ is obtained from $\Phi_i$ by deleting every term for which at least one block at level $i$ is light and projecting the remaining terms by $\rho_i$. It is a DNF on the node values at level $i$, determined by $\Phi$ and $\rho_0,\dots,\rho_i$, and it does not depend on the root value. By Claim~\ref{claim:truncate}, $\width(\Phi_{i+1})\le\lfloor W_i/(t_i+1)\rfloor=:W_{i+1}$, and
\begin{equation}\label{eq:monotone}
 \Phi_{i+1}(z)\le\Phi_i(\rho_i(z))\qquad\text{for every }z.
\end{equation}
Since nested floors satisfy $\lfloor\lfloor x/a\rfloor/b\rfloor=\lfloor x/(ab)\rfloor$ for positive integers $a,b$, we get $W_h=\lfloor W_0/P(n,h)\rfloor=0$, because $W_0<P(n,h)$. A DNF of width zero has no literals, so $\Phi_h$ is a constant $c\in\{0,1\}$ that depends on the projections but not on the root value.

For a root value $b$ let $Y^b_i$ be the vector of node values at level $i-1$ of the instance with root value $b$, so that $Y^b_0=X_b$, $Y^b_i=\rho_i(Y^b_{i+1})$ for $i<h$, and $Y^b_h=(b)$. By~\eqref{eq:monotone}, $\Phi_h(Y^b_h)\le\Phi_{h-1}(Y^b_{h-1})\le\dots\le\Phi_0(Y^b_0)=\Phi(X_b)$, so $\Phi(X_b)=1$ whenever $c=1$. Let $\widetilde E^b_i$ be the event that $\Phi_i(Y^b_i)=1$ and $\Phi_{i+1}(Y^b_{i+1})=0$. By the last assertion of Claim~\ref{claim:truncate}, $\widetilde E^b_i$ is contained in the event that $\Phi_i(\rho_i(Y^b_{i+1}))=1$ and at least one block at level $i$ is light for the first satisfied term of $\Phi_i$ at $\rho_i(Y^b_{i+1})$. If $\Phi(X_b)=1$ and $c=0$, then the sequence $\Phi_0(Y^b_0),\dots,\Phi_h(Y^b_h)$ starts at one and ends at zero, so some $\widetilde E^b_i$ occurs.

We fix $i<h$ and condition on the projections of the levels below $i$, which determine $\Phi_i$, and on $Y^b_{i+1}$, which is determined by the root value and the projections of the levels above $i$. These are independent of $\rho_i$ by the argument of Claim~\ref{claim:proj}(b). For every fixed value of $\Phi_i$ and $z=Y^b_{i+1}$, Theorem~\ref{thm:winner} with $K=N_i$, $W=W_i$, $t=t_i$, and $\ell=1$ bounds the conditional probability of $\widetilde E^b_i$ by
\[
 2N_iW_i\binom{m_i}{\le dt_i}2^{-m_i}\le2W_i\,2^{m_i/15+\log n/8}\,2^{H_2(1/16)m_i-m_i}\le2W_i\,2^{-0.59\,m_i+\log n/8},
\]
using~\eqref{eq:blocks} for $N_i$, $dt_i\le m_i/16$ with the entropy bound of Section~\ref{sec:prelim}, and $1-0.34-1/15>0.59$. If $W_i=0$, the event $\widetilde E^b_i$ is empty. Since $W_i\le W_0<P(n,h)\le(n+1)^h\le2^{(\log n)^2/4}$ and $m_i\ge8\sqrt n$ by~\eqref{eq:mi-bounds}, the bound is at most $2^{1+(\log n)^2/4+\log n/8-4.7\sqrt n}\le2^{-4\sqrt n}$ for all sufficiently large $n$, uniformly in $i$ and $h$. Averaging over the conditioning and summing over the $h\le\log n$ levels,
\[
 \Pr\bigl[\Phi(X_b)=1\ \text{and}\ c=0\bigr]\le\sum_{i<h}\Pr[\widetilde E^b_i]\le2^{-3\sqrt n}.
\]
Therefore $\E\,\Phi(X_1)\le\Pr[c=1]+2^{-3\sqrt n}$ and $\E\,\Phi(X_0)\ge\Pr[c=1]$, so $\E\,\Phi(X_1)-\E\,\Phi(X_0)\le2^{-3\sqrt n}$, and the same bound holds with the root values exchanged. For $h=0$ the leaf table is the root bit itself, $P(n,0)=1$, and a DNF of width zero is constant, so the difference is zero.
\end{proof}

\subsection{From protocols to DNFs}

In the remainder of this appendix the input of a protocol is a valid instance of size $n$ and height $h\le h_{\max}(n)$, given by its leaf table $X$, and every other oracle answer is fixed. A verifier may compute arbitrarily. Only its queries to $X$ and its messages are counted. The value of a protocol on $X$ is the maximum, over all provers, of the probability that the verifier accepts. A protocol is correct on $X$ if its value is at least $2/3$ when the root value of $X$ is one and at most $1/3$ when the root value is zero.

\begin{lemma}[Two-message protocols]\label{lem:am}
Consider a protocol in which the verifier sends a uniformly random string of a fixed length, the prover replies with a string of a fixed length, and the verifier makes at most $q$ queries to $X$ and decides as a deterministic function of its random string, the reply, and the answers. If the protocol is correct on every valid instance of size $n$ and height $h$, and $n$ is beyond the threshold of Theorem~\ref{thm:widthzero}, then $q\ge P(n,h)$.
\end{lemma}

\begin{proof}
We fix the verifier's random string $c$. For each reply $w$ the verifier's decision is computed by an adaptive decision tree of depth at most $q$ over the leaves, and an input that satisfies the conjunction of the answers along an accepting path of this tree is accepted with the same reply. Let $D_c(X)$ be the disjunction, over all replies $w$ and all accepting paths, of these conjunctions. After merging identical conjunctions, $D_c$ is a DNF of width at most $q$, and $D_c(X)=1$ exactly when some reply makes the verifier accept $X$. The value of the protocol on $X$ is $\E_c D_c(X)$, since the prover may choose its reply after seeing $c$. Correctness on all valid instances gives $\E_cD_c(X)\ge2/3$ when $X$ has root value one and $\E_cD_c(X)\le1/3$ when it has root value zero. Averaging over the coupled pair,
\[
 \E_c\bigl[\E\,D_c(X_1)-\E\,D_c(X_0)\bigr]\ge\frac13 .
\]
If $q<P(n,h)$, Theorem~\ref{thm:widthzero} bounds the bracketed difference by $2^{-2\sqrt n}<1/3$ for every $c$, a contradiction.
\end{proof}

In the next lemma a public-coin protocol proceeds in rounds. In round $i$ the verifier sends a uniformly random string $c_i$ of a fixed length $\ell_i\ge0$ and the prover replies with a string $a_i$ of a fixed length $s_i\ge0$. After the last round the verifier makes at most $q$ queries to $X$ and decides as a deterministic function of the transcript and the answers. The number $r$ of rounds and the lengths $\ell_i$ and $s_i$ do not depend on $X$.

\begin{lemma}[Round reduction]\label{lem:collapse}
Consider a public-coin protocol with $r\ge1$ rounds as above, and let $S=\sum_is_i$ and $\Delta=\bigl\lceil10^4r^2(S+r+1)^2\bigr\rceil$. There is a two-message protocol of the form of Lemma~\ref{lem:am}, with at most $q\Delta^r$ queries, whose verifier is determined by the given one, and whose value on every $X$ is at least $5/6$ if the value of the given protocol on $X$ is at least $2/3$, and at most $1/6$ if that value is at most $1/3$.
\end{lemma}

\begin{proof}
The new verifier samples the edge labels of a complete $\Delta$-ary tree of depth $r$. Every edge from a vertex at depth $i-1$ to one of its $\Delta$ children is labelled by an independent uniformly random string of length $\ell_i$, which serves as a round-$i$ challenge. The new verifier sends all labels in one message. The prover replies with a round-$i$ reply of length $s_i$ for every vertex at depth $i\ge1$. Every root-to-leaf path now carries a complete transcript $(c_1,a_1,\dots,c_r,a_r)$. The new verifier accepts if the given verifier accepts on at least half of the $\Delta^r$ paths. Each evaluation of the given verifier makes at most $q$ queries, so the new verifier makes at most $q\Delta^r$ queries in total. A malformed reply is rejected. We show that the new protocol has the stated values on every fixed $X$.

We fix $X$. For a transcript prefix $\pi_i=(c_1,a_1,\dots,c_{i-1},a_{i-1})$ before round $i$, we define the continuation values of the given protocol by backward induction, with $v_{r+1}(\pi_{r+1})\in\{0,1\}$ the given verifier's decision on the complete transcript,
\[
 w_i(\pi_i,c)=\max_a v_{i+1}(\pi_i,c,a),\qquad v_i(\pi_i)=\E_c\,w_i(\pi_i,c),
\]
where $a$ ranges over strings of length $s_i$ and $c$ is uniform of length $\ell_i$. Then $v_1(\emptyset)$ is the value of the given protocol on $X$. Indeed, a prover that chooses each reply to maximize the continuation value attains $v_1(\emptyset)$. No prover exceeds $v_1(\emptyset)$, since the behavior of any prover is a mixture of deterministic strategies and, by induction on the rounds, no deterministic strategy exceeds the continuation value. In the labelled tree, we define, for a vertex $u$ at depth $i-1$ and a prefix $\pi_i$ whose challenges are the labels on the path to $u$, the empirical values $\widehat v_{r+1}=v_{r+1}$ and
\[
 \widehat w_i(\pi_i,c;u')=\max_a\widehat v_{i+1}(\pi_i,c,a;u'),\qquad
 \widehat v_i(\pi_i;u)=\frac1\Delta\sum_{j=1}^{\Delta}\widehat w_i(\pi_i,c_{u,j};u_j),
\]
where $u'$ is a vertex at depth $i$ and $c$ is the label of the edge into $u'$, and where $u_1,\dots,u_\Delta$ are the children of $u$ and $c_{u,j}$ is the label of the edge to $u_j$. The reply at a vertex affects only the paths through that vertex, so $\widehat v_1(\emptyset;\mathrm{root})$ is the maximum, over all replies of the prover, of the fraction of accepting paths. It remains to compare $\widehat v_1$ with $v_1$.

We put $\delta=1/(12r)$. We fix a vertex $u$ at depth $i-1$ and a reply prefix $(a_1,\dots,a_{i-1})$, which together with the labels on the path to $u$ determine a prefix $\pi_i$. Conditional on the labels on the path to $u$, the labels $c_{u,1},\dots,c_{u,\Delta}$ are independent and uniform, so $w_i(\pi_i,c_{u,1}),\dots,w_i(\pi_i,c_{u,\Delta})$ are independent random variables in $[0,1]$ with mean $v_i(\pi_i)$, and Hoeffding's inequality~\cite{Hoe63} gives
\[
 \Pr\Bigl[\Bigl|\frac1\Delta\sum_{j}w_i(\pi_i,c_{u,j})-v_i(\pi_i)\Bigr|>\delta\Bigr]\le2e^{-2\Delta\delta^2}=2e^{-\Delta/(72r^2)}.
\]
There are fewer than $\Delta^r$ vertices of depth less than $r$ and at most $2^{S}$ reply prefixes at each, so with probability at least $1-\vartheta$, where $\vartheta\le2^{S+1}\Delta^re^{-\Delta/(72r^2)}$, the displayed deviation is at most $\delta$ for every vertex and every reply prefix. We call this the good event. It is determined by the labels alone, before the prover replies.

On the good event we claim that $|\widehat v_i(\pi_i;u)-v_i(\pi_i)|\le(r-i+1)\delta$ for every vertex $u$ at depth $i-1$ and every prefix $\pi_i$ as above. For $i=r+1$ there is nothing to prove. Assuming the claim for the vertices at depth $i$, for every child $u_j$ of $u$ and every reply $a$ we have $|\widehat v_{i+1}(\pi_i,c_{u,j},a;u_j)-v_{i+1}(\pi_i,c_{u,j},a)|\le(r-i)\delta$, so the maxima over $a$ differ by at most $(r-i)\delta$, that is, $|\widehat w_i(\pi_i,c_{u,j};u_j)-w_i(\pi_i,c_{u,j})|\le(r-i)\delta$. Averaging over $j$ and using the good event gives the claim for the vertices at depth $i-1$. At the root, $|\widehat v_1(\emptyset;\mathrm{root})-v_1(\emptyset)|\le r\delta=1/12$. Hence if the value of the given protocol is at least $2/3$, the prover can make at least a $7/12$ fraction of the paths accept, and if that value is at most $1/3$, no reply makes more than a $5/12$ fraction accept. The threshold $1/2$ separates the two cases on the good event.

Finally, we show that $\vartheta<1/6$. We put $U=S+r+1\ge2$, so that $\Delta\le2\cdot10^4r^2U^2$ and $\ln\Delta\le10+2\ln r+2\ln U$. Using $r\le U$, $\ln U\le U/2$, and $(S+1)\ln2\le U^2$,
\[
 \ln\vartheta\le(S+1)\ln2+r\ln\Delta-\frac{10^4}{72}U^2\le U^2+\bigl(10U+2U^2\bigr)-138U^2<-\ln6 .
\]
Therefore the value of the new protocol is at least $1-\vartheta>5/6$ in the first case and at most $\vartheta<1/6$ in the second.
\end{proof}

We put a polynomial-time public-coin verifier into the form of Lemma~\ref{lem:collapse} as follows. Suppose that on input $1^n$ it runs in time at most $n^a$ with every oracle, prover, and random string, and has $M$ alternating communication slots, counting empty messages as in Section~\ref{sec:prelim}. The schedule and the lengths of its random challenges are fixed by the input. Adding a formal empty first challenge if the prover speaks first and a formal empty last reply if the verifier speaks last gives $r\le\lfloor M/2\rfloor+1$ rounds. These formal entries are notation, not additional communication slots. For $M=0$ we take one round with both entries empty. Each genuine challenge stays in its original slot, with its original fixed length. We encode each genuine prover reply by its length followed by padding to $\lfloor n^a\rfloor$ bits, and we reject malformed encodings. For $n\ge2$ this uses at most $2n^a$ bits per reply, so $s_i\le2n^a$ and $S\le2rn^a$. Formal replies have length zero. This changes only the encoding of replies and preserves their order relative to all random challenges. The verifier's final decision is a deterministic function of the transcript and of at most $q\le n^a$ oracle queries. Thus the protocol has the form required by Lemma~\ref{lem:collapse} and has the same value on every $X$.

\begin{theorem}[Finite form of the lower bound]\label{thm:finite-messages}
For every $a\ge1$ there is a threshold on $n$, depending only on $a$ and $d$, beyond which the following holds for every $h\le h_{\max}(n)$ with $h\ge42a$. Every public-coin verifier that runs in time at most $n^a$ on input $1^n$ with every oracle, prover, and random string, and whose protocol is correct on every valid instance of size $n$ and height $h$, uses at least $h/(14a)$ communication slots.
\end{theorem}

\begin{proof}
Let $M$ be the number of communication slots and suppose that $M<h/(14a)$. We put the protocol into the form of Lemma~\ref{lem:collapse} with $r\le\lfloor M/2\rfloor+1\le n^a$, $S\le2rn^a$, and $q\le n^a$. The two-message protocol of Lemma~\ref{lem:collapse} is correct on every valid instance of size $n$ and height $h$, with completeness at least $5/6$ and soundness error at most $1/6$, so Lemma~\ref{lem:am} gives $q\Delta^r\ge P(n,h)$. Since $S+r+1\le4n^{2a}$, we have $\log\Delta\le\log(2\cdot10^4)+2\log r+2\log(4n^{2a})\le7a\log n$ once $\log n\ge19$. We take logarithms in $q\Delta^r\ge P(n,h)$ and use the bounds $q\le n^a$ and $\log\Delta\le7a\log n$ on the left and~\eqref{eq:P-lower} with $\log n\ge4\log(32d)$ on the right. This gives
\[
 a\log n+7ar\log n\ge\log P(n,h)\ge\frac12\,h\log n,
\]
so $r\ge h/(14a)-1/7$ and $M\ge2r-2\ge h/(7a)-3$. For $h\ge42a$ this is at least $h/(14a)$, contradicting the assumption.
\end{proof}

Private coins are handled by the transformation of Goldwasser and Sipser~\cite[Section~4.2]{GS86} (see also~\cite{Gol08}). Suppose a polynomial-time verifier $V$ is supplied with a polynomially bounded, polynomial-time computable bound $M(n)$ on its number of communication slots. If $V$ halts early, we record its decision and fill the remaining slots with dummy messages, whose contents do not affect that decision. We encode each message by its length followed by padding within its slot, as for the replies above. The resulting fixed-schedule private-coin protocol has the same value as $V$. The Goldwasser--Sipser transformation turns this protocol into a polynomial-time public-coin verifier $V'$ with at most $c_0(M(n)+1)$ slots, for an absolute positive integer $c_0$. For every fixed input and oracle, an original value at least $2/3$ gives a transformed value at least $2/3$, and an original value at most $1/3$ gives a transformed value at most $1/3$.

In the full Goldwasser--Sipser protocol, the public challenges specify random hash functions and random targets, while the final check simulates the original verifier on a proposed transcript. We pad the challenges within their slots to input-dependent fixed lengths, using fresh independent bits and ignoring unused suffixes, and we defer all consistency checks to the final decision. This gives the public-coin form of Section~\ref{sec:prelim}. No later challenge is revealed before an earlier reply. The construction relativizes because its oracle-dependent checks are simulations of $V$. Its running time is polynomial in a time bound for $V$ and for computing $M(n)$. We use this polynomial bound when we freeze the oracle below.

\subsection{The oracle}

\begin{proof}[Proof of Theorem~\ref{thm:interaction}]
We modify the construction of Section~\ref{sec:oracleproof} as follows. In addition to the machines of that construction, we enumerate all triples $(V,a,k)$ in which $V$ is a verifier with oracle access, run under a clock of $n^a$ steps on input $1^n$ and rejecting when the clock runs out, $a\ge1$ is an integer, and $k\ge1$ is a repetition index. Thus every pair $(V,a)$ occurs infinitely often.

For a fixed positive integer $K$, we cap a simulation of $V$ at $\lfloor h(n)/K\rfloor$ communication slots. If the simulation requests another slot, we reject before sending or receiving that message. If the simulation decides earlier, we record that decision. We fill any unused slots up to the cap with dummy messages and retain the recorded decision. This padded verifier has the same value as the capped simulation. We choose an integer $a_1\ge\max\{a,2\}$ so that, for every fixed $K$, this verifier and the computation of its slot bound take time at most $n^{a_1}$ for all sufficiently large $n$. The exponent can be chosen independently of $K$, since computing $h(n)$, dividing by a fixed integer, counting slots, and padding take polynomial time. The threshold on $n$ may depend on $K$. The Goldwasser--Sipser transformation has a uniform polynomial overhead in the time bound of the verifier, so we can choose an integer $a'\ge a_1$, depending only on $a_1$ and not on $K$, such that the transformed verifier runs in time at most $n^{a'}$ for all sufficiently large $n$. We put
\[
 \beta_V=\frac{1}{28a'c_0},\qquad H_V=\max\{42a',\,56a'c_0\}.
\]
Let $V_\beta$ be the padded capped verifier just defined with the positive integer $K=28a'c_0$. It has exactly $M(n)=\lfloor\beta_Vh(n)\rfloor$ slots, and $M(n)$ is computable in polynomial time. Let $V'_\beta$ be the public-coin verifier obtained from $V_\beta$ by the Goldwasser--Sipser transformation. It runs in time at most $n^{a'}$ and uses at most $c_0(M(n)+1)\le c_0(\beta_Vh(n)+1)$ slots for all sufficiently large $n$.

The stages for the machines of Section~\ref{sec:oracleproof} are unchanged. At the stage of a triple $(V,a,k)$, we choose $n\ge3$ with $h(n)\ge H_V$ that exceeds the current length bound of the construction, the threshold of Theorem~\ref{thm:finite-messages} for $a'$, and the thresholds for the stated running-time bounds of $V_\beta$ and $V'_\beta$. Such $n$ exists because $h$ is unbounded. We assign to every unassigned instance of size other than $n$ whose addresses have length at most $n^{a'}$ its default tree of root value zero, and we assign zero to the unassigned malformed strings of length at most $n^{a'}$, exactly as before. The target instance, the instance of size $n$, is entirely unassigned, because its addresses have length at least $n+1$, greater than the previous length bound. Its addresses have length at most $3n\le n^{a'}$. Every oracle query of $V'_\beta$ on input $1^n$ is therefore either to the target or to a string that is now fixed.

Thus $V'_\beta$ on input $1^n$ is a public-coin protocol of the kind considered above, with the target leaf table $X$ as its input. It runs in time at most $n^{a'}$ and uses at most
\[
 c_0(\beta_Vh(n)+1)\le\frac{h(n)}{28a'}+c_0
 \le\frac{3h(n)}{56a'}<\frac{h(n)}{14a'}
\]
slots, because $h(n)\ge56a'c_0$. By Theorem~\ref{thm:finite-messages}, since $h(n)\ge42a'$, $V'_\beta$ is not correct on some valid target instance $X$. We assign this instance, assign zero to every remaining unassigned string of length at most $n^{a'}$, and replace the length bound by $n^{a'}$. Every string that either $V_\beta$ or $V'_\beta$ can query on input $1^n$ is now fixed, since $a'\ge a_1$. The guarantee of the transformation for each fixed input implies that $V_\beta$ is not correct on $1^n$ either. If the root of $X$ is one, no prover makes $V_\beta$ accept with probability at least $2/3$. If the root of $X$ is zero, some prover makes $V_\beta$ accept with probability greater than $1/3$. These values do not change at later stages.

We complete the oracle and define $\calL'_h$ as in Section~\ref{sec:oracleproof}. All stored instances are valid, so the quantum algorithm, the protocol, and the resource bounds of Theorem~\ref{thm:slow} hold as before, and the stages of the machines of Section~\ref{sec:oracleproof} give $\calL'_h\notin\PH^{O'_h}$ by Claim~\ref{claim:freeze}.

Now let $V$ be a polynomial-time verifier such that $\calL'_h$ has an interactive proof system relative to $O'_h$ with verifier $V$, completeness at least $2/3$, and soundness error at most $1/3$. Some clocked verifier in the enumeration, with an exponent $a$, has the same behavior as $V$ on every input $1^n$ with $n\ge2$, with every oracle, prover, and random string. We use the constant $\beta_V$ of that pair. Let $n$ be the length chosen at the stage of a triple $(V,a,k)$, and suppose that $V$ uses at most $\beta_Vh(n)$ slots on input $1^n$ with every prover and random string. Then $V$ never requests a slot beyond the cap, so its value on $1^n$ agrees with that of $V_\beta$, since dummy padding after a recorded decision does not affect the value. But $V_\beta$ is incorrect on $1^n$ relative to the final oracle, a contradiction. Hence $V$ uses more than $\beta_Vh(n)$ slots on input $1^n$ for every such length $n$. There are infinitely many of these lengths, and all satisfy $h(n)\ge H_V$.
\end{proof}

For the fixed height function $h$, the constant $\beta_V$ depends on the running-time bounds used above through $a'$. The lower bound holds on infinitely many lengths rather than on all large lengths. At other lengths the stored instances are chosen at other stages or are default trees. For $h=h_{\max}$ the theorem excludes every proof system that uses $o(\log n)$ messages at all sufficiently large lengths, while Theorem~\ref{thm:slow} gives one with $2h(n)\le\log n/4$ messages. The lower bound makes no assumption on the computational power of the provers, and it applies in particular to the public-coin protocol of Theorem~\ref{thm:slow} itself. Therefore, on infinitely many lengths and up to a factor that depends on the verifier, the number of messages of that protocol is the least possible.

\section{Codes from bent functions}\label{app:codes}

This appendix gives nonlinear code families of Definition~\ref{def:code}, whose codewords are the shifts of a bent function, and in particular a family of degree three. It also records the connection between the addressing function of Section~\ref{sec:codes} and Forrelation.

A Boolean function $g$ on $\F^m$ is bent if all Fourier coefficients of $(-1)^g$ have the same magnitude $2^{-m/2}$, and its dual $\widetilde g$ is the Boolean function that records their signs, so that
\[
 2^{-m}\sum_{x\in\F^m}(-1)^{g(x)+x\cdot\alpha}=2^{-m/2}(-1)^{\widetilde g(\alpha)}
 \qquad\text{for all }\alpha\in\F^m .
\]
Bent functions were introduced by Rothaus~\cite{Rot76}, and Carlet~\cite{Car21} is a general reference. The code $C_m(s,x)=g(x+s)$ has as codewords the shifts of $g$. The Hadamard transform of its phase state is the state with amplitude
\[
 2^{-m/2}(-1)^{s\cdot\alpha+\widetilde g(\alpha)}\qquad\text{at }\alpha\in\F^m ,
\]
so when $\widetilde g$ is efficiently computable, the unitary $U_m=H^{\otimes m}D_{\widetilde g}H^{\otimes m}$, where $D_f$ multiplies $\ket x$ by $(-1)^{f(x)}$ for a Boolean function $f$, recovers $s$ exactly. This is R\"otteler's hidden-shift algorithm~\cite{Rot10} with the dual computed rather than queried. For every fixed $x$, the coordinate $s\mapsto g(x+s)$ is a multilinear polynomial in $s$ of degree at most the degree of $g$, since translation by a constant does not increase the degree. Hence a family $(g_m)$ of bent functions of degree at most $d$, one for every positive multiple $m$ of eight, with $g_m$ and $\widetilde g_m$ computable in time polynomial in $m$ uniformly in $m$, gives the code family $C_m(s,x)=g_m(x+s)$ of degree $d$.

An explicit cubic example exists at every $m=8r$ and comes from the Maiorana--McFarland construction~\cite{McF73,Car21}. We write the input as $(u,w)$ with $u,w\in\F^{4r}$ and let $\pi$ act on every group of four bits of $w$ by the permutation
\[
 \pi(w_1,w_2,w_3,w_4)=(w_1+w_2w_3,\ w_2+w_3w_4,\ w_3,\ w_4),
\]
whose inverse is
\[
 \pi^{-1}(w_1,w_2,w_3,w_4)=(w_1+w_2w_3+w_3w_4,\ w_2+w_3w_4,\ w_3,\ w_4).
\]
The function $g_m(u,w)=u\cdot\pi(w)$ is bent, and summing its Fourier transform first over $u$ shows that its dual is
\[
 \widetilde g_m(\alpha,\beta)=\beta\cdot\pi^{-1}(\alpha)\qquad(\alpha,\beta\in\F^{4r}).
\]
Both functions have degree three and polynomial-size circuits that implement their phase oracles exactly, so the codes whose codewords are the shifts of the functions $g_m$ form a code family of degree three.

For the Hadamard code and $\eta=0$, the value $R_s$ of the addressing function of Section~\ref{sec:codes} is determined by a Forrelation, since $(-1)^{R_s}$ is the $3$-fold Forrelation of the functions $(-1)^{L}$, $(-1)^{R}$, and $(-1)^{L}$ in the sense of Aaronson and Ambainis~\cite{AA15}. Indeed, if $L$ is the codeword of $s$, if $\ket L=\ket{C_s}$ is its phase state, and if $D_R$ multiplies $\ket y$ by $(-1)^{R_y}$, then the $3$-fold Forrelation equals $\bra{+}^{\otimes m}D_LH^{\otimes m}D_RH^{\otimes m}D_L\ket{+}^{\otimes m}=\bra L\,H^{\otimes m}D_RH^{\otimes m}\ket L$, and
\[
 \bra L\,H^{\otimes m}D_RH^{\otimes m}\ket L=(-1)^{R_s},
\]
since the Hadamard transform maps $\ket L$ to $\ket s$.

\end{document}